\documentclass[11pt]{amsart}
\usepackage{amsmath,amssymb,times}

\newtheorem{theorem}{{Theorem}}[section]
\newtheorem{proposition}[theorem]{{Proposition}}
\newtheorem{propositiondefinition}[theorem]{{Proposition and definition}}
\newtheorem{lemma}[theorem]{{Lemma}}
\newtheorem{sublemma}[theorem]{{Sub-lemma}}
\newtheorem{corollary}[theorem]{{Corollary}}
\newtheorem{fact}[theorem]{{Fact}}
\newtheorem{factdefinitions}[theorem]{{Fact and definitions}}
\newtheorem{claim}[theorem]{{Claim}}

\theoremstyle{definition}
\newtheorem{definition}[theorem]{{Definition}}
\theoremstyle{remark}
\newtheorem{remark}[theorem]{{Remark}}
\newtheorem{remarks}[theorem]{{Remarks}}

\newcommand{\fin}{\hfill$\square$\medskip}

\renewcommand{\AA}{\mathbb{A}}

\newcommand{\CC}{\mathbb{C}}
\newcommand{\DD}{\mathbb{D}}

\newcommand{\HH}{\mathbb{H}}

\newcommand{\NN}{\mathbb{N}}

\newcommand{\PP}{\mathbb{P}}

\newcommand{\RR}{\mathbb{R}}
\renewcommand{\SS}{\mathbb{S}}

\def\cB{{\mathcal B}}    
\def\cC{{\mathcal C}}  \def\cI{{\mathcal I}}  
\def\cD{{\mathcal D}}   \def\cP{{\mathcal P}} 
   \def\cQ{{\mathcal Q}} 
  \def\cL{{\mathcal L}}

\newcommand{\op}{\operatorname}
\newcommand{\wt}{\widetilde}

\newcommand{\AdS}{\operatorname{AdS}_{3}}
\newcommand{\dS}{\operatorname{dS}_{3}}
\newcommand{\Min}{\operatorname{Min}_{3}}
\newcommand{\ADS}{\operatorname{\AA\DD\SS}_{3}}
\newcommand{\DS}{\operatorname{\DD\SS}_{3}}

\newcommand{\Met}{\operatorname{Met}}
\newcommand{\Teic}{\operatorname{Teich}}

\title[Surfaces with constant K-curvature in 3d spacetimes]{Prescribing Gauss curvature of surfaces in 3-dimensional spacetimes\\ ~\\Application to the Minkowski problem in the Minkowski space}

\author[T. Barbot]{Thierry Barbot$^\star$}
\address{${}^\star$CNRS, UMPA, \'Ecole Normale Sup\'erieure de Lyon.}
\email{Thierry.Barbot@umpa.ens-lyon.fr}

\author[F. B\'eguin]{Fran\c cois B\'eguin$^\dagger$}
\address{${}^\dagger$Laboratoire de
Math\'ematiques, Univ. Paris Sud.}
\email{Francois.Beguin@math.u-psud.fr} 

\author[A. Zeghib]{Abdelghani Zeghib$^\ddagger$}
\address{${}^\ddagger$ CNRS, UMPA, \'Ecole
Normale Sup\'erieure de Lyon.}
\email{zeghib@umpa.ens-lyon.fr}

\thanks{Work supported by ANR project GEODYCOS}

\keywords{}
\subjclass{}

\begin{document}

\setlength{\leftmargini}{16pt}

\sloppy

\begin{abstract} 
We study the existence of surfaces with constant or prescribed Gauss curvature in certain Lorentzian spacetimes. We prove in particular that every (non-elementary) 3-dimensional maximal globally hyperbolic spatially compact spacetime with constant non-negative curvature is foliated by compact spacelike surfaces with constant Gauss curvature. In the constant negative curvature case, such a foliation exists outside the convex core. The existence of these foliations, together with a theorem of C.~Gerhardt, yield several corollaries. For example, they allow to solve the Minkowski problem in $\Min$ 
for datas that are invariant under the action of a co-compact Fuchsian group.
\end{abstract}

\maketitle


\section{Introduction}

K-hypersurfaces --- i.e. hypersurfaces with constant Gauss curvature --- have always played a central role in Riemannian geometry and geometric analysis. It seems however that, unlike CMC-hypersurfaces --- i.e. hypersurfaces with constant mean curvature ---, they have not attracted much attention in mathematical relativity. The main reason might be that they are rigid and rare in dimension $4$ and higher. Nevertheless, there is no reason not to manipulate K-surfaces in 3-dimensional gravity. For example, one can consider the Einstein equation in the gauge defined by a K-surface. 

In the present paper, we study the existence of K-surfaces on 3-dimensional maximal globally hyperbolic spatially compact vacuum spacetimes. Roughly speaking, we prove that each such spacetime admits a unique foliation by K-surfaces. This foliation will provide us with a canonical time function on the spacetime under consideration. 

Let us start by some general introduction, first on the problematic of existence of privileged time functions on spacetimes,  and then on the class of spacetimes we are interested in.


\subsection{Geometric times on MGHC spacetimes} 

\subsubsection*{MGHC spacetimes.} 
Recall that a Lorentz manifold $M$ is \emph{globally hyperbolic} if it admits a \emph{Cauchy hypersurface}, i.e. a spacelike hypersurface which intersects every inextendible causal curve at exactly one point. A classical result of R. Geroch states that the existence of a single Cauchy hypersurface implies the existence of a foliation by such hypersurfaces. More precisely, recall that a \emph{time function} on a spacetime is a submersion  $t: M \to {\mathbb R}$ strictly increasing along every future-oriented causal curves. Geroch result states that a globally hyperbolic spacetime admits a  time function whose levels  are Cauchy hypersurfaces. 

A globally hyperbolic spacetime is said to be \emph{spatially compact} if its Cauchy hypersurfaces are compact. This is equivalent to having a time function which is a proper map, or equivalently, a time function with compact levels. A globally hyperbolic spacetime $(M,g)$ is \emph{maximal} if every isometric embedding of $M$ in another globally hyperbolic spacetime of the same dimension is onto. For  short,  we shall write MGHC for ``maximal globally hyperbolic spatially compact". MGHC spacetimes 
are the tamest Lorentz manifolds from the geometric analysis viewpoint. These spacetimes appear as cosmological models in mathematical Relativity (MGHC spacetimes satisfying the strong positivity energy condition are sometimes called \emph{cosmological spacetimes}).

\subsubsection*{Geometric time functions} 
The simplest examples of MGHC spacetimes are metric products, i.e. spacetimes of the type $M= ({\RR}, -dt^2)\oplus (\Sigma, h)$, where $(\Sigma, h)$ is a compact Riemannian manifold.

Let $(M,g)$ be a MGHC spacetime. From a topological viewpoint,  $M$ is always homeomorphic to the product $\RR\times \Sigma$, where $\Sigma$ is a Cauchy hypersurface of $M$. More precisely, a MGHC spacetime $M$ can always be written as a topological product $\RR\times \Sigma$, where the first projection $T: (t, x) \to t$ is a time function, and $(\{t\} \times \Sigma)_{t\in\RR}$ is a foliation of $M$ by spacelike hypersurfaces. 

>From a metrical viewpoint, $(M,g)$ is in general far from being isometric to a direct product. In the $(t,x)$ coordinates given by the topological splitting $M\simeq \RR\times \Sigma$, the metric on $M$ has the ADM form: $g= -N(t, x) dt^2 \oplus (\omega_t dt+ h_t)$, where $h_t$ is a one parameter family of Riemannian metrics on $\Sigma$, $\omega_t$ a one parameter family of 1-forms on $\Sigma$, $N(t, x)$ a function (called the lapse function),  and $X_t$ the dual (with respect to $h_t$) of $(1/2) \omega_t$ is the shift (non-autonomous) vector field.  (One can roughly say at this stage that a Lorentz structure on a MGHC space with topology ${\RR}\times \Sigma$,  is a curve in ${\mathcal M}et(\Sigma)$, the space of Riemannian metrics of $\Sigma$ up to isotopy).

The topological splitting $M\simeq \RR\times \Sigma$, in particular the time function $T:(t, x) \to t$,  are by no means unique. Nonetheless, it is natural and worthwhile to ask if there are privileged splittings (or time function) for a given MGHC spacetime? 
It is specially exciting to ask what remains from the couple of the orthogonal foliations after perturbation of a direct 
product metric $-dt^2 \oplus h$? 

So the general question we are posing is to produce canonical geometric foliations by Cauchy hypersurfaces, or equivalently geometric time functions, which 
yield a kind of measurement of the default for $(M,g)$ to be a metric product.  Actually, asymptotic behaviour, singularities, shocks, and similar  questions are meaningful only in a ``natural'' coordinates  system (which does not create artificial singularities).

\subsubsection*{Rigid time functions.} Let us first give examples of local geometric conditions on times: 
\begin{itemize}
\item[--]{\it Static time:} $\frac{\partial }{\partial t}$ is a Killing vector field and is orthogonal to ${\mathcal F}$. In this case $g = -N(x)dt^2 \oplus h_0$. \\
\item[--] {\it Static geodesic time:} that is $\frac{\partial}{\partial t}$ has furthermore geodesic trajectories. This characterize the 
direct product case, $g = -dt^2 \oplus h_0$. \\
\item[--] {\it Homothetically static-geodesic case:} $M$ is a warped product
$g= -dt^2 \oplus  w(t)h_0$.
\end{itemize}
However,  all these  situations are very rigid and happen only in a highly symmetric case. Worse, sometimes, a local existence of such times does   imply neither existence  of a global one, nor its uniqueness. 
The reason of rigidity of the last times is that they correspond to solutions of a {\it system} of 
PDE describing the extrinsic geometry of the leaves of the spatial foliation ${\mathcal F}$. However, 
the reasonable situation is that of a single (scalar) geometric PDE, since solving systems involve compatibility (roughly comparable  to the integrability conditions in Frobenius Theorem).

\subsubsection*{F-time functions.} One defines a  (scalar) F-curvature for a 
hypersurface $\Sigma$, where $F$ is is a real valued function on the space of symmetric $n$ by $n$
matrices invariant by orthogonal conjugacy: such a data is equivalent to a map 
$F: {\RR}^n \to {\RR}$ invariant by permutation of the coordinates (here the space $M$ has dimension $1+n$). The F-curvature of $\Sigma$ is the function obtained by evaluating $F$ on the eigenvalues of the second fundamental form of $S$.

The uniformization (or geometrisation) problem can be stated as follows: find a foliation ${\mathcal F}$ on $M$ by Cauchy hypersurfaces each of which has a constant $F$-curvature. We will furthermore require as an additional condition that the $F$-curvature is increasing along causal curves, i.e. that the map associating to $x$ in $M$ the $F$-curvature of the leaf ${\mathcal F}_x$ through $x$ is a time function (that we may call a \emph{F-time function} or a \emph{F-time}). Such functions are automatically unique (see \S\ref{sub.de}).

In the particular case where $F$ is the (arithmetic) mean  of eigenvalues, the $F$-curvature is the \emph{mean curvature}. We are going here to consider the case where $F$ is the opposite of the product of eigenvalues~; in the particular case, the $F$-curvature is called the Gauss-Killing-Kronecker-Lipschitz-curvature or K-curvature.

\subsubsection*{Constant mean curvature versus constant K-curvature}  A CMC-hypersurface is a  hypersurface with a constant mean curvature. A K-hypersurface is a  hypersurface with a constant K-curvature. 

>From a PDE point of view, since one takes a linear sum in the definition of the mean curvature, the equation defining CMC hypersurfaces is the simplest one (among all the equation defining hypersurfaces with constant F-curvature). It is quasi-linear. In all the other cases, the PDE is fully non-linear of Monge-Amp\`ere type. 

Following G. Darboux,  CMC surfaces   are very important in physics, and equally are 
K-surfaces in geometry!\footnote{...``On peut dire que la courbure totale a plus d'importance en G\'eom\'etrie ; comme elle ne d\'epend que de l'\'el\'ement lin\'eaire, elle intervient dans toutes les questions relatives \`a la d\'eformation des surfaces. En Physique math\'ematique, au contraire, c'est la courbure moyenne qui para\^{\i}t jouer le r\^ole pr\'epond\'erant'', G. Darboux, ``Le\c cons sur la th\'eorie g\'en\'erale des surfaces'', Livre V, chapitre II. } 
Despite the marriage  of geometry and physics via the Theory of General Relativity 
(some decades after Darboux),  K-hypersurfaces remain mostly ignored by physicists, who are still investigating CMC hypersurfaces. In fact one of the central and natural question in Relativity is whether a given class of spacetimes admits CMC foliations? 

 One explanation of this CMC success is that the Cauchy problem for Einstein equations (in vacuum)  can be formulated as  a hyperbolic-elliptic well posed PDE system, in  a CMC gauge. Roughly speaking, one can incorporate the CMC condition in the ADM representation of the metric, and set a well posed system, that is if initial data satisfy ``CMC constraints'', then, the system has a solution. This solution is  a Ricci flat Lorentz manifold extending the given initial  Riemannian manifold, together with a {\it canonical} CMC foliation on a neighbourhood of it.  In other words, one has a kind of a CMC flow. 
 
 It is natural to try to extend this consideration to the K-curvature case, that is to write Einstein equation in a K-gauge. This does not seem easy because of the the non-linear coupling of  equations. 
  
One beautiful corollary  of CMC gauges is the regularity of global CMC foliations.  If  one knows that the ambient Lorentz metric is (real) analytic, then the locally defined CMC foliation given by the CMC flow is analytic (by  analyticity  of solutions  
of analytic  hyperbolic systems with analytic initial data).  Therefore,   if one knows that the spacetime 
has a CMC foliation, and moreover,  any CMC hypersurface is a leaf of it, then this foliation must be analytic (since it is locally defined by the CMC flow). One achievement of the present article is to show existence of K-foliations as well as  uniqueness of K-hypersurfaces, but we can not yet deduce its analyticity, since we were not able to treat  Einstein equation in a K-gauge. 

\subsubsection*{Other geometric time functions.} A time function $T:(x,t)\mapsto t$ is \emph{of Gauss type} if the  metric has the form $-dt^2 \oplus g_t$. If a level $\{t\}\times \Sigma$ of $T$ is given, then the others levels  are obtained  by pushing along orthogonal geodesics. The cosmological time (CT {for short}) enjoys this property. It will be one important tool in our study of K-times.

\subsection{MGHC spacetimes of  constant curvature.} 

${}$

In the present paper, we restrict ourselves to MGHC spacetimes of constant curvature. Such spacetimes have locally trivial geometry, being locally isometric to Minkowski space $\Min$, de Sitter space $\dS$ or 
anti-de Sitter space $\AdS$. However, the topology and the global geometry of these spacetimes may be highly nontrivial. Actually, all the questions studied in the present paper concern the global geometry of the spacetimes under consideration.

Focussing our attention on spacetimes with constant curvature is clearly an important restriction. Nevertheless, let us recall that, 
from a physical viewpoint, every $3$-dimensional solution of the vacuum Einstein equation has constant curvature. 
Actually, $3$-dimensional spacetimes with constant curvature have received much attention in the last fifteen years because 
of the role they play in quantum gravity (see for instance \cite{carlip}), and since the discovery of the so-called BTZ 
black holes models (\cite{BTZ}). Moreover, from a purely mathematical viewpoint, 3-dimensional MGHC spacetimes with 
constant curvature are the exact Lorentzian analogs of the Riemannian \emph{hyperbolic ends}, which play a fundamental 
role in 3-dimensional topology. More formally, there exists a duality between hyperbolic ends and 3-dimensional MGHC spacetimes 
of positive constant curvature, which allows to translate some of our results on MGHC spacetimes into results on hyperbolic ends 
(see \S\ref{s.hyperbolic-ends}).

The theory of 3-dimensional MGHC spacetimes of constant curvature naturally splits into two cases:
the spacetimes whose Cauchy surfaces have genus $0$ or $1$ are called \emph{elementary}, whereas
the spacetimes whose Cauchy surfaces have genus at least $2$ are called \emph{non-elementary}. As suggested by the terminology, 
the global geometry of elementary spacetimes is much easier to understand than those of non-elementary spacetimes. In some sense, 
elementary spacetimes can be considered as ``particular" or ``exact" solutions of the Einstein equation, for which all geometrical problems can be solved explicitely ``by hand" (see e.g.~\cite[\S10.2 and \S10.3]{ABBZ2}). Nevertheless, since tools are quite different, we will restrict ourselves to non-elementary spacetimes.

All the spacetimes we shall consider in the sequel are not only time-orientable, but actually time-oriented. If $M$ is a non-elementary 3-dimensional MGHC spacetime of constant curvature $\Lambda\geq 0$, then $M$ is always either past complete (all past directed causal geodesic rays are complete), or future complete. Reversing the time-orientation if necessary, we will always assume we are in the second case. Note that a non-elementary 3-dimensional MGHC spacetime of constant curvature $\Lambda<0$ is always neither past nor future complete.


 \subsection{Some  definitions: K-surfaces, K-slicings, K-times.}
 \label{sub.de} 
 
 ${}$
 
 Let $(M,g)$ be a (time-oriented) 3-dimensional spacetime. 
 Given a spacelike surface $\Sigma$ in $M$, the second fundamental form $k_x^\Sigma$ of 
 $\Sigma$ at $x$ is defined by $k_x^\Sigma(X, Y) = -<\nabla_X n,Y>$ where $n$ is the future oriented unit normal vector of $\Sigma$. The \emph{K-curvature} (or Gauss-Killing-Kronecker-Lipschitz curvature) of $\Sigma$ at $x$ is $\kappa^\Sigma(x) = - \lambda_1(x)\lambda_2(x)$, where $\lambda_1(x)$ and $\lambda_2(x)$ are the principal curvatures of $\Sigma$ at $x$, i.e. the eigenvalues of the second fundamental form $k_x^\Sigma$. Note that, in the case where the spacetime $M$ has constant (sectional) curvature $\Lambda$, the Gaussian (i.e. sectional) curvature of $\Sigma$ is $R^\Sigma = \Lambda + \kappa^\Sigma$, and its scalar curvature is $2R^\Sigma$.

A \emph{K-surface} in $M$ is a spacelike surface with constant K-curvature. Observe that, in the case where $M$ has constant curvature, being a K-surface is equivalent to having constant Gaussian curvature.

>From now on, we assume $(M,g)$ to be globally hyperbolic and spatially compact.
We define a \emph{K-slicing} of $M$ as a foliation of $M$ by compact K-surfaces. Note that a K-slicing of $M$ is always a trivial foliation, all the leaves of which are Cauchy surfaces of $M$. Equivalently, the leaves of a K-slicing of $M$ are always the level sets of a certain time function.

\begin{remark}
 ({\it Slicing versus Foliation}). Our choose of the term ``slicing'' instead of the familiar one ``foliation'' (in the mathematical literature) is to emphasize the fact that we are  dealing   only with topologically trivial foliations, e.g. with all leaves compact. 
However, one could imagine to study general K-foliations   with non-trivial dynamics.  Although  we will 
not investigate   the question here, we guess that, in our context of MGHC spacetimes of constant curvature, any K-foliation   is in fact a slicing. A similar question about 
CMC-foliations of the 3-dimensional Euclidean space is handled in \cite{Meeks}.
 \end{remark}

A K-slicing need not be unique in general. In order to get some uniqueness, we need to define a more restrictive notion.  A $C^2$ spacelike surface $\Sigma$ in $M$ is said to be \emph{convex} (resp. \emph{strictly convex}) if it has non-positive (resp. negative) principal curvatures. Similarly, $\Sigma$ is \textit{concave} or \emph{strictly concave} if the principal curvatures are non-negative or positive. 
A \emph{K-time} on $M$ is a time function $\tau: M \to \RR$ such that, for every $a\in \tau(M)$, the 
level set $\tau^{-1}(a)$ is a compact strictly convex K-surface with K-curvature equal to $a$. 
Of course, the level sets of any K-time define a K-slicing of $M$. Nevertheless, the K-slicings defined by K-times are quite specific since all the leaves have negative principal curvatures, and since the K-curvature of the leaves increases with time. Note that the range $\tau(M)$ of a K-time is always included in $(-\infty,0)$.

The maximum principle implies that K-times are unique. More precisely, if $M$ admits a K-time $\tau:M\to\RR$, then, for every $a\in\tau(M)$, the only compact K-surface of K-curvature $a$ in $M$ is the
level set $\tau^{-1}(a)$. In particular, the only K-slicing on $M$ is the one defined by the level sets of $\tau$ (see \S\ref{ss.maximum}).

\subsection{Purpose of the paper.}

${}$ 

In the present paper, we are going here to consider  the following questions:
\begin{itemize}
\item[--] the existence of K-time (or K-slicings) in 3-dimensional MGHC spacetimes of constant curvature, 
\item[--] the existence of Cauchy surfaces of prescribed K-curvature in 3-dimensional MGHC spacetimes of constant curvature, 
\item[--] the Minkowski problem in the 3-dimensional Minkowski space.
\end{itemize}
The results concerning the last two items are essentially application of the first one.


\section{Statements of results}

\subsection{K-slicings of MGHC spacetimes with constant curvature.} 

${}$

The following is our main result:

\begin{theorem} 
\label{theo.main} 
Let $M$ be a $3$-dimensional non-elementary MGHC spacetime with constant curvature $\Lambda$. If $\Lambda\geq 0$, reversing the time orientation if necessary, we assume that $M$ is future complete.
\begin{itemize}
\item If $\Lambda\geq 0$ (flat case or locally de Sitter case), then $M$ admits a unique K-slicing. The leaves of this slicing are the level sets of a K-time ranging over $(-\infty, -\Lambda)$. 
 
\item If $\Lambda<0$ (locally anti-de Sitter case),  then $M$ does not admit any global K-slicing, but each of the two connected component of the complement of the convex core\footnote{About the notion of convex core, see \S \ref{sub.core}.} of $M$ admits a unique K-slicing. The leaves of the K-slicing of the past of the convex core are the level sets of a K-time ranging over $(-\infty, 0)$.  The leaves of the K-slicing of the future of the convex core are the level sets of a reverse K-time\footnote{A reverse K-time on $N$ is a function $\tau:N\to\RR$ which is strictly decreasing along every future oriented causal curve, and such that, for every $a$, the level set $\tau^{-1}(a)$ is a compact locally strictly concave surface of constant K-curvature equal to $a$.} ranging over $(-\infty, 0)$.  
\end{itemize}
\end{theorem}

Let us make a few comments on this result.

\subsubsection*{Gaussian  curvature.}
Since the Gaussian  curvature of a surface is $R = \Lambda + \kappa$, the Gaussian  curvature of the leaves of the K-slicings provided by 
Theorem~\ref{theo.main} varies in $(-\infty, 0)$ when $\Lambda\geq 0$, and in 
$(-\infty, \Lambda)$ when $\Lambda<0$. 

\subsubsection*{CMC times.} 
It is our interest on CMC-times that led us to extend our attention  to more general geometric times. Existence of CMC times on MGHC spacetimes of constant non-positive curvature and any dimension was proved in \cite{ABBZ1, ABBZ2}. For spacetimes locally modelled on the de Sitter space, there are some restrictions but not in dimension 3. 

\subsubsection*{Regularity.}
The slicings provided by Theorem~\ref{theo.main} are
  continuous. It follows from their uniqueness (i.e. they are    canonical). Extra smoothness is not automatic (e.g. the cosmological time is $C^{1,1}$, but not $C^2$). 
   Here, we can hope that our K-slicings are 
 (real) analytic, and even more, they depend analytically  on the spacetime (within the space of MGHC spacetimes of curvature $\Lambda$ and fixed topology). All this depends on consideration of Einstein equations in a K-gauge.

\subsubsection*{Non-standard isometric immersions of $\HH^2$ in $\Min$} \label{nonstandard}
It was observed by Hano and Nomizu \cite{Nomizu}, that the hyperbolic plane $\HH^2$ admits  non standard isometric immersions in the Minkowski space $\Min$ (i.e. different from the  hyperbola, up to a Lorentz motion). 

Theorem \ref{theo.main}, applied in the flat case, yields a K-slicing with (exactly) one  isometric  copy of $\HH^2$ in $\RR^{1, 2}$, invariant by the holonomy group ${\bf \Gamma} \subset SO(1, 2) \ltimes \RR^3$ (since the curvature varies from $-\infty$ to 0). This isometric immersion is different from the standard one as soon as ${\bf\Gamma}\not\subset SO(1, 2) \ltimes \{0\}$. 

Observe  that, in higher dimension,  ${\HH}^n$ is rigid in $\hbox{Min}_{n+1}$.  Indeed,  the theory of rigidity   of submanifolds in the Euclidean space extends straightforwardly to the Minkowski space. From this theory, a hypersurface with a second fundamental form of rank $\geq 3$ is rigid. By Gauss equation, any isometric immersion of a hyperbolic space in a Minkowski space is non-degenerate, and hence has the required condition on the rank, whence $n \geq 3$.

\subsubsection*{Spacetimes as curves in the Teichm\"{u}ller space} 
For the sake of simplicity, let us restrict ourselves here to the case $\Lambda = 0$, i.e. that of flat spaces. Denote by $\Sigma$ the abstract topological surface homeomorphic to a Cauchy surface of $M$, and by ${\mathcal M}et(\Sigma)$ the space of Riemannian metrics on $\Sigma$ up to isotopy. Any time function of $M$ gives rise to a family $(\Sigma_t, g_t)$ of Riemannian metrics on $\Sigma$ (well-defined up to isotopy) parametrized by the given time. We have in particular, associated to our three favourite, K, CMC and CT times, three curves in ${\mathcal M}et(\Sigma)$: 
$$\Met^K\; \hbox{(resp.}\; \Met^{cmc}\;  \hbox{and}\;  \Met^{ct}):\;   t \to g_t \in {\mathcal M}et(\Sigma)$$ 

Now, let $Teich(\Sigma)$ be the Teichm\"{u}ller space of $\Sigma$, i.e.  the space of hyperbolic metrics (of curvature$-1$) on $\Sigma$, up to isotopy. Then, we have associated curves:
$$\Teic^K\; \hbox{(resp.}\; \Teic^{cmc}\;  \hbox{and}\;  \Teic^{ct}):\;   t \to [g_t] \in Teich(\Sigma)$$ 
Thus, $\Met(t)$ is the Riemannian metric of the $t$-level, and 
$\Teic(t)$ is its underlying complex structure.  

Let us  note the following facts (but we cannot give here a complete overview):
\begin{enumerate} 
\item  The curves $\Met^K,\; \Met^{cmc},\; \Met^{ct},\; \Teic^K,\; \Teic^{cmc},\;\Teic^{ct}$ encode the metric properties and the geometry of the spacetime $M$. 
\item These curves all coincide exactly  when the spacetime $(M,g)$ is static (i.e. its universal cover is the  solid lightcone). 
 \item In the vein of V. Moncrief's work on the ``reduction of Einstein equations to the Teichm\"{u}ller space'' in a  CMC gauge in dimension $2+1$ \cite{Moncrief}, one can  show in general that all the above  curves in $Teich(\Sigma)$ have canonical lift in  $T^*(Teich(\Sigma))$ or alternatively $T (Teich(\Sigma))$. There, they define (semi-)flows. In the CT-case, this is nothing but the grafting \cite{bonwick}.
 \item  It is an essential question to study the behaviour of these curves in both ${\mathcal M}et(\Sigma)$ and  $Teich(\Sigma)$ when the parameter tends  to one or the other  extremity  of the existence interval. A delicate point is that there are many compactifications for $Teich(\Sigma)$
 and also many notions of  convergences for sequences 
 in ${\mathcal M}et(\Sigma)$. We have for instance,  equivariant Gromov topology, convergence 
 of spectrum, where in general the limit is a real tree. For existing  asymptotic  study,  
 see  for instance   \cite{bonthese} in the  CT-case and \cite{Lars} in the CMC-case, and a forthcoming paper. 
 \item One advantage of the K-slicing is that $\Teic^K(t)$ is directly given by $\Met^K(t)$, that is we don't need to uniformize it since it has constant curvature $t$ (by definition of the parameter). In this case, all notions of convergence 
 in ${\mathcal M}et (\Sigma)$ (up to scaling) and 
 in $Teich(\Sigma)$, coincide.
 \end{enumerate}

\subsubsection*{Affine foliations} 
It is interesting to compare our results (existence of CMC-slicings and K-slicings) with
similar results in the context of (equi-)affine geometry, in particular with the beautiful
theory of affine spheres (see \cite{lizhao}): for every hypersurface $\Sigma$ in the
affine space $\RR^{n+1}$, endowed with a parallel volume form, one can define its affine principal 
curvatures $\lambda_1, \lambda_2, \ldots, \lambda_n$. The hypersurface is an affine 
hypershere if all these affine principal curvatures are
equal and constant along $\Sigma$. It is equivalent to require that all the affine normals either
intersect at one point, or are mutually parallel. If $\lambda_1 = \lambda_2 = \ldots = \lambda_n < 0$,
the affine sphere is hyperbolic. Calabi's conjecture states that:

\begin{enumerate}
\item Every hyperbolic affine sphere which is complete for its ``affine'' metric is asymptotic 
to the boundary of a proper open convex cone,
\item Conversely, for every $\lambda < 0$, every proper open convex contains a unique 
hyperbolic affine sphere of affine principal curvatures $\lambda$.
\end{enumerate}

The first part has been solved by Cheng and Yau under the hypothesis that the affine sphere
is complete for an Euclidean metric on $\RR^{n+1}$, hypothesis removed later by Li.
Most interesting for us is the second part, proved by Sasaki and Gigena. It shows
that any proper open cone admits a natural foliation by hyperbolic affine spheres,
which can be considered as hypersurfaces of (affinely) constant scalar curvature, 
or as well of constant mean curvature. 

For more details, see \cite[Chapter 2]{lizhao}, where the authors
also discuss the method of C. P. Wang for constructing all hyperbolic affine 2-spheres 
which admit the action of a discrete subgroup of the equiaffine 
group with compact quotient.


\subsection{Duality hyperbolic-de Sitter, K-slicings of hyperbolic ends.}
\label{duality.statements}

${}$

The duality between the de Sitter space $\dS$ and the hyperbolic space $\HH^3$ (see \S \ref{s.hyperbolic-ends}) will  allow us to deduce the existence of K-surfaces in hyperbolic ends from the existence of K-surfaces in MGHC spacetimes of constant positive curvature. So we will be able to recover the following result of F. Labourie:

\begin{theorem}[\cite{Lab}]
\label{theo.ends} 
A 3-dimensional hyperbolic end possesses a K-slicing with K-curvature ranging in~$(0, 1)$. 
\end{theorem}

Some explanations on hyperbolic ends can be found in \S\ref{s.hyperbolic-ends}. Note that, in the Riemannian setting, the K-curvature of a surface is just the product $\lambda_1\lambda_2$ of the principal curvatures.

\begin{remarks}\-
\begin{enumerate}
\item It was observed by Mess that the $\dS$-case of our main Theorem~\ref{theo.main} can be deduced from Labourie's Theorem, using the duality between $\dS$ and $\HH^3$. 
\item Our proof here of Theorem~\ref{theo.main} in the $\dS$-case is completely different from Labourie's one for hyperbolic ends. 
Actually, we think that our method gives the natural framework for approaching many results of  existence ``geometric" surfaces or foliation, 
even in the hyperbolic setting. 
 \end{enumerate}
 \end{remarks}


\subsection{Surfaces with prescribed K-curvature.} 

${}$

The barriers method (see  \S \ref{ss.local-foliation}) allows one to find a surface with constant K-curvature $\kappa$ in between two surfaces with (non-constant) K-curvatures bounded respectively from below and above by $\kappa$. This is an important ingredient in our proof of Theorem~\ref{theo.main}. This method generalizes to the case where $\kappa$ is not a constant any more, but rather a function (see \cite{Ger1} or section~\ref{s.prescribed}). Using this generalisation and Theorem~\ref{theo.main}, we will get the following existence result:

\begin{theorem}
\label{cor.prescri}
Let $M$ be a $3$-dimensional non-elementary MGHC spacetime with constant curvature $\Lambda$. If $\Lambda\geq 0$, reversing the time-orientation if necessary, we assume that $M$ is future complete. Let $f$ be a smooth function on $M$ admitting a range contained in a compact interval $[a,b]\subset ]-\infty\,,\,\min(0,-\Lambda)[$.  Then, there is a Cauchy surface $\Sigma$ is $M$ with K-curvature $\kappa^\Sigma(x)=f(x)\mbox{ for every }x \in \Sigma.$
\end{theorem}

In particular: 
 
\begin{corollary}
\label{coro.prescri}
Let $M$ be as above. Let $\Sigma_0$ be the abstract topological surface homeomorphic to the Cauchy surfaces of $M$, and let $f_0$ be a smooth real-valued function on $\Sigma_0$, such that the range of $f_0$ is a contained in a compact interval $[a,b]\subset ]-\infty\,,\,\min(0,-\Lambda)[$. Then, there exists an embedding of $\phi:\Sigma_0\hookrightarrow M$ such that $\Sigma=\phi(\Sigma_0)$ is a Cauchy surface in $M$, and the K-curvature of $\Sigma$ at $\phi(x)$ is equal to $f_0(x)$ for every $x\in \Sigma_0$.
\end{corollary}

\subsection{The Minkowski problem.} 

${}$

Let us recall the classical formulation of the Minkowski problem (see for instance \cite{Yau, Ger2, Lab, Nirenberg}). If $S$ is a closed convex smooth surface in the Euclidean space $\RR^3$, then its Gauss map $\nu_S: S \to \SS^2$ is a diffeomorphism, and one can consider the map $K^S:=\kappa^S  \circ \nu_S^{-1}: \SS^2 \to (0, +\infty)$, where $\kappa^S$ is the K-curvature of $S$. The Minkowski problem consists in characterizing the functions on the sphere $\SS^2$ which have the form $K^S$ for some surface $S$. 

This problem can be transposed in the Lorentzian setting, by replacing the Euclidean space $\RR^3$ by the Minkowski space $\Min$, requiring the surface $S$ to be spacelike (but not compact!), and replacing the sphere $\SS^2$ by the hyperbolic space $\HH^2$. Using Theorem~\ref{theo.main} and a barrier theorem of Gerhardt (\cite{Ger1}), we are able to solve the Minkowski problem in $\Min$ in the particular case where the function $K^S$ is invariant under a co-compact Fuchsian group $ \Gamma \subset \mbox{SO}(1,2)=\mbox{Isom}(\HH^2)$:

\begin{theorem}  
\label{Minkowski} 
Let $\Gamma$ be a co-compact Fuchsian subgroup of $SO(1,2)$, and $f:\HH^2\to (-\infty,0)$ be a $\Gamma$-invariant smooth function. Then there exists a  strictly convex spacelike surface $S$ in $\Min$ such that $f=\kappa^S\circ (\nu^S)^{-1}$, where $\nu^S:S\to\HH^2$ is the Gauss map of $S$ and $\kappa^S:S\to(0,+\infty)$ is the K-curvature (= Gauss curvature) of $S$. Moreover, if $\mathbf{\Gamma}$ is a subgroup of $\mbox{SO}(1,2)\ltimes\RR^3=\mathrm{Isom}(\Min)$ which projects bijectively on $\Gamma$, then there exists a unique such (convex) surface $S$ which is  $\mathbf{\Gamma}$-invariant.
\end{theorem}

A few comments on the Minkowski problem in $\Min$:
\begin{enumerate}
\item There is no uniqueness in general  for the Minkowski problem in $\Min$. For example, for $f = - 1$, any isometric copy of ${\HH}^2$ in $\Min$ is a solution of the Minkowski problem.
The  uniqueness in Theorem~\ref{Minkowski} strongly relies on the convexity and ${\bf \Gamma}$-invariance hypotheses.
\item Despite the  obvious and natural interest of the Minkowski problem in the Lorentzian setting, there seems to be very little work on this problem. We know essentially two substantial contributions: \cite{Li} by A. M. Li and \cite{Schoen} by B. Guan, H-Y Jian and R. Schoen. In the last paper, the authors proved a result similar to ours (equally in dimension $2+1$),  with the invariance condition replaced by prescribing the asymptotic behaviour at infinity. 
\item There are generalisations of the Minkowski problem in other directions, still in a Riemannian setting, but for  
ambient spaces different from ${\RR}^n$. The point here is to define a substitute of the Gauss map. This is indeed possible for universal spaces of constant curvature ${\SS}^{n+1}$
and ${\HH}^{n+1}$. One  can quote here recent works by C. Gerhardt which solve the corresponding Minkowski problem 
(for convex bodies)\cite{Ger3, Ger2},
and a previous work of 
F. Labourie on an  equivariant Minkowski problem for 
surfaces in the hyperbolic space ${\HH}^3$ \cite{Lab}. 
Finally, the Minkowski problem admits generalizations to other 
curvatures (see for instance \cite{Guan-Guan, Sch1}).
\end{enumerate}

\subsection{Related works}

${}$

The general mathematical framework unifying our contributions here is that of prescribing (extrinsic) curvatures of hypersurfaces in $M$. Such a 
hypersurface $N$   
  is sometimes called of Weingarten type,    i.e. it is an 
  F-hypersurface in the sense that:   $ F (\overrightarrow{\lambda}(x)) = F(\lambda_1(x), \ldots,  \lambda_n(x)) = f(x)$, where $f$ is a given function on $N$, and $F$   is a function of the principal curvatures.  For instance, one asks the following general questions about them:   \\ 
-- A Dirichlet problem, \\
-- A Dirichlet problem at infinity,  also   said ``entire hypersurfaces problem'' (i.e. existence of complete  hypersurfaces with a prescribed asymptotic behaviour), \\ 
--  Existence of F-slicings. This requires $M$ to be topologically trivial, or that the slicing is defined on an end of $M$, \\
-- A natural generalisation consists in considering curvature functions $F( \overrightarrow{\lambda}, \nu)$, i.e  $F$ also depends on a normal vector $\nu$. \\
-- Finally,  regarding any of these questions, one specifies  whether $M$ is Riemannian or Lorentzian.  

As  said previously, there is a wide literature  on the CMC case (also called the H-curvature).  Let us quote   some recent achievements concerning the K-curvature: \\ 
- The Dirichlet problem for the K-curvature in the Minkowski space was studied in particular  by P. Delano\"{e} \cite{Delanoe} and 
B. Guan \cite{Guan}. \\
- Entire K-hypersurfaces were  
studied by  B. Guan, H-Y Jian and R. Schoen \cite{Schoen}. (The corresponding CMC case was considered by A. Treibergs \cite{treibergs}, and that on the ``finite'' Dirichlet problem, by Bartnik and Simon \cite{Bartnik}). \\
- In \cite{Bayard1, Urbas} P. Bayard and afterwards    J. Urbas solved the Dirichlet problem 
  for the scalar curvature 
in the Minkowski space  of  any dimension.
 Remark  that for  our dimension $2+1$, the scalar an K-curvatures coincide. Also \cite{Bayard2} solved the entire hypersurface problem for the scalar curvature. \\
- C Gerhardt and O. Schn\"{u}rer \cite{Ger0, Sch2} 
proved criteria for the solvability of the Dirichlet problem on general Lorentz manifolds, and for various curvature functions, \\
- Existence of K-slicings of hyperbolic ends in 3 dimension 
was proved by F. Labourie (see \S \ref{duality.statements} ), and generalised 
by  G. Smith for  higher dimensional  quasi-Fuchsian ends, where the K-curvature is replaced by a variant, governed by   the   ``special Lagrangian'' equation \cite{Smith}, \\ 
- Recently, R. Mazzeo and F. Pacard \cite{Maz-Pac} considered manifolds that are merely asymptotically hyperbolic and admitting a conformal compactification. They prove existence of  various  geometric foliations near their boundary.


\section{Ingredients of the proofs}
\label{s.ingredients}

Before going into the details, we want to the main ingredients of the proofs of our results. 
Let $(M,g)$ be a non-elementary 3-dimensional MGHC spacetime with constant curvature $\Lambda$. 
If $\Lambda\geq 0$, reversing the time-orientation if necessary, we can (and we do) assume that $M$ is future complete. To prove our main Theorem~\ref{theo.main}, we have to construct a K-slicing on $M$ (or on the complement of the convex core of $M$ if $\Lambda<0$). Here are the main ingredients of this construction:


\subsection{Existence of barriers}
\label{ss.local-foliation}

${}$

A very general principle states that the existence of a surface with prescribed curvature should follow from the existence of so-called \emph{barriers}. In our context, the suitable result was proved by C. Gerhardt:

\begin{definition}
\label{def.kappa-barrier}
For $\kappa\in\RR$, a \emph{pair of $\kappa$-barriers} is a pair of disjoint strictly convex Cauchy surfaces $\Sigma^-,\Sigma^+$ in $M$ such that:
\begin{enumerate}
\item[a.] $\Sigma^-$ is in the past of $\Sigma^+$, 
\item[b.] the K-curvature of $\Sigma^-$ is bounded from above by $\kappa$, 
\item[c.] the K-curvature of $\Sigma^+$ is bounded from below by $\kappa$. 
\end{enumerate}
\end{definition}

\begin{theorem}[Gerhardt, \cite{Ger0}\footnote{The main result of~\cite{Ger0} is stronger than the result stated above (see~\S\ref{s.prescribed}).} ]
\label{t.barriers}
Given a real number $\kappa<-\Lambda$, if $(M,g)$ admits a pair of $\kappa$-barriers $(\Sigma^-,\Sigma^+)$, then $(M,g)$ admits a strictly convex Cauchy surface $\Sigma$ with constant  K-curvature $\kappa$. Moreover, the Cauchy surface $\Sigma$ is in the future of $\Sigma^-$ and in the past of $\Sigma^+$.
\end{theorem}

\begin{remark}
\label{r.sign-convention-Gerhardt}
The sign convention of Gerhardt for the K-curvature is the opposite of ours (this is the reason why the main result of~\cite{Ger0} asserts  that one can find a Cauchy surface with constant Gauss curvature $k$ for any $k>0$ provided that there is a pair of barriers). 

Also note Gerhardt uses the \emph{past} directed unit normal vector to define the principal curvatures; it follows that a surface which is  called convex in~\cite{Ger0,Ger1,Ger2} is called concave here, and vice-versa. This is not a problem since all the result are valid both for convex and concave surfaces (one needs to be careful with the definition of the barriers).
\end{remark}

We will prove the following result:

\begin{theorem}
\label{th.asymptotic-barriers}
Assume $\Lambda$ is non-negative\footnote{See\S\ref{s.barriers} for some statements in the case $\Lambda<0$.}. There exists $\epsilon>0$ such that there exists a pair of $\kappa$-barriers for every $\kappa\in (-\Lambda-\epsilon,-\Lambda)$. 
\end{theorem}

This result will be obtained by using previously known results on the existence of locally strictly convex 
Cauchy surfaces in flat spacetimes, and some estimates on the behaviour of the K-curvature when one 
pushes such a Cauchy surface along the orthogonal geodesics. Theorem~\ref{th.asymptotic-barriers} 
together with Gerhardt's theorem show the existence of a convex Cauchy surface $\Sigma_\kappa$ 
with constant K-curvature $\kappa$ in $M$ for every $\kappa\in (-\Lambda-\epsilon,-\Lambda)$. Using 
classical arguments and some informations on our barriers, we will prove that the family of Cauchy surfaces $(\Sigma_\kappa)_{\kappa\in (-\Lambda-\epsilon,-\Lambda)}$ is a foliation a neighbourhood of the future end of $M$. This will provide us with a ``local K-slicing" on a neighbourhood  of the future end of $M$ (in the case $\Lambda\geq 0$).


\subsection{Systole and distance to the past singularity}

${}$

Let $\tau$ be the \emph{cosmological time} on $M$ (see  \S\ref{s.cosmological-time} for more details). A major (and the most original) ingredient of our proof of Theorem~\ref{theo.main} is the following 
result, which relates the systole of a Cauchy surface $\Sigma$ in $M$ and the ``distance from $\Sigma$ to the past singularity of $M$":

\begin{theorem}
\label{th.systole}
For every $\epsilon > 0$, there exists a constant $\alpha>0$ such that, for any Cauchy surface $\Sigma$, if $\inf\tau_{|\Sigma}$ is smaller than $\alpha$ then the systole of  $\Sigma$ is smaller than $\epsilon$.
\end{theorem}

This statement should be understood as follows: \emph{if the systole of a Cauchy surface $\Sigma$ in $M$ is not too small, then no point of $\Sigma$ is close to the past singularity of $M$}. 
We will use this to prove that a sequence $(\Sigma_n)_{n\in\NN}$ of Cauchy surfaces with K-curvature bounded away $-\infty$ remains far from the initial singularity (see~\S\ref{ss.sequence-constant}).

The proof of Theorem~\ref{th.systole} strongly relies on some fine knowledge of the geometry of MGHC spacetimes with constant curvature. 
We think that this Theorem~\ref{th.systole} is not only a crucial step in the proof of our main result, but is also interesting in its own right. 
It can be used in various type of situations to prove that a sequence of Cauchy surfaces is relatively compact.


\subsection{Decreasing sequences of convex Cauchy surfaces} 

${}$

Another ingredient of our proof of Theorem~\ref{theo.main} is the fact that the limit of a decreasing sequence of convex Cauchy surfaces is always ``spacelike". More precisely: 

 \begin{theorem} 
 \label{th.spacelike}  
 Let $(\Sigma_n)_{n\in\NN}$ be a sequence of convex Cauchy surfaces\footnote{or generalized Cauchy surfaces, see \S\ref{sub.achrodansgh}.}  in $M$. 
 Assume that this sequence is decreasing (i.e. $I^+(\Sigma_{n+1})\supset I^+(\Sigma_n)$ for every $n$), and that the set  $\Omega = \bigcup_{n\geq 0}  I^+(\Sigma_n)$ 
 is not the whole spacetime $M$. Let $\Sigma_\infty = \partial \Omega$ (note that the $\Omega$ is a locally geodesically convex set, and thus $\Sigma_\infty$ is a convex topological surface). Then all the support planes of $\Sigma_\infty$ are spacelike. 
 \end{theorem}

Note that $\Sigma_\infty$ is not a Cauchy surface in general. An easy (but important) corollary of Theorem~\ref{th.spacelike} is the fact that a decreasing sequence of convex Cauchy surfaces is  always "uniformly spacelike". To make this precise, let us denote by $T^{-1}M$ the set of all couples $(x,P)$ where $x$ is a point of $M$ and $P$ is a totally geodesic spacelike plane passing through $x$. The set $T^{-1}M$ is naturally identified with the subset of the tangent bundle $TM$ of $M$ made of all couples $(x,v)$ where $v$ is a future-directed tangent vector of norm $-1$ (the identification is of course through the correspondance between a totally geodesic spacelike plane and its future-directed unit normal vector). This endows $T^{-1}M$ with a natural topology. Clearly, $T^{-1}M$ is not compact (a sequence of spacelike planes may converge to a null plane!).

\begin{definition}
\label{d.uniformly-spacelike}
A sequence of spacelike surfaces $(\Sigma_n)_{n\in\NN}$ is  \emph{uniformly spacelike} if, for every sequence  $(x_n)_{n\in\NN}$ with $x_n\in \Sigma_n$:
\begin{itemize}
\item[--] either the sequence $(x_n)_{n\in\NN}$ escapes from any compact subset of $M$,
\item[--] or the sequence $(x_n,P_n)_{n\in\NN}$, where $P_n$ is the tangent plane of the surface $\Sigma_n$ at $x_n$, stays in a compact subset of~$T^{-1}M$.
\end{itemize} 
\end{definition}

\begin{corollary}
\label{c.uniformly-spacelike}
Let $(\Sigma_n)_{n\in\NN}$ be a sequence of convex Cauchy surfaces in $M$. Assume that this sequence is decreasing, and that the set  $\Omega = \bigcup_{n\geq 0}  I^+(\Sigma_n)$ is not the whole spacetime $M$. Then $(\Sigma_n)_{n\in\NN}$  is uniformly spacelike.
\end{corollary}


\subsection{Sequences of convex Cauchy surfaces with constant K-curvature}
\label{ss.sequence-constant}

${}$

Using the two results discussed above, we will get a quite precise description of the possible asymptotic behaviour of a decreasing sequence of convex Cauchy surfaces with constant K-curvature:

\begin{theorem} 
\label{th.sequence-constant}
Let $(\Sigma_n)_{(n \in \NN)}$ be a sequence of strictly convex Cauchy surfaces in $M$, such that, for every $n\in\NN$, the surface $\Sigma_n$ has constant K-curvature $\kappa_n$. Assume that this sequence is bounded away from the future end of $M$ (i.e. the cosmological time is bounded from above on $\bigcup_n \Sigma_n$). 

If $\kappa_n\to \kappa$ with $-\infty<\kappa<  \,\min(0,-\Lambda)$, then the sequence $(\Sigma_n)_{n\in\NN}$ is precompact in the $C^\infty$ topology. In particular, $(\Sigma_n)$ is bounded away from  the past  singularity of $M$ (there exists a Cauchy surface $\Sigma$  such that  $\Sigma_n \subset I^+(\Sigma)$ for all $n$), and there exists a subsequence of  $\Sigma_n$ converging to a smooth surface $\Sigma_\infty$ with  constant K-curvature $\kappa$. 

On the contrary, if $\kappa_n \to -\infty$, then $\Sigma_n$ is covering, \emph{that is} the union over $n\in\NN$ of the sets $I^+(\Sigma_n)$ equals the whole spacetime $M$.
\end{theorem}

The proof of this theorem involves several ingredients~:
\begin{itemize}
\item[--] an easy lemma which provides a uniform upper bound for the diameter of the $\Sigma_n$'s,
\item[--] Margulis Lemma which tells that a uniform upper bound for the diameter of the $\Sigma_n$'s yields some uniform lower bound for the systole of the $\Sigma_n$'s (in the case where the sequence $(\kappa_n)_{n\in\NN}$ is  bounded from below),
\item[--] Theorem~\ref{th.systole} which implies that the $\Sigma_n$'s remain far from the initial singularity (in the case where he sequence $(\kappa_n)_{n\in\NN}$ is  bounded from below),
\item[--] Theorem~\ref{th.spacelike} which roughly says that any limit point $\Sigma_\infty$ of the sequence $(\Sigma_n)_{n\in\NN}$ is uniformly spacelike,
\item[--] a result of Schlenker which says that a $C^0$ limit of convex spacelike surfaces either is a smooth surface or contains a complete geodesic ray,
\item[--] the geometric description of non-elementary MGHC spacetimes of constant curvature which entails the fact that such spacetimes never contain any complete geodesic ray.
\end{itemize}


\subsection{Perturbation of Cauchy surfaces with constant K-curvature}

${}$

 A classical calculation yields:

\begin{proposition}
\label{p.perturbation}
Assume that $(M,g)$ admits a strictly convex Cauchy surface $\Sigma$ with constant K-curvature $\kappa<-\Lambda$.  Then, one can find a strictly convex Cauchy surface $\Sigma^-$ in the past of $\Sigma$ such that the K-curvature is strictly bounded from above by $\kappa$ (i.e. such that $(\Sigma^-,\Sigma)$ is a pair of $\kappa'$-barriers for every $\kappa'<\kappa$ such that $\kappa'$ close enough to $\kappa$).
\end{proposition}

The surface $\Sigma^-$ can be obtained simply by pushing slightly $\Sigma$ along the geodesics orthogonal to $\Sigma$. Proposition~\ref{p.perturbation} and Theorem~\ref{t.barriers} show that, if $M$ admits a convex Cauchy surface $\Sigma$ with constant K-curvature $\kappa<-\Lambda$, then some neighbourhood of $\Sigma$ in the past of $\Sigma$ is foliated by convex Cauchy surfaces 
with constant K-curvature. 
Together with Theorem~\ref{th.sequence-constant}, this will allow us to prove that any local K-slicing on $M$ can be extended towards the past untill it reaches the past singularity of $M$ (i.e. until one gets a K-slicing on a neighbourhood of the past end of $M$). Since, in the case $\Lambda\geq 0$ (flat and de Sitter case), we have already explained how to get a local K-slicing on a neighbourhood of the future end of $M$ (see~\S\ref{ss.local-foliation}), this will complete the proof of Theorem~\ref{theo.main} in the flat case and in the de Sitter case.

\subsection{Duality between convex and concave surfaces in the anti-de Sitter space}

${}$

All the arguments developped above in the case $\Lambda\geq 0$ (flat and de Sitter case) apply in the case $\Lambda<0$ (anti-de Sitter case), except for one: if $\Lambda<0$, the $M$ is neither future complete nor past complete, and pushing along orthogonal geodesics does not provide, as explained in \S\ref{ss.local-foliation},  a family of $\kappa$-barriers foliating a nieghbourhood of the future end of $M$. 

So, in the anti-de Sitter case, one needs an alternative argument to complete the proof of Theorem~\ref{theo.main}. The proof goes as follows (see \S\ref{s.proof-main-AdS}): from \cite{BBZ} we know that the spacetime contains a maximal surface Cauchy surface $\Sigma$. The set of points at Lorentzian distance
$\pi/4$ from $\Sigma$ is an union of two Cauchy surfaces $\Sigma^{\pi/4}$, $\Sigma^{-\pi/4}$, one in the future of $\Sigma$, the other in the past of $\Sigma$. These two surfaces have
both constant K-curvature $-1$, and are thus $\kappa$-barriers. As discussed just above, we cannot prove that the future of $\Sigma^{-\pi/4}$ is foliated by K-surfaces.
However, the arguments used in the flat and de Sitter cases ensure that the past $J^-(\Sigma^{-\pi/4})$ admits a K-time ranging over $(-\infty, -1)$. Similarly,
the future $J^+(\Sigma^{\pi/4})$ of $\Sigma^{\pi/4}$ admits a reverse K-time. Now the key point is that there is a natural duality between convex and concave 
spacelike surfaces in $\AdS$, preserving the property of having constant K-curvature. Therefore, the dual of the K-slicing of $J^+(\Sigma^{\pi/4})$
is a K-slicing of of the future of $\Sigma^{-\pi/4}$ in the past of the convex core, extending the K-slicing of $J^-(\Sigma^{-\pi/4})$.


\subsection{Ingredients of the proofs of Theorems~\ref{theo.ends},~\ref{cor.prescri} and~\ref{Minkowski}}

${}$

As already explained in the introduction, Theorem~\ref{theo.ends} is an easy corollary of the existence of K-slicing for non-elementary 3-dimensional MGHC spacetimes of de Sitter type, and of the duality between the de Sitter space $\dS$ and the hyperbolic space $\HH^3$. Theorems~\ref{theo.ends} and~\ref{cor.prescri} will follow from a generalized version of the Gerhardt's barriers theorem cited above and from Theorem~\ref{theo.main}: the barriers needed to apply Gerhardt's Theorem will be certain leaves of the K-slicing provided by Theorem~\ref{theo.main}.


\subsection{Organisation of the paper.} 

${}$

Section~\ref{section.preli1} is devoted to some general preliminaries on time functions, surfaces with contant curvature, etc. 
In Section,~\ref{section.preli2}, \ref{s.dS-spacetimes} and \ref{sub.core}, 
we recall the main geometrical properties of MGHC spacetimes with constant curvature. 
In Section~\ref{section.systole}, we establish the desired relations between the systole of a Cauchy surface and the distance 
from this Cauchy surface to the initial singularity (Theorem~\ref{th.systole}). In Section~\ref{section.convex}, we prove that the limit of a decreasing sequence of convex Cauchy surfaces is always uniformly spacelike (Theorem~\ref{th.spacelike}). In Section~\ref{section.sequence}, we prove  Theorem~\ref{th.sequence-constant} concerning decreasing sequence of Cauchy surfaces with constant K-curvature. In Section~\ref{s.barriers}, we explain how to construct barriers; in particular, we prove Theorem~\ref{th.asymptotic-barriers} and Proposition~\ref{p.perturbation}. In Section~\ref{s.proof-main-1} and~\ref{s.proof-main-AdS}, we prove our main Theorem~\ref{theo.main} concerning the existence of K-slicings of  3-dimensional non-elementary MGHC spacetimes of constant curvature. In Section~\ref{s.hyperbolic-ends}, we explain the duality between $\dS$ and $\HH^3$, and how to deduce Theorem~\ref{theo.ends} from Theorem~\ref{theo.main}. In Section~\ref{s.prescribed}, we deduce Theorem~\ref{cor.prescri} from Theorem~\ref{theo.main} and Gerhardt's barrier theorem. In Section~\ref{s.Minkowski}, we consider the Minkowski problem and prove Theorem~\ref{Minkowski}.


\section{Cosmological time, Cauchy surfaces, maximum principle}
\label{section.preli1}

For all basic Lorentzian notions such as chronological orientation, past/future, causal past/future, achro\-nal subsets, acausal subsets, edgeless achronal subsets, Cauchy surfaces, global hyperbolicity, and maximal globally hyperbolic space, we refer to \cite{oneill, beem}. 
Recall that MGHC is the acronym for ``maximal globally hyperbolic spatially compact''.


\subsection{Cosmological time} 
\label{s.cosmological-time}

${}$

In any spacetime $(M,g)$, we can define the \textit{cosmological time\/} (see \cite{cosmic}):

\begin{definition}
The cosmological time of a spacetime $(M,g)$ is the function
$\tau:M\rightarrow [0,+\infty]$ defined by
$$\tau(x)=\mbox{Sup}\{ \mbox{Length}(\gamma) \mid \gamma \in {\mathcal R}^-(x) \},$$
where ${\mathcal R}^-(x)$ is the set of all past-oriented  causal curves
starting at $x$, and $\mbox{Length}(\gamma)$ the Lorentzian length of the causal curve
$\gamma$.
\end{definition}

In general, this function may have a very bad behavior: for example, if $(M,g)$ is
Minkowski space or the de Sitter space, then $\tau(x)=+\infty$ for every $x$. 

\begin{definition}
\label{d.regular}
A spacetime $(M,g)$ is said to  have \textit{regular cosmological time,\/} if
\begin{enumerate} 
\item $M$ has \textit{finite existence time,\/} i.e. $\tau(x) < +\infty$ for every  $x$ in $M$,
\item for every past-oriented inextendible causal curve $\gamma: [0, +\infty)
 \rightarrow M$, $\lim_{t \to \infty} \tau(\gamma(t)) = 0$.
\end{enumerate} 
\end{definition}

Regular spacetimes admit many interesting properties (\protect{\cite[Theorem 1.2]{cosmic}}): in particular, they are globally hyperbolic, and their cosmological time is a locally Lipschitz time function. A common and important feature of all non-elementary MGHC spacetimes of constant curvature is that
up to reversal of the time orientation, they have regular cosmological time (\cite{ABBZ1, ABBZ2, bonwick}).

\begin{remark}
For every spacetime $(M,g)$ one can also introduce the \emph{reverse cosmological time} $\check\tau: M \to [0, +\infty]$ of a spacetime $(M,g)$, defined by $\check\tau(x)=\mbox{Sup}\{ \mbox{Length}(\gamma) \mid \gamma \in {\mathcal R}^+(x) \}$, where ${\mathcal R}^+(x)$ is the set of all future-oriented  causal curves starting at $x$. In other words, $\check\tau$ is the cosmological time of the spacetime obtained by reversing the time-orientation of $(M,g)$. 
\end{remark}


\subsection{Generalized Cauchy surfaces}
\label{sub.achrodansgh}

${}$

Let $(M,g)$ be a globally hyperbolic spacetime. 

First of all, we recall how Geroch proved that $M$ is diffeomorphic to a product $\Sigma \times \RR$ (see \cite{dependence}). Select a Cauchy-time $t: M \to \RR$, i.e. a time function such that every level $\{ t=Cte \}$ is a Cauchy surface. Let $X$ be the vector field $-\nabla t$, i.e. minus the gradient of $t$, and let $\phi^{t}$ be the flow of $X$. Let finally $\Sigma$ be the level set $\{ t=0 \}$. Then, the map $F: \Sigma \times \RR \to M$ defined by $F(x,t) = \phi^{t}(x)$ is the required diffeomorphism.

It follows then nearly immediately from the definitions that every achronal subset $E$ of $M$ is the
image by $F$ of the graph of a locally Lipschitz function $u: \Lambda \to \RR$, where $\Lambda$ is a subset of $\Sigma$:
$$
E = \{ (x,t) \in \Lambda \times \RR \mid t = u(x) \} 
$$
If the achronal subset is edgeless then $\Lambda$ is open and $u$ is proper. Furthermore, if $E$ is edgeless achronal and compact, then $\Lambda = \Sigma$. In particular, $\Sigma$ is compact too (i.e. $M$ is spatially compact), and every inextendible causal curve intersects $E$. 

Following the spirit of \cite{dependence}, we thus can define:

\begin{definition}
\label{def.Cauchy}
A \emph{generalized Cauchy surface} in a  MGHC spacetime is a compact achronal edgeless subset.
\end{definition}

Now, let us observe that, in the above discussion, $X$ can be replaced by any future-oriented Lipschitz-regular timelike vector field on $M$. The only difference is that the associated flow 
$\phi^{t}$ might be non-complete, but this does not matter: applying the process above we obtain
an (bilipschitz) identification $F: \Omega \to M$ where $\Omega$ is an open domain in
$\Sigma \times \RR$ defined by:
$$
\Omega = \{ (x,t) \in \Sigma \times \RR \mid u_{-}(x) < t < u_{+}(x) \}
$$
where $\Sigma$ is any Cauchy surface in $M$ and $u_{-}$ (respectively $u_{+}$) is upper semi-continuous (respectively lower semi-continuous). We still have at our disposal a nice description of compact achronal edgeless subsets, i.e. generalized Cauchy surfaces: they are graphs of locally Lipschitz functions $u: \Sigma \to \RR$ satisfying $u_{-} < u < u_{+}$. In the sequel, we will use this later remark with $X$ being the opposite of the gradient of a cosmological time function, and the surface $\Sigma$ being the level set $\tau^{-1}(1)$.


  \subsection{Convex surfaces} 
 \label{sub.concan}

${}$

Let $(M,g)$ be a time-oriented spacetime. Recall that a $C^2$ spacelike surface $\Sigma\subset M$ is said to be \emph{convex} (resp. \emph{strictly convex}) if its second fundamental form is a negative (resp. negative definite) quadratic form. It is not hard to prove that $\Sigma$ is  convex if and only if its future $I^+(\Sigma)$ is locally geodesically convex (every point in $M$ has an arbitrarly small neighbourhood $U$ such that $U \cap I^+(\Sigma)$ is geodesically convex). The advantage of the later characterization is that its extends to non-smooth hypersurfaces:

 \begin{definition}
 An achronal edgeless set  $\Sigma\subset M$ (e.g. a generalized Cauchy surface) is said to be \emph{convex} if its future $I^+(\Sigma)$ is locally geodesically convex.
\end{definition}

The three model spacetimes $\Min$, $\dS$ and $\AdS$ admits some locally projective models (usually called \emph{Klein models}). These models allow to reformulate the definition of convexity in the case where $(M,g)$ is a $3$-dimensional spacetime with constant curvature. Indeed in this particular case, every point $p_0\in M$ admits arbitrarly small \emph{convex charts centered at $p_0$,\/} i.e. causally convex and geodesically convex neighborhood $U_0$ such that there is a diffeomorphism $\varphi_0: U_0 \to V_0$ where $V_0$ is an open domain in the vector space ${\mathbb R}^{3}$ with coordinates
$(x, y, z)$ satisfying the following properties:
\begin{enumerate}
\item $\varphi_0(p_0) = 0$,

\item $\varphi_0$ maps nonparametrized geodesics in $U_0$ to affine segments in $V_0$,

\item $\varphi^\ast\frac{\partial}{\partial{x_0}}$ is a future oriented timelike vector field,

\item $V_0$ is the domain $\{ x^2 + y^2 < 1 , -1 < z < 1 \}$,

\item for every $t \in ]-1, +1[$ the hypersurface $\varphi_0^{-1}(\{ z = t \})$ is a totally
geodesic spacelike surface acausal in $U_0$,

\item the image by $\varphi_0$ of $I^{+}_{U_0}(p_0)$ (resp. $I^-_{U_0}(p_0)$) is the open
domain $\{ z > 0, z^2 > x^2 + y^2 \}$ (resp. $\{ z < 0, z^2 > x^2 + y^2 \}$).
\end{enumerate}
Moreover, for any spacelike Cauchy surface $\Sigma$,
\begin{enumerate}
\item[\textbf (7)]  for any convex chart centered at $p_0\in \Sigma$, the image under $\varphi_0$ of the intersection $\Sigma \cap U_0$ is the graph of a Lipschitz map $(x, y) \mapsto f(x, y)$,
\item[\textbf (8)] if $\Sigma$ is convex (resp. concave) if and only if for any sufficiently small convex chart centered at a point $p_0$ of $\Sigma$ the function $f$ is concave (resp. convex).
\end{enumerate}

Another important property of spacetimes with constant curvature is the abundance of totally geodesic subspaces. Let $M$ be a $3$-dimensional spacetime with constant curvature, and $\Sigma$ be a spacelike surface in $M$. A \emph{lower support plane} of $\Sigma$ is a $2$-dimensional (necessarily spacelike) totally geodesic space $P\subset M$ such that:
\begin{enumerate}
\item $P$ intersects $\Sigma$ at sme point $p$;
\item $\Sigma$ is locally contained in the future of $P$; more precisely, there exists a convex neighbourhood $U$ of $p$ such that $\Sigma\cap U\subset J^+_U(P)$.
\end{enumerate}
One defines similarly the notion of \emph{upper support plane} (by replacing ``future" by ``past" in the above definition). Using convex charts, it is easy to see that a spacelike surface $\Sigma$ is convex (resp. concave) if and only if it admits a lower (resp. upper) support plane at each of its points.


 \subsection{Maximum principle and K-times.} 
 \label{ss.maximum}

${}$

Let  $(M,g)$ be a (time-oriented) MGHC spacetime. Recall that a K-time on $M$ is a time function $\tau:M\to \RR$ such that every level set of $\kappa$ is a hypersurface 
One of the most important features of the K-times is their uniqueness. This property is a consequence of the so-called \emph{maximum principle}. 

\begin{proposition}[Maximum Principle]
Let $\Sigma_1,\Sigma_2$ be two convex spacelike surfaces in $M$. Assume that $\Sigma_1,\Sigma_1$ meet at some point $x$, and that $\Sigma_1$ is in the future of $\Sigma_1$.  Then, the principal curvatures of $\Sigma_1$ are greater than or equal to those of $\Sigma_1$. 
\end{proposition}

To prove this, one only needs to write the surfaces $\Sigma_1,\Sigma_1$ as the graphs of some functions $f_1,f_2$ over the exponential of their common tangent plane at $x$, and to compare the Hessians of $f_1$ and $f_2$ at $x$. See, e.g.,~\cite[Lemma 2.3]{BBZ}. 

\begin{corollary}[Uniqueness of K-times]
Assume that $M$ admits a K-time $\kappa:M\to\RR$. Then, for every $a\in\tau(M)$, the level set $C_a:=\kappa^{-1}(a)$ is the unique compact K-surface of K-curvature $a$ in $M$. In particular, the unique K-slicing on $M$ is the one defined by the level sets of $\kappa$.
\end{corollary}

\begin{proof}
Let $\Sigma\subset M$ be a compact K-surface with K-curvature $a_0$. For every $a\in\kappa(M)$, let $C_a:=\kappa^{-1}(a)$. Define 
$$a^-:=\inf_S\kappa, \;\; a^+=\sup_S\kappa$$
Since $\Sigma$ is compact, $\Sigma$ meets $C_{a^-}$ and $C_{a^+}$. Moreover, $\Sigma$ is in the future of  $C_{a^-}$, and in the past of $C_{a^+}$.

Let $x^-$ be a point where $\Sigma$ meets $C_{a^-}$. The maximum principle implies that the principal curvatures $\Sigma$ at $x^-$ are smaller than or equal to those of $C_{a^-}$. By definition of a K-time, the surface $C_{a^-}$ is strictly convex, i.e. has negative principal curvatures. It follows that the principal curvatures of $\Sigma$ at $x^-$ are negative. Moreover, the product of the principal curvatures of 
$\Sigma$ (resp. $C_{a^-}$) at $x^-$ is equal to $-a_0$ (resp. $-a^-$). It follows that $a_0\leq a^-<0$. 

Now, we see that the principal curvature of $\Sigma$ are negative at every point of $\Sigma$. Indeed, they are negative at $x^-$, and their product is 
everywhere equal to $-a_0\neq 0$. Let $x^+$ be a point where $\Sigma$ meets $C_{a^+}$. By definition of $a^+$ and since $\kappa$ is a 
K-time, $\Sigma$ is in the past of $C_{a^+}$. Hence, the maximum principle implies that the principal curvatures curvatures of $\Sigma$ at $x^+$ 
are bigger than those of $C_{a^+}$. Now, recall that the principal curvatures of $\Sigma$ and $C_{a^+}$ are negative, and the product of principal curvatures 
of $\Sigma$ (resp. $C_{a^+}$) is equal to $a_0$ (resp. $a^+$). It follows that $a_0\geq a^+$.

So we have $a^+\leq a\leq a^-$ and $a^-\leq a^+$. It follows that $a=a^-=a^+$. Since $\Sigma$ is in the future of $C_{a^-}$ and in the past of $C_{a^+}$, we obtain the equality $\Sigma=C_{a^-}=C_{a^+}$.
\end{proof}

\begin{remark}
The strict  convexity of the level sets of $\kappa$ is crucial in the above proof. 
\end{remark}


\section{Geometry of 3-dimensional MGHC flat spacetimes}
\label{section.preli2}

Before starting the proof of Theorem~\ref{theo.main}, we need to recall some general facts on the geometry 3-dimensional MGHC spacetimes with constant curvature. In particular, we need to describe the decomposition of every such spacetime into simple ``building blocks". We start here by discussing the case of flat spacetimes.

\subsection{Mess' description of non-elementary MGHC flat spacetimes}

${}$

In his celebrated preprint~\cite{Mes}, Mess has proved that every MGHC flat spacetime can be obtained as a quotient of domain of the Minkowski space $\Min$ by a discrete group of isometries. More precisely:

\begin{definition}
A \emph{future complete regular domain} in $\Min$ is a domain of the form
$$E=\bigcap_{P\in\Lambda} I^+(P)$$
where $\Lambda$ is a set of lightlike affine planes in $\Min$, and $I^+(P)$ denotes the future of the lightlike plane $P$.  Such a domain $E$ is called \emph{non-elementary} if $\Lambda$ contains three pairwise non-parallel planes. The notion of \emph{(non-elementary)} \emph{past complete regular domain} is defined analogously.
\end{definition}

Observe that every future (resp. past) complete regular domain $E\subset\Min$ is a convex set in $\Min$. Moreover, if $E$ is non-elementary, then $E$ is a \emph{proper} convex set, i.e.  $E$ does not contain any entire affine line of $\Min$.

\begin{theorem}[Mess, \cite{Mes}]
\label{th.Mess-flat}
Every non-elementary 3-dimensional MGHC flat spacetime is isometric to the quotient ${\bf \Gamma}\setminus E$ of a non-elementary future or past complete regular domain 
$E\subset \Min$ by a torsion-free discrete subgroup $\bf{\Gamma}$ of  $\operatorname{Isom}_0(\Min)=\operatorname{SO}_{0}(1, 2)\ltimes\RR^3$. Moreover, 
the linear part of ${\bf\Gamma}$ is a co-compact Fuchsian subgroup $\Gamma$ of $\operatorname{SO}_{0}(1, 2)$, and the projection $p:{\bf\Gamma}\to\Gamma$ is one-to-one.

Conversely, for every torsion-free discrete subgroup $\bf{\Gamma}$ of $\operatorname{Isom}_0(\Min)=\operatorname{SO}_{0}(1, 2)\ltimes\RR^3$, such that the linear part $\Gamma$ of ${\bf\Gamma}$ is a co-compact Fuchsian subgroup of $\operatorname{SO}_{0}(1, 2)$, and such that the projection $p:{\bf\Gamma}\to\Gamma$ is one-to-one, there exists a non-elementary future (resp. past) complete regular domain $E\subset \Min$ such that $\bf\Gamma$ preserves $E$, acts properly discontinuously on $E$, and such that ${\bf\Gamma}\setminus E$ is a non-elementary 3-dimensional MGHC flat spacetime.
\end{theorem}

The index 0 in $\operatorname{Isom}_0(\Min)$ and $\operatorname{SO}_{0}(1, 2)$ means "identity component of the Lie group''.

\begin{remark}
\label{rk.futurepast}
This theorem entails the fact that every non-elementary 3-dimensional MGHC flat spacetime is either future complete (and past incomplete) or past complete (and future incomplete).
\end{remark}

\subsection{Cosmological time of a regular domain}

${}$

Let $E$ be a non-elementary future complete regular domain in $\Min$, and 
$\wt\tau:E\to (0,+\infty]$ its cosmological time. Recall that, for every $x\in E$, the cosmological time $\wt\tau(x)$ is defined by
$$\wt\tau(x)=\sup\{\mbox{length}(\gamma) \mid \gamma \in {\mathcal R}^-(x) \},$$
where ${\mathcal R}^-(x)$ is the set of all past-oriented causal curves in $E$ starting at $x$. Using the mere definition of a regular domain, Bonsante has given a much more concrete description of $\wt\tau$:

\begin{propositiondefinition}[Bonsante, \cite{Bon}]
\label{p.retraction}
Let $\partial E$ be the topological boundary of the domain $E$ in $\Min$.

\smallskip

\noindent 1. For every $x\in E$, there exists a unique point $r(x)\in\partial E$ such that $\wt\tau(x)$ is the 
length of the geodesic segment $[x,r(x)]$ (in particular $\wt\tau(x)<+\infty$). The map $x\mapsto r(x)$ is called the \emph{retraction on the singularity} of the domain $E$.

\smallskip

\noindent  2. The set $T_E=\{r(x), x\in E\}$ is made of all points of $\partial E$ where $E$ admits at least two distinct support planes (or equivalently, admits spacelike support planes). This set $T_E$ is called the \emph{past singularity} of $E$. 

\smallskip

\noindent  3. For every $x\in E$, $r(x)$ is the unique point of $\partial E$ such that the geodesic segment $[r(x),x]$ is orthogonal to a spacelike support plane of $\partial E$ at $r(x)$.
\end{propositiondefinition}

\begin{remark}
The set $T_E$, equipped with the metric induced by the Lorentzian metric of $\Min$, is a $\RR$-tree. 
\end{remark}

Using Proposition~\ref{p.retraction}, one can prove that the cosmological time $\wt\tau$ satisfies nice regularity properties:

\begin{proposition}[Bonsante, \cite{Bon}]
\label{p.cosmo-flat-regular}
The cosmological time $\widetilde{\tau}:E\to (0,+\infty)$ is a $C^{1,1}$ function, that is $\wt\tau$ is a $C^1$ function, whose gradient $\nabla \tilde{\tau}$ is a (locally) Lipschitz vector field. The  levels of $\tilde{\tau}$ are $C^{1, 1}$ Cauchy surfaces in $E$.  
\end{proposition}

\subsection{Decomposition of a regular domain in building blocks}
\label{ss.decompostion-domain}

${}$

We consider as above a non-elementary future complete regular $E\subset\Min$. We will now describe a natural partition of $E$ in simple ``building blocks". We use the notations $\wt\tau$, $r$, $T_E$ defined in the previous paragraph. Some proofs of all the facts stated below can be found in~\cite{Bon}.

For every $p\in T_E$, we consider the set $E(p):=r^{-1}(p)$. By item~1 of Proposition~\ref{p.retraction}, $\{E(p)\}_{p\in T_E}$ is a partition of $E$. For every $p\in T_E$, let $\Delta(p)$ be the set of all spacelike support planes of $E$ at $p$. Item~3 of Proposition~\ref{p.retraction} implies that $E(p)$ is the union of all future directed geodesic rays starting at $p$ that are orthogonal to an element of $\Delta(p)$. Identifying every spacelike affine plane with its future directed unit normal vector, we can see $\Delta(p)$ as a subset of $\HH^2$. Moreover, since $E$ is obtained as the intersection of the futures of some lightlike planes, $\Delta(p)$ is an \emph{ideal} convex subset of $\HH^2$, that is $E(p)$ is the convex hull in $\HH^2$ of a subset of $\partial\HH^2$. We distinguish two cases:
\begin{enumerate}
\item[\textbf{1)}] if $p$ is not a vertex of the $\RR$-tree $T_E$, then $\Delta(p)$ is a geodesic of $\HH^2$, and $E(p)$ is isometric to $\RR\times (0,+\infty)$ endowed with the metric $t^2\operatorname{d\theta}^2-\operatorname{dt}^2$.
\item[\textbf{2)}] if $p$ is a vertex of the $\RR$-tree $T_E$, then $\Delta(p)$ is a ideal convex subset of $\HH^2$ with non-empty interior, and $E(p)$ is isometric to $\Delta(p)\times (0,+\infty)$ endowed with the metric $t^2\operatorname{ds_{hyp}}^2-\operatorname{dt}^2$. 
\end{enumerate}
Moreover, we observe that:
\begin{enumerate}
\item[\textbf{3)}] if $\gamma$ is an edge of length $\ell$ in the $\RR$-tree $T_E$, then $E(\gamma):=\bigcup_{p\in\gamma} E(p)$ is isometric to $\RR\times (-\ell/2,\ell/2)\times (0,+\infty)$ endowed with the Lorentzian metric $t^2\operatorname{d\theta}^2+dx^2-\operatorname{dt}^2$, where $\ell$ is the length of $\gamma$. The set $E(\gamma)$ is foliated by sets of the form $E(p)$ and is maximal for this property. 
\end{enumerate}

\begin{factdefinitions}
\label{f.decomposition}
The partition $E=\bigcup_{p\in T_E}E(p)$ is called the \emph{canonical decomposition} of $E$. Each element of this decomposition is:
\begin{enumerate}
\item[\textbf{1)}] either a \emph{thin cone} isometric to $(\RR\times (0,+\infty)\;,\;t^2\operatorname{d\theta}^2-\operatorname{dt}^2)$,
\item[\textbf{2)}] or a \emph{truncated solid cone} isometric to $(\Delta\times (0,+\infty)\;,\;t^2\operatorname{ds_{hyp}}^2-\operatorname{dt}^2)$ for some ideal convex set with non-empty interior $\Delta\subset\HH^2$.
\end{enumerate}
Every maximal subset of $E$ foliated by thin cones of the canonical decomposition is isometric to $(\RR\times (-\ell/2,\ell/2)\times (0,+\infty)\;,\;t^2\operatorname{d\theta}^2+dx^2-\operatorname{dt}^2)$ for some $\ell$, and is called a \emph{Misner sector of width $\ell$}. 
\end{factdefinitions}

\begin{remark}
In general, $T_E$ is not a simplicial tree. In particular, there might be some points in $T_E$ that neither are vertices, nor belong to a non-trivial edge. Therefore, in the canonical decomposition of $E$, there might be some Minkowski thin cones that do not belong to any Misner sector. 
\end{remark}    

For every $a\in (0,+\infty)$, let $\wt C_a:=\wt\tau^{-1}(a)$ be the $a$-level set of the cosmological time $\wt\tau$. Recall that $\wt C_a$  is a Cauchy surface in $E$. The canonical decomposition of the domain $E$ induces a decomposition of the surface $\wt C_a$. It is easy to see that the elements of this decomposition are the leaves of a lamination and the connected components of the complement of this lamination. More precisely:

\begin{fact}
There is a lamination $\wt\cL_a$ in the surface $\wt C_a$ such that:
\begin{enumerate}
\item[\textbf{1)}]  The leaves of $\wt\cL_a$ are the intersections of $\wt C_a$ with the thin Minkowski cones of the canonical decomposition of $E$,
\item[\textbf{2)}] The connected components of $\wt C_a\setminus\wt\cL_a$ are the intersections of $\wt C_a$ with the truncated solid cones of the canonical decomposition of $E$.
\end{enumerate}
The lamination $\wt\cL_a$ supports a transverse measure defined as follows~: for every absolutely continuous path $c: [a,b] \to \HH^2$ transverse to $\widetilde{\mathcal L_a}$ in $\wt C_a$, the measure of $c$ is the length of the path $r\circ c$ in the $\RR$-tree $T_E$. Note that:
\begin{enumerate}
\item[\textbf{3)}] The intersection of $\wt C_a$ with a Misner sector of width $\ell$ associated with the canonical decomposition of $E$ is a strip (homeomorphic to $\RR\times [-\ell/2,\ell/2]$) in $\wt C_a$ foliated by leaves of the lamination $\wt\cL_a$. The width of this strip, measured with respect to the transverse measure of $\wt\cL_a$, is exactly $\ell$.
\end{enumerate}
\end{fact}

The canonical decomposition of the domain $E$ is invariant under the gradient flow of the cosmological time $\wt\tau$. It follows that, for every $a,a'\in (0,+\infty)$, the time $a'-a$ of the gradient flow of the cosmological time $\wt\tau$ maps the lamination $\wt\cL_{a'}\subset \wt C_{a'}$ on the lamination $\wt\cL_a\subset \wt C_a$.

\bigskip

Now, we can consider the map $\wt N:E\to \HH^2$ which associates to each point $x\in\wt E$ the opposite of the gradient of $\wt\tau$ at $x$. Note that, for every $a\in (0,+\infty)$ and every $x\in\wt C_a$,  the vector $\wt N(x)$ is the future-directed unit normal vector of $\wt C_a$ at $x$. The image under $\wt N$ of a thin Minkowski cone $E(p)$ of the canonical decomposition of $E$ can be naturally identified with the geodesic $\Delta(p)\subset\HH^2$. If $E(p),E(p')$ are two different Minkowski cones, the corresponding geodesics in $\HH^2$ are clearly disjoint. Therefore:

\begin{fact}
The image under $\wt N$ of the union of the thin Minkowski cones of the canonical decomposition of $\cL$ is a geodesic lamination  $\wt\cL$ in $\HH^2$.
\end{fact}

For every $a\in (0,+\infty)$, the image under $\wt N$ of the lamination $\wt\cL_a$ is the geodesic lamination $\wt\cL$. 
Therefore, one can push the transverse measure of $\wt\cL_a$ to get a transverse measure on $\cL$. This transverse measure is independant of $a$.

Since $\wt\cL$ is a geodesic lamination in a hyperbolic surface, the arcwise connected components of $\wt\cL$ are its leaves. It follows that:  

\begin{fact} 
The pre-image (for the map $\wt N$) of a leaf of $\wt\cL$ is an arcwise connected component of the union of the thin Minkowski cones of the 
canonical decomposition of $E$. Therefore:
\begin{enumerate}
\item[{\bf 1)}] The pre-image  of a leaf of $\wt\cL$ which carries a zero weight is a thin Minkowski cone of the canonical decomposition of $E$ which is not included in a Misner sector.
\item[{\bf 2)}] The pre-image of a connected component of $\HH^2\setminus\wt\cL$ is a truncated solid cone  of the canonical decomposition of $E$.
\item[{\bf 3)}] The pre-image of a leaf of $\wt\cL$ which carries a weight $\ell>0$  is a Misner sector of width $\ell$ associated with the canonical decomposition of $E$.
\end{enumerate}
\end{fact}

\begin{remark}
The $\RR$-tree $T_E$ is the dual of the measured lamination $\wt\cL$ (cf. Proposition 3.7.2 and equality (3.6) 
in \cite{bonwick}).
\end{remark}

\subsection{Decomposition of a 3-dimensional MGHC flat spacetime}
\label{ss.decomposition-flat}

${}$

Let $(M,g)$ be a non-elementary MGHC flat spacetime. By Theorem~\ref{th.Mess-flat}, reversing the time orientation if necessary, $(M,g)$ is isometric to the quotient of a non-elementary future  complete regular domain $E\subset \Min$ by a torsion-free discrete subgroup ${\bf\Gamma}$ of  $\operatorname{SO}_{0}(1, 2)\ltimes\RR^3$. 

Clearly, the cosmological time $\wt\tau$ of the domain $E$ is ${\bf\Gamma}$-invariant, and the projection of $\wt\tau$ in $M\simeq {\bf\Gamma}\backslash E$ 
is the cosmological time of the spacetime $M$. Therefore, the cosmological time $\tau$ of $M$ enjoys the same nice regularity properties as $\wt\tau$. In particular, 
Propositon~\ref{p.cosmo-flat-regular} implies that $\tau$ is a $C^{1,1}$ function, and, for every $a\in (0,+\infty)$, the level set $C_a:=\tau^{-1}(a)$ is a $C^{1,1}$ Cauchy surface in $M$ (in particular, $C_a$ is a compact surface).

The canonical decomposition of $E$ is ${\bf \Gamma}$-invariant. Therefore, it induces a 
decomposition of the spacetime $M$. More precisely:

\begin{factdefinitions}
The canonical decomposition of $E$ induces a \emph{canonical decomposition} of $M$. Each element of this decomposition is:
\begin{enumerate}
\item[{\bf 1)}] either a \emph{thin block} isometric to $\RR \times (0, +\infty)$ or $\SS^1\times (0,+\infty)$ endowed with the metric $t^{2}\operatorname{d\theta}^{2} - \operatorname{dt}^{2}$,
\item[{\bf 2)}] or a \emph{solid block} isometric to $\Sigma \times (0, +\infty)$ with the metric $t^{2}\operatorname{ds}_{hyp}^{2} - \operatorname{dt}^{2}$ where $(\Sigma, \operatorname{ds}_{hyp}^{2})$ is the interior of a hyperbolic surface with boundary. 
\end{enumerate}
Every maximal subset of $M$ foliated by thin blocks of the canonical decomposition is isometric to $\SS^{1} \times (-\ell/2, \ell/2) \times (0, +\infty)$ with the metric $\operatorname{t^{2}d\theta}^{2} + \operatorname{dx}^{2} - \operatorname{dt}^{2}$, and is called a \emph{Misner block} of width $\ell$. 
\end{factdefinitions}

Now, for every $a\in (0,+\infty)$, the measured lamination $\wt\cL_a$ projects to a measured lamination $\cL_a$ 
in the surface $C_a={\bf\Gamma}\backslash\wt C_a=\tau^{-1}(a)$. Moreover:
\begin{enumerate}
\item[\textbf{1)}] The leaves of $\cL_a$ are the intersections of $C_a$ with the thin blocks of the canonical decomposition of $M$. More precisely, the closed leaves of $\cL_a$ are the intersections of $C_a$ with the thin blocks homeomorphic to $\SS^1\times\RR$ and the non-closed leaves of $\cL_a$ are the intersections of $C_a$ with the thin blocks homeomorphic to $\RR\times\RR$.
\item[\textbf{2)}] The connected components of $C_a\setminus\cL_a$ are the intersections of $C_a$ with the solid blocks of the canonical decomposition of $M$. 
\item[\textbf{3)}] The intersection of $C_a$ with a Misner block of width $\ell$ associated with  the canonical decomposition of $M$ is a annulus in $M$ foliated by closed leaves of $\cL_a$. The width of this annulus, measured with respect to the transverse measure of $\cL_a$, is exactly $\ell$.
\end{enumerate}

Now,  the map $\wt N:E\to\HH^2$ is ${\bf\Gamma}$-equivariant, where ${\bf\Gamma}$ acts on $\HH^2$ through its linear part $\Gamma\subset \operatorname{SO}_0(1,2)$. 
It follows that $\wt N$ induces a Lipschitz map $N:M\to C:=\Gamma\backslash\HH^2$, that the measured geodesic lamination $\wt\cL$ induces a measured geodesic 
lamination $\cL$ on the compact hyperbolic surface $C:=\Gamma\backslash\HH^2$, and that $\cL$  is the image of the union of all the thin blocks in the canonical 
decomposition of $M$ under the map $N$. Moreover:
\begin{enumerate}
\item[{\bf 1)}] The pre-image (for the map $N:M\to C$) of a leaf of $\cL$ which carries a zero weight is a thin block of the canonical decomposition of $M$ which is not included in a Misner block.
\item[{\bf 2)}] The pre-image of a connected component of $C\setminus\cL$ is a solid block of the canonical decomposition of $E$.
\item[{\bf 3)}] The pre-image of a leaf of $\cL$ which carries a weight $\ell>0$ is a Misner block of width $\ell$ associated with the canonical decomposition of $E$.
\end{enumerate}
Note that, since $C$ is compact, a leaf of $\cL$ which carries a non-zero weight is necessarily closed, and that there are only finitely many such leaves in $\cL$. It follows that there are only finitely many Misner blocks associated with the canonical decomposition of $M$.

\begin{remark}
In \cite{Mes} G. Mess observed that the pair $(C, \cL)$ parametrizes the set of $3$-dimensional non-elementary MGHC 
flat spacetimes of prescribed topology. In \cite{bonthese, Bon}, F. Bonsante studied in detail this parametrization, 
discussed its extension to higher dimensions, and proved in particular that this parametrization is continuous at many respect. 
\end{remark}

\section{Geometry of 3-dimensional MGHC $\dS$-spacetimes} 
\label{s.dS-spacetimes}

\subsection{The 3-dimensional de Sitter space.}
\label{s.dS}

${}$

Consider the quadratic form $Q_{1,3}=-x_1^2+ x_2^2 + x_3^2 + x_4^2$ on $\RR^4$. The \emph{linear model} $\dS$ of 
the 3-dimensional de Sitter space is the hyperboloid $\{Q_{1,3}=1\}$ endowed with the Lorentzian metric induced by $Q_{1,3}$. 
This spacetime is time-orientable; we choose the time orientation for which the  curve $t\mapsto (\sinh t,\cosh t,0,0)$ is future-oriented.  
Moreover, $\dS$ is globally hyperbolic and that the coordinate $x_1$ is a time function on $\dS$. 
The group of orientation and time orientation preserving isometries of $\dS$ is the group $\mbox{O}_0(1,3)$.

It is sometimes convenient to consider the \emph{Klein model} $\DS$ of the de Sitter space. By definition, 
$\DS$ is the image of $\dS$ by the radial projection $\pi:\RR^4\setminus\{0\}\to\SS^3$ (endowed with the push-forward 
of the Lorentzian metric of $\dS$). The projection $\pi:\dS\to\DS$ is a diffeomorphism. The boundary of $\DS$ in $\SS^3$ 
is the image under $\pi$ of the cone $\{Q_{1,3}=0\}$. This boundary is the disjoint union of two round $2$-spheres: 
$\SS^2_-:=\pi(\{Q_{1,3}=0\}\cap\{x_1<0\})$ and $\SS^2_+:=\pi(\{Q_{1,3}=0\}\cap\{x_1>0\})$. Note that every future 
oriented inextendible causal curve $\gamma\subset\DS$ ``goes from $\SS^2_-$ to $\SS^2_+$'' (more precisely: the $\alpha$-limit 
and $\omega$-limit sets of $\gamma$  are single points lying respectively in $\SS^2_-$ and $\SS^2_+$).

Now denote by  $H^3_-$ (resp. $H^3_+$) the connected components of the hyperboloid $\{Q_{1,3}=-1\}$ contained respectively in the half-space 
$\{x_1<0\}$ (resp. $\{x_1>0\}$), endowed with the Riemannian metric induced by $Q_{1,3}$. Note that $H^3_-$ and $H_+^3$ are two copies of the 3-dimensional hyperbolic space. The Klein models of $H^3_-$ (resp. $H^3_+$) is the image $\HH_-^3$ (resp. $\HH^3_+$) under the projection $\pi$. Observe that 
$\SS^3=\HH^3_-\sqcup \SS^2_-\sqcup \DS\sqcup \;\SS^2_+\sqcup \HH^3_+$, and that $\SS^2_-$ (resp. $\SS^2_+$) is the topological boundary of $\HH^3_-$ (resp. $\HH^3_+$) in $\SS^3$.

The group $\mbox{O}_0(1,3)$ is simultaneously the isometry group of the Lorentzian space $\DS$, the isometry group of the hyperbolic spaces $\HH^3_-$ and $\HH^3_+$, and the conformal  group of the spheres $\SS^2_-$ and $\SS^2_+$. 

A key ingredient in the sequel is the fact that the de Sitter space $\DS$
can be thought of as the space of (non-trivial open) round balls in
$\mathbb S^2_+$. Indeed, for every point $x\in\DS$, let us denote by
$\partial^+ I^+(x)$ the set of the future endpoints in $\SS^{2}_+$
of all the future oriented timelike geodesic rays starting at $x$. 
Then, for every $x\in\DS$, the set $\partial^+I^+(x)$ is an open
round ball in $\SS^{2}_+$. One can easily check that the map
associating to $x$ the round ball $\partial^+ I^+(x)$ establishes a
one-to-one correspondance between  the points in $\DS$ and the
(non-trivial open) round balls in $\SS^2_+$.
Of course, there is a similar identification between the points of
$\DS$ and the round balls in ${\mathbb S}^2_-$.

\subsection{Scannell's description of MGHC $\dS$ spacetimes}
\label{ss.general-case}

${}$

We now sketch the correspondance between MGHC $\dS$ spacetimes and compact M\"obius surfaces. For more details (and for proofs of the facts stated below), we refer to~\cite{scannell} or~\cite{ABBZ2}.

Let us consider a simply connected M\"obius surface $S$ (i.e. a surface endowed with a $(\mbox{O}_0(1,3),\SS^2)$-structure).  We will construct a simply connected spacetime $\cB_0^+(S)$ associated to $S$ and locally isometric to $\dS$. For this purpose, we consider a developping map $d:S\to\SS^2_+$ (such a map $d$ does exists since $S$ is a M\"obius surface). 

An (open) \emph{ round ball} $U$ in $S$ is an open domain in $S$ such
that the developing map $d$ to $U$ is one-to-one in restriction to
$U$, and such that $d(U)$ is an open round ball in $\SS^2_+$. A
round ball $U\subset S$ is said to be \emph{proper} if the image under
$d$ of the closure $\overline{U}$ of $U$ in $S$ is the closure of
$d(U)$ in $\SS^2_+$.  

We will denote by $\cB_0(S)$ the set of proper round balls. 
The set $\cB_0(S)$ is naturally ordered by the
inclusion. For every element $U$ of $\cB_0(S)$, we denote by $W(U)$
the subset of $\cB_0(S)$ made of the proper round balls $U'$ such that
$\overline{U'}\subset U$. Given two elements $U,V$ of $\cB_0(S)$ such
that $\overline{U}\subset V$, we denote by $W(U,V)$ the set of all
proper round balls $U'$ in $S$ such that $\overline{U}\subset U'$ and
$\overline{U'}\subset V$. The sets $W(U,V)$ generate a topology on
$\cB_0(S)$ that we call the \textit{Alexandrov topology}.
It can be proved that the set $\cB_0(S)$, equipped with the Alexandrov topology, is a manifold.

We already observed that the de Sitter space $\DS$, as a set, is
canonically identified with the space ${\mathcal B}_0({\mathbb
  S}^{2}_+)$ of all open round balls
in the sphere $\SS^{2}_+$. Actually, the identification between $\DS$ and ${\mathcal B}_0({\mathbb S}^{2}_+)$ is an homeomorphim, once ${\mathcal B}_{0}({\mathbb S}^{2}_+)$ is endowed with the Alexandrov topology.

The map $\cD^+:\cB_0(S) \rightarrow\DS$, composition
of the developing map $d:\cB_0(S)\rightarrow\cB_0(\SS^{2})$ and the 
identification of $\cB_0(\SS^{2})$ with $\DS$, is a local homeomorphism. Hence, we can consider the pull-back by $\cD^+$ of the de Sitter metric on $\cB_0(S)$. We will denote by $\cB_0^+(S)$ the manifold $\cB_0(S)$ equipped with the pull-back by $\cD^+$ of the de Sitter metric. By construction, $\cB_0^+(S)$ is a spacetime locally isometric to $\DS$.

Now suppose that the simply connected M\"obius surface $S$ considered above is the universal cover of a compact M\"obius surface $\Sigma$. Denote by $\Gamma$ 
the fundamental group of $\Sigma$. Then the developping map $d:S\to\SS^2_+$ is $\Gamma$-equivariant: there exists a representation $\rho :\Gamma\to \mbox{O}_0(1,3)$ such 
that $d\circ\gamma=\rho(\gamma)\circ d$ for every $\gamma\in\Gamma$. It follows  that $\rho(\Gamma)$ acts by isometries on the spacetime 
$\cB_0^+(S)$. This action is free, and properly discontinuous. So, we can consider the spacetime $M^+(\Sigma):=\rho(\Gamma)\backslash\cB_0^+(S)$. Up to isometry, this spacetime does not depend on the choice of the developing map $d$. The important result for us is the following:

\begin{theorem}[Scannell]
\label{teo.dscompact}
Every MGHC $\dS$-spacetime is past or future complete. Every future complete MGHC $\dS$-spacetime $M$ is isometric to the spacetime $M^+(\Sigma)$ for some compact M\"obius surface $\Sigma$. 
\end{theorem}

\begin{remark}
\label{rk.functorial}
This correspondance between M\"obius surfaces and MGH $\dS$-spacetimes is an equivalence of category. In particular,
any isometric map $\cB_0^+(S_1) \to \cB_0^+(S_2)$ induces a conformal projective map $S_1 \to S_2$.
\end{remark}

\subsection{Canonical decomposition of MGHC $\dS$-spacetimes}
\label{sub.blocstaticds}

${}$

Let $M$ be a future complete non-elementary MGHC $\dS$-spacetime. Analysing the cosmological time $\tau$ of $M$, and following the same train of thoughts as in the flat case (see \S\ref{ss.decompostion-domain} and \ref{ss.decomposition-flat}), one gets:

\begin{factdefinitions}
The spacetime $M$ admits a \emph{canonical decomposition}. Each element of this decomposition is:
\begin{enumerate}
\item[{\bf 1)}] either a \emph{thin block} isometric to $\RR \times (0, +\infty)$ or $\SS^1\times (0,+\infty)$ endowed with the metric  $\sinh(t)^{2}\operatorname{dx}^{2} - \operatorname{dt}^{2}$,
\item[{\bf 2)}] or a \emph{solid block} isometric to $\Sigma \times (0, +\infty)$ endowed with the Lorentzian metric $\sinh(t)^{2}\operatorname{ds}_{hyp}^{2} - \operatorname{dt}^{2}$ where $(\Sigma, \operatorname{ds}_{hyp}^{2})$ is the interior of a hyperbolic surface with boundary.
\end{enumerate}
Every maximal subset of $M$ foliated by thin blocks of the canonical decomposition is isometric to $\SS^{1} \times (-\ell/2, \ell/2) \times (0, +\infty)$ endowed with the metric $\cosh(t)^{2}\operatorname{d\theta}^{2} + \sinh(t)^{2}\operatorname{dx}^{2} - \operatorname{dt}^{2}$, and is called a \emph{Misner block} of width $\ell$.
\end{factdefinitions}

\begin{fact}
For every $a\in (0,\pi/2)$, there exists a measured lamination $\cL_a$ in the surface $C_a=\tau^{-1}(a)$, such that: 
\begin{enumerate}
\item[\textbf{1)}] The leaves of $\cL_a$ are the intersections of $C_a$ with the thin blocks of the canonical decomposition of $M$.
\item[\textbf{2)}] The connected components of $C_a\setminus\cL_a$ are the intersections of $C_a$ with the solid blocks of the canonical decomposition of $M$. 
\item[\textbf{3)}] The intersection of $C_a$ with a Misner block associated with the canonical decomposition of $M$ is a annulus in $C_a$ foliated by closed leaves of $\cL_a$. 
\end{enumerate}
\end{fact}

\begin{fact}
The gradient of the cosmological time of $M$ defines a Lipschitz map $N: M \to C_\infty$ where $C_\infty$ is a hyperbolic surface. The image under $N$ of the union of the thin blocks of the canonical decomposition of $N$ is a geodesic lamination $\cL_\infty$ on $C_\infty$. The transverse measures of the laminations $\cL_a$ induce a transverse measure on $\cL_\infty$. Moreover:
\begin{enumerate}
\item[{\bf 1)}] The pre-image (for $N$) of a non-weighted leaf of $\cL_\infty$ is a thin block of the canonical decomposition of $M$ which is not included in a Misner block.
\item[{\bf 2)}] The pre-image of a connected component of $C_\infty \setminus\cL_\infty$ is a solid block of the canonical decomposition of $M$.
\item[{\bf 3)}] The pre-image of a leaf of $\cL_\infty$ which carries a weight $\ell>0$ is a Misner block of width $\ell$ associated with the canonical decomposition of $M$.
\end{enumerate}
\end{fact}


\section{Geometry of 3-dimensional MGHC $\AdS$-spacetimes}
\label{sub.core}

\subsection{The 3-dimensional anti-de Sitter space}

${}$

Consider the quadratic form $Q_{2,2}=-x_1^2-x_2^2+x_3^2+x_4^2$ on $\RR^4$. The 3-dimensional anti-de Sitter space $\AdS$ is the quadric $\{Q_{2,2}=-1\}$, 
equipped with the Lorentzian metric induced by $Q_{2,2}$. We choose the time-orientation for which the timelike curve $\theta \to (\cos\theta, \sin\theta, 0, 0)$ 
is future-oriented. The group of isometries of $\AdS$ preserving the orientation and the time-orientation is the connected component $\op{O}_0(2,2)$ of the identity in $\op{O}(2,2)$.

Observe that the vector space of $2$ by $2$ matrices $\mathfrak{gl}(2, \RR)$ endowed with the quadratic form $-\det$ is isometric to $(\RR^4, Q_{2,2})$.
Hence $\AdS$ is naturally identified with $\op{SL}(2, \RR)$ endowed with its Killing form. The actions of $\op{SL}(2, \RR)$ on itself by left and right translations
preserves this Lorentzian metric and commutes each other. Hence we have a natural isomorphism $\op{O}_{0}(2,2) \approx (\op{SL}(2, \RR) \times \op{SL}(2, \RR))/I$ where
$I=\{(Id, Id),(-Id, -Id)\}$.

We will also consider the \emph{Klein model} $\AA\DD\SS_3$ of the anti-de Sitter space. By definition, $\AA\DD\SS_3$ is the image of $\AdS$ under the radial projection 
$\pi:\RR^4\setminus \{0\}\to\SS^3$. The projection $\pi:\AdS\to\AA\DD\SS_3$ is a diffeomorphism. 

We denote by $\cQ$ the boundary of $\AA\DD\SS_3$ in $\SS^3$. The projection of $\cQ$ in $\RR\PP^3=\SS^3/\{\mbox{Id},-\mbox{Id}\}$ is a one-sheeted hyperboloid, 
which notoriously admits two transverse rulings by projective lines. This provides an identification between $\cQ/\{\mbox{Id},-\mbox{Id}\}$ and $\RR\PP^1 \times \RR\PP^1$ 
(equivalently, between $\cQ$ and a two fold cover of $\RR\PP^1 \times \RR\PP^1$). The isometric action of $(\op{SL}(2, \RR) \times \op{SL}(2, \RR))/\cI$ on $\ADS$ 
extends to the boundary $\cQ$. The identification between $\cQ/\{\mbox{Id},-\mbox{Id}\}$ and $\RR\PP^1 \times \RR\PP^1$ can be chosen so that 
$\op{SL}(2, \RR) \times \op{SL}(2, \RR)$ acts on $\cQ/\{\mbox{Id},-\mbox{Id}\}$ coordinate by coordinate:
$$ 
(\gamma_L, \gamma_R).(\theta_L, \theta_R) = (\gamma_L\theta_L, \gamma_R\theta_R)
$$

\subsection{Mess' description of non-elementary MGHC $\AdS$ spacetimes}
\label{ss.AdS-spacetimes}

${}$

We briefly recall the correspondance between non-elementary MGHC $\AdS$ spacetimes and pairs of Fuchsian 
representations of a surface group. For more details, see~\cite{Mes} or~\cite{BBZ,bonwick,barBTZ}.

Let $\Gamma$ be the fundamental group of a closed surface of genus $g\geq 2$, and consider a pair of Fuchsian representations $\rho_L,\rho_R: \Gamma\to \op{PSL}(2,\RR)$.
Two such representations are always topologically conjugate: there is a (unique) homeomorphism $f: \RR\PP^1 \to \RR\PP^1$ such that:
$$
 \forall \gamma \in \Gamma, \quad \rho_R(\gamma) \circ f = f \circ \rho_L(\gamma) 
 $$
The graph of $f$ is a closed topological circle $\bar{\Lambda}_\rho$ embedded in $\cQ/\{\mbox{Id},-\mbox{Id}\} \approx \RR\PP^1 \times \RR\PP^1$ which is invariant 
under the action of $\Gamma$ defined by $\bar{\rho} = (\rho_L, \rho_R)$. Moreover, this topological circle 
$\bar{\Lambda}_\rho\subset\cQ/\{\mbox{Id},-\mbox{Id}\}\subset\RR\PP^3$ is contained in an 
affine chart  of $\RR\PP^3$.

The preimage in $\cQ$ of $\bar{\Lambda}$ is the union of two acausal topological circles $\Lambda_{\rho}^-$, $\Lambda_{\rho}^+$, one image of the other by the antipodal map.
For each of them one can define the \emph{invisible domain} $E(\Lambda^\pm_\rho)$: it is the set of all points $p\in\AA\DD\SS_3$ such that there is no causal 
curve $\gamma$ starting at $p$ and ending at a point $x\in\Lambda^\pm_\rho$, such that $\gamma\setminus\{x\}$ is contained  and in an affine chart, 
and such that $\gamma$ is causal. The intersection between the quadric $\cQ$ and the closure of $E(\Lambda^\pm_\rho)$ in $\SS^3$ is exactly the topological circle $\Lambda^\pm_\rho$, and 
$E(\Lambda^\pm_\rho) \setminus \Lambda^\pm_\rho$ is contained in $\AA\DD\SS_3$. Moreover, $E(\Lambda^+_{\rho})$ and $E(\Lambda_{\rho}^-)$ are disjoint one from the other.

The representation $\bar{\rho}$ lifts to a faithfull representation $\rho: \Gamma \to (\op{SL}(2, \RR) \times \op{SL}(2, \RR))/\cI$ preserving $\Lambda^-_{\rho}$
and $\Lambda^+_{\rho}$, hence also the invisible domains $E(\Lambda^\pm_\rho)$. These actions are free and properly discontinuous.
The quotient space $M(\rho) = \rho(\Gamma)\backslash{E}(\Lambda^+_\rho)$ is a MGHC spacetime with closed Cauchy surfaces of genus $g$.
 It turns out that every non-elementary MGHC $\AdS$-spacetime can be obtained in this way:

\begin{theorem}[Mess]
\label{t.ads-spacetimes}
For $g\geq 2$, let $\Gamma_g$ be the fundamental group of the compact surface of genus $g$. Then every MGHC $\AdS$-spacetime with Cauchy surface of genus 
$g$ is isometric to $M(\rho)= \rho(\Gamma)\backslash{E}(\Lambda_\rho^+)$ for some Fuchsian representation 
$\rho = (\rho_L, \rho_R): \Gamma_g \to \op{SL}(2, \RR) \times \op{SL}(2, \RR)$.
\end{theorem}

\begin{remark}
One may consider the quotient $\rho(\Gamma)\backslash{E}(\Lambda_\rho^-)$ instead of $\rho(\Gamma)\backslash{E}(\Lambda_\rho^+)$,
but it would lead to the same result up to isometry since ${E}(\Lambda_\rho^-)$ is the image of ${E}(\Lambda_\rho^+)$ under the antipodal map.
\end{remark}

One can also define the \emph{convex hull} $\op{Conv}(\Lambda_\rho^+)$: it is simply
the convex hull in the usual meaning in any affine chart $U$ of $\SS^3$ containing
$\Lambda_\rho^+$ (it does not depend on the choice of $U$). It turns out that $\op{Conv}(\Lambda_\rho^+)\setminus\Lambda_\rho^+$ is contained in $\AA\DD\SS_3$.
 The complement of $\Lambda_\rho^+$ in the convex hull $\op{Conv}(\Lambda_\rho^+)$ is contained in $E(\Lambda_\rho^+)$. The projection of $\op{Conv}(\Lambda_\rho^+) \setminus \Lambda_\rho^+$ in $M(\rho)$ is a closed region $\cC(\rho)$, diffeomorphic to $S \times [-1, +1]$,  that we call \emph{convex core} of the spacetime $M(\rho)$. The complementary of $\cC(\rho)$ has two connected components: the past and the future of $\cC(\rho)$, denoted respectively by $M_-(\rho)$ and $M_+(\rho)$, called respectively \emph{past tight region} and \emph{future tight region} of $M(\rho)$.

The regions $\cC(\rho)$, $M_-(\rho)$ and $M_+(\rho)$ can also be defined using the cosmological time and the reverse cosmological time of $M(\rho)$:
\begin{itemize}
\item[--] The spacetime $M(\rho)$ has a regular cosmological time $\tau: M(\rho) \to (0, \pi)$, as well as  a regular reverse cosmological time $\check\tau: M(\rho) \to (0, \pi)$.
\item[--] The convex core $\cC(M)$ corresponds to the set $\{ \tau \leq \pi/2, \; \check\tau \leq \pi/2 \}$. The past tight region $M_-(\rho)$ corresponds to the set $\{ \tau \leq \pi/2\}$.  The future tight region $M_+(\rho)$ corresponds to the set $\{ \check\tau \leq \pi/2\}$. 
\item[--] The cosmological time $\tau$ (resp. the reverse cosmological time $\check\tau$) is a $C^{1,1}$ function in restriction to $M_-(\rho)$ (resp. $M_+(\rho)$).
\end{itemize}

\subsection{Canonical decomposition of MGHC $\AdS$-spacetimes}
\label{sub.blocstaticads}

${}$

We have seen in the two previous sections that every non-elementary MGHC flat or $\dS$ spacetime admits a canonical decomposition in so-called thin blocks and solid blocks. A similar decomposition exists in the $\AdS$-case, except that it does not concern the whole spacetime but only its past (or future) tight region.

Let $M=M(\rho)$ be a non-elementary MGHC $\AdS$-spacetime, and $M_-$ be the past tight region of $M$. Analysing the cosmological time of $M$, and following the same train of thoughts as in the flat case, one gets:

\begin{factdefinitions}
The past tight region $M_-$ admits a \emph{canonical decomposition}. Each element of this decomposition is:
\begin{enumerate}
\item[{\bf 1)}] either a \emph{thin block} isometric to $\RR \times (0, +\infty)$ or $\SS^1\times (0,+\infty)$ endowed with the metric  $\sin(t)^{2}\operatorname{dx}^{2} - \operatorname{dt}^{2}$,
\item[{\bf 2)}] or a \emph{solid block} isometric to $\Sigma \times (0, +\infty)$ endowed with the Lorentzian metric $\sin(t)^{2}\operatorname{ds}_{hyp}^{2} - \operatorname{dt}^{2}$ where $(\Sigma, \operatorname{ds}_{hyp}^{2})$ is the interior of a hyperbolic surface with boundary.
\end{enumerate}
Every maximal subset of $M_-$ foliated by thin blocks of the canonical decomposition is isometric to $\SS^{1} \times (-\ell/2, \ell/2) \times (0, +\infty)$ endowed with the metric $\cos(t)^{2}\operatorname{d\theta}^{2} + \sin(t)^{2}\operatorname{dx}^{2} - \operatorname{dt}^{2}$, and is called a \emph{Misner block} of width $\ell$.
\end{factdefinitions}

\begin{fact}
For every $a\in (0,\pi/2)$, there exists a measured lamination $\cL_a$ in the surface $C_a=\tau^{-1}(a)$, such that: 
\begin{enumerate}
\item[\textbf{1)}] The leaves of $\cL_a$ are the intersections of $C_a$ with the thin blocks of the canonical decomposition of $M_-$.
\item[\textbf{2)}] The connected components of $C_a\setminus\cL_a$ are the intersections of $C_a$ with the solid blocks of the canonical decomposition of $M_-$. 
\item[\textbf{3)}] The intersection of $C_a$ with a Misner block associated with the canonical decomposition of $M_-$ is a annulus in $M_-$ foliated by closed leaves of $\cL_a$. 
\end{enumerate}
\end{fact}

\begin{fact}
The gradient of the cosmological time of $M$ defines a Lipschitz map $N: M_- \to C^-_\infty$ where is a hyperbolic surface. The image under $N$ of the union of the thin blocks of the canonical decomposition of $N$ is a geodesic lamination $\cL^-_\infty$ on $C^-_\infty$. The transverse measures of the laminations $\cL_a$ induce a transverse measure on $\cL^-_\infty$. Moreover:
\begin{enumerate}
\item[{\bf 1)}] The pre-image (for $N$) of a non-weighted leaf of $\cL^-_\infty$ is a thin block of the canonical decomposition of $M_-$ which is not included in a Misner block.
\item[{\bf 2)}] The pre-image of a connected component of $C\setminus\cL^-_\infty$ is a solid block of the canonical decomposition of $M_-$.
\item[{\bf 3)}] The pre-image of a leaf of $\cL^-_\infty$ which carries a weight $\ell>0$ is a Misner block of width $\ell$ associated with the canonical decomposition of $M_-$.
\end{enumerate}
\end{fact}

\begin{remark}
The hyperbolic surface $C^-_\infty$ admits a (non-smooth) isometric embedding in $M$, as a pleated surface. This pleated surface is actually the topological boundary of the past tight region $M_-$ in $M$, i.e. the $(\pi/2)$-level set of the cosmological time $\tau$. The pleating lamination is the measured geodesic lamination $\cL^-_\infty$ defined above.  
\end{remark}

\begin{remark}
The future tight region $M^+$ of $M$ also admits a canonical decomposition associated with a hyperbolic surface $C_\infty^+$ and a measured geodesic lamination 
$\cL_\infty^+$ which are in general different from the hyperbolic surface $C_\infty^-$ and the measured geodesic laminations $\cL_\infty^-$.
\end{remark}

\begin{remark}
\label{rk.LebesgueL}
Supports of measured geodesic laminations in hyperbolic surfaces have zero Lebesgue measure, but it is
not true that the complement in $\cL_{a}$ of the Misner blocks has zero Lebesgue measure. Actually,
the area of $C_a$ for the induced metric depends continuously on the measured lamination $\cL^-_\infty$.
In the case where $\cL_a$ is simplicial, i.e. a simple closed geodesic of length $L$ with a weight $l$, this area
is the sum of the area of $C_\infty$ and the area of the inserted annulus, i.e. $alL$. It follows
that in general the area of $C_a$ is $2\pi(2g-2) + al(\cL_\infty)$, where $l(\cL_\infty)$ is the
\textit{length of the measured lamination} (cf. \S~7 of \cite{Bonlength}). In particular, even if $C_a$ has no Misner block,
the lamination $\cL_a$ has nonzero Lebesgue measure.
This remark applies of course in the flat, de Sitter and anti-de Sitter cases as well.
\end{remark}


\section{Upper bound of the systole}
\label{section.systole}

The purpose of this section is to prove Theorem~\ref{th.systole}. All over the section, $(M,g)$ is a MGHC spacetime of constant curvature $\Lambda$. Rescalling the metric if necessary, we may (and we do) assume that $\Lambda=0$ (flat case), $+1$ ($\dS$ case) or $-1$ ($\AdS$ case). In order to treat simultaneously the three possible cases, we define $M_0$ to be:
\begin{itemize}
\item[--] the whole spacetime $M$ in the flat and in the $\dS$ case, 
\item[--] the past tight region of $M$ in the anti-de Sitter case.
\end{itemize}
We use the notations defined in the preceeding section. In particular, we denote by $\tau:M_{0}\to (0,+\infty)$ the cosmological time of $M_{0}$. We denote by $N$ the opposite of the gradient of $\tau$, and by $(\phi^t)_{t\in\RR}$ the flow associated to $N$. We denote by $\tau_{\max}$ the supremum of $\tau$ over $M_0$ (recall that $\tau_{\max}=+\infty$ if $\Lambda=0$ or $+1$, and $\tau_{\max}=\pi/2$ if $\Lambda=-1$).  For every $a\in (0,\tau_{\max})$, we denote by $C_a$ the $a$-level of $\tau$. As explained in \S\ref{sub.achrodansgh}, there is a Lipschitz parametrization of $M_0$ obtained simply by pushing the surface $C_1$ along the flow $\phi^t$. More precisely, the map
$$
\begin{array}{rrcl}
\Phi\;: &C_1 \times (0, \tau_{\max}) & \to & M_0\\
& (x,t) & \mapsto & \phi^{t-1}(x)
\end{array}
$$
is a global bi-Lipschitz homeomorphism. Recall that the cosmological time $\tau$ defines a measured lamination $\cL_1$ on the surface $C_1$. The lamination $\cL_1$ induces a decomposition of $M_0$ as a disjoint union of ``buildings blocks", each of which being~: 
\begin{itemize}
\item[--] a solid block $\Phi(U\times (0, \tau_{\max}))$ associated to a connected component $U$ of~$C_1\setminus\cL_1$, 
\item[--] or a thin block $\Phi(\gamma\times (0,\tau_{\max}))$ associated to a non-closed leaf  $\gamma$ of~$\cL_1$,
\item[--] or a Misner block $\Phi(A\times (0,\tau_{\max}))$ associated to a maximal annulus $A\subset C_1$ foliated by closed leaves of $\cL_1$ (such a  Misner block can itself be decomposed as a disjoint union of thin blocks associated to the closed leaves of $\cL_1$ foliating $A$).
\end{itemize}

\subsection {The expanding character of the cosmological flow}
\label{ss.expansion}

${}$

Here, we consider a Lipschitz curve $c: [a,b] \to M_0$ which is spacelike (i.e. the velocity $c'(s)$ is almost everywhere spacelike). We denote by $c_1:[a,b]\to M_0$ the projection of curve $c$ on the Cauchy surface $C_1$ along the flow $\phi^t$. Observe that
\begin{equation}
c_1(s) = \phi^{1-\tau(c(s))}(c(s)).
\end{equation}
It follows that $c_1$ is a Lipschitz curve.  In particular, the length of the spacelike curve $c_1$ is well defined.

\begin{proposition}
\label{pro.ok1}
Assume that the curve $c$ is contained in the past of the Cauchy surface $C_{1}$ (i.e. the restriction of the cosmological time $\tau$ to the curve $c$ is bounded from above by~$1$). Then:
\begin{itemize}
\item[(i)] the length of $c$ is smaller than or equal to the length of $c_1$, 
\item[(ii)] the cosmological time variation $|\tau(c(b))-\tau(c(a))|$ is smaller than or equal to the length of $c_1$.
\end{itemize}
\end{proposition}

\begin{proof}
For $s\in[a,b]$, let $t(s)=\tau(c(s))$, so that $c(s) = \phi^{t(s)-1}(c_1(s))$.
Since the curve $c$ is spacelike, one has, for almost every $s\in [a,b]$, 
$$
0\leq  | c'(s) |^{2}
$$
On the other hand, the equality $c(s) = \phi^{t(s)}(c_1(s))$ yields almost everywhere:
\begin{eqnarray*}
 | c'(s) |^{2} & = & \left | t'(s) N(c(s)) + d\left(\phi^{t(s)-1}\right)(c_1(s)).c'_1(s) \right |^{2} \\
                          & = & \left |d\left(\phi^{t(s)-1}\right)(c_1(s)).c'_1(s)\right|^{2} + t'(s)^{2} | N(c(s)) |^2 \\
                          &=& \left |d\left(\phi^{t(s)-1}\right)(c_1(s)).c'_1(s)\right|^{2} - t'(s)^{2} 
\end{eqnarray*}
Indeed,  $d\left (\phi^{t(s)-1}\right )(c_1(s)).c'_1(s)$ is tangent to  the level  $C_{t(s)}$, which is
orthogonal to $N = -\nabla \tau$, and $|N|^2 = -1$. 
In particular, for almost every $s$, one has
\begin{equation}
\label{e.majoration-1}
| c'(s) |^{2} \leq \left |d\left (\phi^{t(s)-1}\right)(c_1(s)).c'_1(s)\right|^{2}.
\end{equation}
Moreover, the map $\phi^{t-1}: C_1\to C_{t}$ is contracting for every $t<1$ (see \cite[Lemma 7.4]{Bon})
\footnote{It does not follow directly from the fact that it is obviously true
on $C_1\setminus\cL_1$  and in every Misner block in $C_1$, since the union of non-closed leaves in $\cL_1$ may have 
non-zero Lebesgue measure (cf. Remark~\ref{rk.LebesgueL}).}.
It follows that, for almost every $s$, one has
\begin{equation}
\label{e.majoration-2}
\left |d\left (\phi^{t(s)-1}\right)(c_1(s)).c'_1(s)\right|^{2}\leq | c_1'(s) |^2.
\end{equation}
Putting together inequalities~\eqref{e.majoration-1} and~\eqref{e.majoration-2}, one gets that $|c'(s)|\leq |c_1'(s)|$ for almost every $s$. Integrating over $[a,b]$, one gets that the length of $c$ is smaller than or equal to the length of $c_1$.

>From the above inequalities, one gets, for almost every $s\in [a,b]$
$$ 
t'(s)^2   \leq \left |d\left(\phi^{t(s)-1}\right)(c_1(s))(c'_1(s))\right|^{2} \leq | c'_1(s)) |^2.
$$
Integrating over the interval $[a,b]$, one gets that $|\tau(c(b))-\tau(c(a))|=|t(b)-t(a)|$ is smaller than or equal to the length of $c_1$, as wanted.
\end{proof}

\subsection{Key estimates}

${}$

As in the previous paragraph, we consider a spacelike Lipschitz curve $c:[a,b]\to M_0$, and we denote by $c_1$ the projection of $c$ on the level set $C_1$ along the flow $\phi^t$. There is one particular situation where Proposition~\ref{pro.ok1} can be dramatically improved~: when the curve $c$ is contained in a solid block or in a thin block.  

\begin{proposition}
\label{pro.okstatic}
Assume that $c$ is contained in a solid block, or in a thin block. Denote by $\ell$ the length of $c_1$. Then, one has 
$$\nu_{\Lambda}^{-1}.\exp(-\ell) \leq \frac{\tau(c(b))}{\tau(c(a))} \leq \nu_{\Lambda}.\exp(\ell)$$
where $\nu_{\Lambda}$ is a constant (depending on  $\Lambda$ but not on $M$), and  
$$ \hbox{length}(c) \leq \mu_{\Lambda}(\tau(c(a)),\ell)$$
where   $\mu_{\Lambda}: (0, 1) \times (0,+\infty) \to (0,+\infty)$ is a universal function (depending on $\Lambda$ but not on $M$) such that: for every $L > 0$, the map $\tau \mapsto \mu_{\Lambda}(\tau, L)$ is increasing and $\displaystyle\mathop{\lim}_{\tau \to 0} \mu_{\Lambda}(\tau, L) = 0$.
\end{proposition}

\begin{remark} 
The last estimate should be interpreted as follows:  \emph{given the length of the projection of the curve $c$ on the cosmological level set $C_1$, if the cosmological time takes a small value somewhere on $c$, then the length of $c$ is small.}
\end{remark}

\begin{proof}[Proof of Proposition~\ref{pro.okstatic}: static blocks] Assume that $c$ is contained in a solid block. This block is of the form $M_0(U):=\Phi(U\times (0,\tau_{\max}))$ where $U$ is a connected component of $C_1\setminus\cL_1$. If $\op{ds}_{hyp}^{2}$ denotes the hyperbolic metric, then $(M_0(U),g)$ is locally isometric to $\HH^2 \times (0, \tau_{\op{max}})$ equipped with the metric:
\begin{itemize}
\item $g_0=\tau^{2}\op{ds}_{hyp}^{2} -d\tau^{2}$ in the flat case (see \S\ref{ss.decomposition-flat}),
\item $g_0=\sinh^{2}(\tau)\op{ds}_{hyp}^{2} -d\tau^{2}$ in the $\dS$ case (see \S\ref{sub.blocstaticds})
\item $g_0=\sin^{2}(\tau)\op{ds}_{hyp}^{2} - \tau^{2}$ in the $\AdS$ case (see \S\ref{sub.blocstaticads})
\end{itemize}

\subsubsection*{The flat case}
Let $T = \log(\tau)$. Then $(M_0(U),g)$ is locally isometric to  $\HH^2 \times (0,+\infty)$ equipped with the Lorentzian metric
$$
g_0=\exp(2T)(\op{ds}_{hyp}^{2} - dT^{2})=t^2(\op{ds}_{hyp}^{2} - dT^{2}).
$$
This metric is conformally equivalent to the metric $k=\op{ds}_{hyp}^{2} - dT^{2}$. In particular, the curve $c$ is spacelike for the metric $k$. Therefore, the same arguments (for the metric $k$) as in the proof of Proposition \ref{pro.ok1} show that $T$-variation $|T(c(a) - T(c(b))|$ is smaller than or equal to the length of $c_1$ (note that the length of $c_1$ for the metric $g_0=\exp(2T)(\op{ds}_{hyp}^{2} - dT^{2})$ and for the metric $k=\op{ds}_{hyp}^{2} - dT^{2}$ are equal, since $T=0$ on $c_1$). Since $T=\log(\tau)$, this yields the desired inequality~:
 $$
 \exp(-\ell) \leq \frac{\tau(c(b))}{\tau(c(a))} \leq \exp(\ell).
 $$
For any $s\in [a,b]$, if we replace $b$ by $s$ in the above arguments, one gets  $\exp(-\ell) \leq \tau(c(s))/\tau(c(a)) \leq \exp(\ell)$. In particular, 
$$
\tau(c(s)) \leq \tau(c(a))\exp(\ell).
$$
Now, the same calculations as in the proof of Proposition \ref{pro.ok1} (for the metric $k$) show that
$|c'(s)|_k^2 \leq  |c'_1(s)|_k^2$ where $|\cdot |_k$ denotes the norm of a vector for the metric $k$. Using 
$|c'(s)|_k^2=\tau(c(s))^2.|c'(s)|_{g_0}^2$ and $|c_1'(s)|^2_k=\tau(c_1(s))^2|c_1'(s)|^2_{g_0}=|c_1'(s)|^2_{g_0}$, we get
$$
|c'(s)|_{g_0} \leq \tau(c(s)) |c'_1(s)|_{g_0}.
$$
Putting together the two last inequalities above, one obtains
\begin{eqnarray*}
\mbox{length}_{g}(c) = \mbox{length}_{g_0}(c)  & = & \int_a^b |c'(s)|_{g_0} ds \\
& \leq & \tau(c(a))\exp(\ell) \int_a^b |c_1'(s)|_{g_0} ds \\
& = & \tau(c(a))\ell\exp(\ell)
\end{eqnarray*}
We have thus proved Proposition~\ref{pro.okstatic} in the case $\Lambda = 0$, with $\nu_{0}=1$ and $\mu_{0}(\tau, l) = l\exp(l)\tau$.

\subsubsection*{The $\dS$ case} 
Let $T = \log(\tanh(\frac{\tau}{2}))$ (note that $\tau\mapsto \log(\tanh(\frac{\tau}{2}))$ is the antiderivative of  $\tau\mapsto \frac{1}{\sinh(\tau)}$). Then $(M(U),g)$ is locally isometric to $\HH^2 \times ]0, T_{\op{max}}[$ endowed with the metric:
$$ 
g_1=\sinh(\tau)^2( \op{ds}_{hyp}^2 - dT^2)
$$
As in the flat case, the $T$-variation $c$ is therefore at most $\ell$.
The inequality 
$$
\exp(-\ell) \leq \frac{\tanh(\tau(c(b))/2)}{\tanh(\tau(c(a))/2)} \leq \exp(\ell)
$$ 
follows. This yields 
$$ 
\nu_{1}^{-1}\exp(-\ell) \leq \frac{\tau(c(b))}{\tau(c(a))} \leq \nu_{1}\exp(\ell)
$$
where $\nu_1=\frac{e+1}{e-1}$  is the upper bound  of $\frac{\tau}{\tanh(\tau/2)}$ on $[0,1]$.
Now, by the same arguments as in the flat case, we get for every $s\in [a,b]$
$$
\tanh(\tau(c(s)/2))\leq \tanh(\tau(c(a)/2))\exp(\ell)
$$
and
$$
|c'(s)|_{g_1} \leq \sinh(\tau(c(s))) |c'_1(s)|_{g_1} = \frac{2\tanh(\tau(c(s))/2)}{1-\tanh^2(\tau(c(s))/2)}|c'_1(s)|_{g_1}.
$$
Putting these two last inequalities together, and integrating over $[a,b]$, we obtain as wanted
$\mbox{length}(c)\leq \mu_1(\tau(c(a)),\ell)$ for 
$$
\mu_{1}(\tau, l) = \frac{2l\exp(l)\tanh(\tau/2)}{1-\exp(2l)\tanh^2(\tau/2)}.
$$

\subsubsection*{The $\AdS$ case.}
Similar the $\dS$ case, after replacing $\sinh$ and $\tanh$ by $\tan$ and $\sin$. 
The constant $\nu_{-1}$ is the supremum of $\frac{\tau}{\tan(\tau/2)}$ over $[0, \pi/2]$,
and the universal map $\mu_{-1}$ is:
$$
\mu_{-1}(\tau, l) = \frac{2l\exp(l)\tan(\tau/2)}{1+\exp(2l)\tan^2(\tau/2)}.
$$
This completes the proof of Proposition~\ref{pro.okstatic} is the case where $c$ is contained in a solid block.
\end{proof}

\begin{proof}[Proof of Proposition~\ref{pro.okstatic}: thin blocks]
Now, we assume that $c$ is contained in a thin block $M_0(\gamma):=\Phi(\gamma\times (0,\tau_{\max}))$ where $\gamma$ is leaf of $\cL_1$. The block $(M_0(\gamma),g)$ is isometric to $\RR \times (0, \tau_{\op{max}})$ (if $\gamma$ is not closed) or $\SS^1 \times (0, \tau_{\op{max}})$ (if $\gamma$ is closed) equipped with the metric:
\begin{itemize}
\item $g_0=\tau^2\op{d\theta}^{2} - d\tau^{2}$ in the flat case (see \S\ref{ss.decomposition-flat}),
\item $g_0=\sinh^{2}(\tau)\op{d\theta}^{2} -d\tau^{2}$ in the $\dS$ case (see \S\ref{sub.blocstaticds})
\item $g_0=\sin^{2}(\tau)\op{d\theta}^{2} - d\tau^{2}$ in the $\AdS$ case (see \S\ref{sub.blocstaticads})
\end{itemize}
The end of the proof is exactly the same as in the case where $c$ is contained in a  solid block, replacing $\op{ds}^2$ by $\op{d\theta}^2$. 
\end{proof}

\subsection{Proof of Theorem \ref{th.systole}} 

${}$

We consider a constant $\epsilon>0$ and a spacelike Cauchy surface $\Sigma$ in $M_0$, such that the systole of $\Sigma$ bigger than or equal to $\epsilon$ (i.e. that any closed curve in $\Sigma$ whose length is less than $\epsilon$ is homotopically trivial). 

The discussion in \S \ref{sub.achrodansgh} shows that we can write $\Sigma$ as a  graph over $C_1$. More precisely, there exists a map $t:C_1\to (0,\tau_{\max})$ such that 
 $$
 \Sigma = \{ \phi^{t(x)-1}(x) \mid x \in C_1 \}.
$$
Observe that this definition of the function $t$ yields 
$$\tau\left(\phi^{t(x)-1}(x)\right)=t(x)$$
for every $x\in C_1$. We will prove that the function $t:C_1\to (0,\tau_{\max})$ is bounded from below by a constant $\alpha_\epsilon>0$. Using the equality above, this will imply that the restriction of the cosmological time $\tau$ to the surface $S$ is bounded from below by $\alpha_\epsilon$. Theorem \ref{th.systole} wil follow.

\begin{remark}
\label{r.past-level-1}
Assume that the systole of the cosmological level set $C_1$ is bigger than or equal to $\epsilon$. Let $\hat t(x)=\min(t(x),1)$. Then the systole of the surface $\widehat\Sigma:=\{\phi^{\hat t(x)-1}(x) \mid x \in C_1 \}$ where $\hat t(x)=\min(t(x),1)$ is also greater than or equal to $\epsilon$. And of course, if $\hat t$ is bounded from below by some constant $\alpha_\epsilon$, then  $t$ is also bounded from below by $\alpha_\epsilon$. Therefore, in the sequel, we can (and we will) assume than that $t$ is bounded from above by $1$. In other words, we will assume that the surface $S$ is in the past of the cosmological level set $C_1$.
 \end{remark}

\subsubsection{Decomposition of  the lamination $\cL_1$.}
We denote by $\cL_1^c$ (resp. $\cL_1^{nc}$) the union of the closed (resp. non-closed) leaves of the lamination $\cL_1$.  Since $\cL^1$ is a measured lamination, $\cL_1^c$ and $\cL_1^{nc}$ are closed. Indeed, if $\cL_1^{nc}$ were not closed, there would be a bunch of leaves of $\cL_1^{nc}$ spiraling on some closed leaf $\gamma$ of $\cL_1^c$. But then, the measure of any small arc transverse to $\gamma$ would  be infinite. Contradiction.

Therefore, $\cL_1^c$ and $\cL_1^{nc}$ are sublaminations of $\cL_1$. So we have a decomposition 
of $\cL_1$ into two disjoint sublaminations $\cL_1^c$ and $\cL_1^{nc}$, the first one made only of closed leaves, the second one without closed leaves. 

Observe that the support of $\cL_1^c$ is exactly the intersection of the surface $C_1$ with the Misner blocks of $M_0$. In particular, this support of $\cL_1^c$ is a finite union of disjoint annuli in $C_1$, each of these annuli being foliated by leaves of $\cL_1^c$. 

\subsubsection{Estimates of $t$ on $\cL_1^c$.} 

\begin{lemma}
\label{le.t-c} 
There exists a constant $\eta^c>0$ such that $t(x) \geq \eta^c$ for every $x$ in the support of the lamination $\cL_1^c$. 
\end{lemma}

\begin{proof}
Let us first recall that the support of the lamination $\cL_1^c$ is exactly the intersection of the surface $C_1$ with the Misner blocks of $M_0$. The intersection of $C_1$ with a Misner block is an annulus foliated by leaves of $\cL_1^c$, all of which have the same length. There are only finitely many Misner blocks in $M_0$. Therefore there are only finitely many possible lengths for the leaves of $\cL_1^c$. We denote these lengths by $\ell_1,\dots,\ell_k$.

Now, let $x\in C_1$ be a point in the support of the lamination $\cL_1^c$. Denote by $\gamma_{1,x}$ the leaf of $\cL_1^c$ containing $x$, and by $\gamma_x$ be the projection of $\gamma_{1,x}$ on the surface $\Sigma$ along the flow $\phi^t$ (of course, it follows that $\gamma_{1,x}$ is the projection of $\gamma_x$ on the surface $C_1$). From Proposition \ref{pro.okstatic}, we get 
$$
\mbox{length}(\gamma_x)\leq \mu_\Lambda(t(x),\ell_i).
$$ 
where $\ell_i$ is the length of the closed leaf $\gamma_{1,x}$. This closed leaf $\gamma_{1,x}$ is not homotopically trivial 
in $C_1$ (since the image of $\gamma_{1,x}$ under the Gauss map $N:C_1\to C_\infty$ is a closed leaf of the geodesic lamination $\cL_\infty$, i.e. a closed geodesic in the hyperbolic surface $C_\infty$). Hence the closed curve $\gamma_{x}$ is not homotopically trivial in $\Sigma$.  So our assumption on the systole of $\Sigma$ and the above estimate on the length of $\gamma_x$ implies 
$$
\mu_\Lambda(t(x),\ell_i)\geq \epsilon.
$$

It remains to choose a constant $\eta^c>0$ such that $\mu_\Lambda(\tau,\ell_i)<\epsilon$ for every $\tau\leq \eta$ and every $i\in\{1,\dots,k\}$. The above inequality implies that $t(x)\geq\eta^c$ for every $x$ in the support of $\cL_1^c$. This completes the proof of Lemma~\ref{le.t-c}.
\end{proof}

 \subsubsection{Estimates of $t$ on $\cL_1^{nc}$.}  We will prove the following:

\begin{lemma}
\label{le.t-nc}
There exists a constant $\eta^{nc}>0$ such that  $t(x) \geq \eta^{nc}$ for every $x\in\cL_1^{nc}$.
\end{lemma}

\begin{proof}
For every $x\in\cL_1$, we denote by $\cL_1(x)$ the leaf of $\cL_1$ containing $x$. The idea of the proof is to construct, for every $x\in\cL_1^{nc}$, a closed homotopicaly non-trivial loop containing $x$, by closing long segment of the leaf of $\cL_1(x)$.  Recall that a lamination $\cL$ is \emph{minimal} if every leaf of  it is dense (in its support). We will use the following fact (see for example Proposition (iii)  in \cite{levitt}): \emph{the lamination $\cL_1^{nc}$ is a union of finitely many minimal sublaminations $\cL_{1}^1,\dots,\cL_1^{k}$.}

Let $i\in\{1,\dots,k\}$. Going to a double cover if necessary, we assume that the lamination $\cL_1^i$ is orientable. 
Since $\cL_1^i$ is minimal (and compact),  there exists $\ell_i\in\RR$ such that every $x\in\cL_1^i$ is 
\emph{$(\ell_i,\frac{\epsilon}{2})$-recurrent,\/}: there is a segment $\alpha_{1,x}$ starting at $x$ in the leaf $\cL_1(x)$, 
such that the length of $\alpha_{1,x}$ is at most $\ell_i$, and such that the ends of $\alpha_{1,x}$ can be joined by a small 
arc $\beta_{1,x}$ transverse\footnote{By such, we mean that $\beta_{1,x}$ always intersects $\cL_1$ with the same orientation.} 
to $\cL_1$ of length at most $\frac{\epsilon}{2}$. 

For any $x\in\cL_1^i$, we consider the concatenation of the arcs $\alpha_{1,x}$ and $\beta_{1,x}$. This is a closed loop $\gamma_{1,x}$. 
This loop $\gamma_{1,x}$ can be perturbed to get a loop transverse to the lamination $\cL_1$. So, by Poincar\'e-Bendixon Theorem, 
$\gamma_{1,x}$ is homotopically non trivial in $C_1$. We denote by $\gamma_x$ the projection of the loop $\gamma_{1,x}$ on $\Sigma$ along 
the flow $\phi^t$. Of course, $\gamma_x$ is the concatenation of the projections $\alpha_x$ and $\beta_x$ of the arcs $\alpha_{1,x}$ and $\beta_{1,x}$.

Since $\alpha_{1,x}$ is contained in a thin geodesic cone, we can apply Proposition \ref{pro.okstatic} to this arc. We get that the length of $\alpha_{x}$ is bounded from above by $\mu_{\Lambda}(t(x),\ell_x)$, where $\ell_x$ is the length of $\alpha_x$. On the other hand, according to the Proposition \ref{pro.ok1}  and Remark~\ref{r.past-level-1}, the length of the arc $\beta_{x}$ is less than the length of the arc $\beta_{1,x}$. In particular, it is less than $\epsilon/2$. So we get an upper bound for the length of the loop $\gamma_x$:
$$\mbox{length}(\gamma_x)\leq\sup_{\ell\leq\ell_i}\mu_{\Lambda}(t(x),\ell)+\epsilon/2.$$

Since $\gamma_x$ is a non-homotopically trivial loop in $\Sigma$, and since the systole is bounded from below by $\epsilon$, we get
$$
\epsilon \leq \epsilon/2 + \sup_{\ell\leq\ell_i}\mu_{\Lambda}(t(x), \ell),
$$
which leads to
$$ 
\epsilon/2 \leq \sup_{\ell\leq\ell_i}\mu_{\Lambda}(t(x), \ell).
$$

Now, let $\eta^{c}$ be a positive number such that, if $\tau\leq \eta^c$, then 
$$
\sup_{\ell\leq\ell_i}\mu_{\Lambda}(\tau, \ell) < \epsilon/2$$
for every $i\in\{1,\dots,k\}$. The inequalities above imply that $t(x) \geq \eta^c$ for every $x\in\cL_1^{nc}$.
\end{proof}

\subsubsection {Estimates on $C_1\setminus \cL_1$}  
\label{nonclosed}
Let $\delta$ be the diameter of the Cauchy surface $C_1$ (for the riemanian metric induced by $g$ on $C_1$).

\begin{lemma}
\label{le.t-static}
Let $\eta=\min(\eta^c,\eta^{nc})$, where $\eta^c,\eta^{nc}$ are the positive numbers defined in Lemma~\ref{le.t-c} and~\ref{le.t-nc}. Then 
$$t(x) \geq \nu_\lambda \exp(\delta) \eta\quad\mbox{for every}\quad x\in C_1\setminus\cL_1.$$
\end{lemma}

\begin{proof}
Any point $x \in C_1\setminus  \cL_1$ can be joined to a point $y\in\cL_1$ by a path $c_1$  of length at most $\delta$. We can assume the whole of $c_1$ but $y$ is contained in $C_1 \setminus \cL_1$, that is the curve $c_1$ minus $y$ is contained in a solid block. The first inequality in Proposition
\ref{pro.okstatic} yields: 
$$t(x) \geq \nu_\Lambda \exp(\delta) t(y).$$ 
Since $y\in\cL_1=\cL_1^c\cup\cL_1^{nc}$, Lemmas~\ref{le.t-c} and~\ref{le.t-nc} yield
$$t(y)\leq \eta.$$
Putting together the two inequalities above, we get $t(x)\geq \nu_\lambda \exp(\delta) \eta$, as wanted.
\end{proof}

\subsubsection{Conclusion} 
Using Lemmas~\ref{le.t-c},~\ref{le.t-nc} and~\ref{le.t-static} together with the decomposition 
$$C_1:=\cL_1^c\cup\cL_1^{nc}\cup(C_1\setminus\cL_1),$$ 
we get a uniform positive lower bound for $t$ on $C_1$. Since $\phi^{t(x)-1}(x)$ ranges over $\Sigma$ when $x$ ranges over $C_1$, and since $\tau\left(\phi^{t(x)-1}(x)\right)=t(x)$, this yields a uniform positive lower bound for the cosmological time $\tau$ on $\Sigma$. This completes the proof of Theorem~\ref{th.systole}. 
\fin


\section{Sequences of convex Cauchy surfaces}
\label{section.convex}

The aim of this section is to prove Theorem \ref{th.spacelike} and Corollary~\ref{c.uniformly-spacelike}.  

Let us recall what are the objects and the hypotheses. We consider a 3-dimensional non-elementary MGHC spacetime $(M,g)$ of constant curvature, and a sequence $(\Sigma_n)_{(n \in \NN)}$ of (locally) convex generalized Cauchy surfaces in $M$ (see \S \ref{sub.achrodansgh} and \ref{sub.concan}). We assume that this sequence is \emph{decreasing} (i.e. $I^+(\Sigma_{n+1})\supset I^+(\Sigma_n)$ for every $n$), and we assume that the set $\Omega = \bigcup_{n\in\NN} I^+(\Sigma_n)$ is not the whole $M$.  

We want to prove that every support plane of $\Omega$ is spacelike (Theorem~\ref{th.spacelike}), and that the $\Sigma_n$'s are uniformly spacelike (Corollary~\ref{c.uniformly-spacelike}; see definition~\ref{d.uniformly-spacelike}).

Observe that $\Omega$ is a future domain,  that is, for  every $x$ in $\Omega$,
we have $I^+(x) \subset \Omega$. It follows that the boundary $\Sigma_\infty = \partial\Omega$ is 
a closed (but non-compact in general) edgeless achronal subset (see \cite[\S 14, Corollary 27]{oneill}), hence a topological surface. 

\begin{lemma}
\label{le.autrelimit}
The set $\Sigma_\infty$ is the set of limits in $M$ of sequences $(p_n)_{n \in {\mathbb N}}$ 
with $p_n \in \Sigma_n$.
\end{lemma}

\begin{proof}
Let $x$ be an element of $\Sigma_\infty$, and let $c: ]-\infty, +\infty[ \to M$ be an inextendible
future timelike curve with $c(0) = p$. Since every $\Sigma_n$ is a Cauchy surface, for every $n$
there is a unique real number $t_n$ such that $c(t_n)$ belongs to $\Sigma_n$. Then $c(]t_n,+\infty[)$
is contained in $I^+(\Sigma_n)$. Since $p$ does not belong to $\Omega_n$ we have $t_n \geq 0$.
Since the sequence $(\Sigma_n)_{n \in {\mathbb N}}$ is decreasing, the sequence 
$(t_n)_{n \in {\mathbb N}}$ is also decreasing and admits a limit $\bar{t}$. Then 
$c(\bar{t})$ belongs to $\Sigma_\infty$: we obtain $\bar{t}=0$ since $\Sigma_\infty$ is achronal.

Conversely, if $p$ is a limit of a sequence $(p_n)_{n \in {\mathbb N}}$ with $p_n \in \Sigma_n$,
it belongs to the closure of $\Omega$. If $p$ belongs to $\Omega$, then it belongs to some
$\Omega_{n_0}$. The same then would be true for all the $p_n$ with $n$ sufficiently big,
but this is impossible for $n > n_0$. Hence, $p$ belongs to the boundary 
$\Sigma_\infty = \partial\Omega$.
\end{proof}

Let $p \in \Sigma_\infty$, and let $\varphi: U \to V$ be a convex chart centered at
$p$. By our discussion in \S \ref{sub.concan}, the surfaces $\varphi(\Sigma_n \cap U)$ are graphs of convex functions. Using Lemma~\ref{le.autrelimit}, it follows that there is a convex map $(x, y) \to z = f_\infty(x, y)$ such that $\varphi(U \cap \Sigma_\infty)$ is the graph of $f_\infty$. In other words, the topological surface $\Sigma_\infty$ is convex.

\begin{proof}[Proof of Theorem~\ref{th.spacelike}]
Since $\Sigma_\infty$ is achronal, it is quite immediate that a support plane of $\Sigma_\infty$ cannot contain a timelike curve. Hence, we must show that no support plane to $\Sigma_\infty$ at $p$
is lightlike.

Assume   by contradiction that $\Sigma_\infty$ admits a lightlike hyperplane $H$ at $p$
in a convex chart  $U$ centered at $p$.  On one hand, $\Sigma_\infty \cap U$ is contained in
$J^{+}_{U}(H)$. On the other hand, $\Sigma_\infty \cap U$ is contained in the causal past of the future $I^{+}_{U}(p)$.
But $H \cap J^{+}_{U}(p) = H^{\perp}$ is a null segment, and  hence   this null segment
is contained in $\Sigma_\infty$.

More formally, we have proved that there is a future-oriented null geodesic segment $c: [0,1] \to M$
with $c(0) = p$ such that every $c(t)$ belongs to $\Sigma_\infty$.
Let $\hat{c}: [0, T[ \to M$ be the inextendible geodesic ray extending $c$ in the future.  
Let $T'$ be the $\sup$ of  times $t$ such that $\hat{c}(t)$ belongs to $\Sigma_\infty$.
Observe that $T' > 0$. 
Assume $T' < T$. Since $\Sigma_\infty$ is closed, $q = \hat{c}(T')$
belongs to $\Sigma_\infty$. 

Since there is a null geodesic segment $[p,q]$ contained in $\Sigma_\infty$, there is only
one support plane to $\Sigma_\infty$ at $q$: the null plane containing $\hat{c}(t)$ for $0 < t$ near $T'$. The argument  above then implies that $\hat{c}(t)$ belongs to $\Sigma_\infty$ for $t>T'$ and close enough to $T'$. This contradicts the definition of $T'$. Therefore, $T' = T$, that is,  the whole future oriented null ray $\hat{c}$ is contained in $\Sigma_\infty$.

Now, for every $n$, the point $p = c(0)$ is in the past of $\Sigma_n$, and  $\Sigma_n$ is a  Cauchy surface, thus, 
the inextendible geodesic ray $\hat{c}$ intersects $I^+(\Sigma_n)$ . Since $I^+(\Sigma_n)\subset I^+(\Sigma_\infty)$, it follows that the geodesic ray $\hat{c}$ intersects $I^+(\Sigma_\infty)$.
This contradicts the fact that  $\Sigma_\infty$ is achronal and $\hat{c}(t)$ is contained in $\Sigma_\infty$.
\end{proof}

 \begin{proof}[Proof of Corollary~\ref{c.uniformly-spacelike}]
Assume that the sequence $(\Sigma_n)_{n\in\NN}$ is not uniformly spacelike. Then one can find a sequence of points $(x_n)_{n\in\NN}$, with $x_n\in \Sigma_n$, such that:
\begin{itemize}
\item[--] extracting a subsequence if necessary, the  sequence $(x_n)_{n\in\NN}$ converges towards a point $x\in M$,
\item[--]  if $P_n$ is the unique support plane of $\Sigma_n$ at $x$, then, extracting a subsequence if necessary, the sequence $(x_n,P_n)_{n\in\NN}$ converges towards a couple $(x,P)$ where $P$ is a null plane at $x$.
\end{itemize}
By Lemma~\ref{le.autrelimit}, the point $x$ is on the topological surface $\Sigma$. For every $n$, since $P_n$ is the support plane of  $\Sigma_n$ at $x_n$, the surface $\Sigma_n$ is in the future of the totally geodesic plane $P_n$. Using again Lemma~\ref{le.autrelimit}, it follows that $\Sigma$ is in the future of the null plane $P$. Since $x\in \Sigma\cap P$, this implies that $P$ is a support plane of $\Sigma$ at $x$, contradicting Theorem~\ref{th.spacelike}.

As a consequence, the sequence $(\Sigma_n)_{n\in\NN}$ is uniformly spacelike.
  \end{proof}


\section{Sequences of Cauchy surfaces of constant K-curvature}
\label{section.sequence}

The present section is devoted to the proof of Theorem~\ref{th.sequence-constant}. Recall that we are considering a $3$-dimensional non-elementary MGHC spacetime $(M,g)$ of constant curvature $\Lambda$, and a decreasing sequence of Cauchy surfaces $(\Sigma_n)_{n \in \NN}$ in $M$, such that, for every $n$, the surface $\Sigma_n$ is locally strictly convex and has a constant K-curvature $\kappa_n <-\Lambda$ (note that this condition is automatically satisfied if $\Lambda\leq 0$ since locally strictly convex surfaces have negative K-curvature). 

We will see later that $(\kappa_n)_{n\in\NN}$ is automatically a decreasing sequence of real numbers (see Remark~\ref{r.unique}). Anyway, we do not really need this here: we only need to assume that 
$$\kappa_n\to \kappa\in [-\infty,\min(-\Lambda,0))\quad\mbox{when}\quad n\to \infty.$$ 

If $\kappa>-\infty$, we have to prove that the sequence $\Sigma_n$ converges to a smooth surface $\Sigma_\infty$ of constant K-curvature $\kappa$. If $\kappa=-\infty$, we have to prove that $M = \bigcup_{n\in\NN} I^+ (\Sigma_n)$.

\subsection{The $\kappa>-\infty$ case.} 

${}$

The main idea of the proof is to apply a result of Schlenker which describes the asymptotic behaviour of a sequence of immersions of a disc in a Lorentzian space of constant curvature, when the sequence of immersions itself does not converge, although the sequence of metrics obtained by the pulling-back the Lorentzian metric of constant curvature by the immersions does converge (see Theorem~\ref{th.Schlenker} below for a precise statement).

We consider an abstract compact surface $\Sigma$ with the same genus as the $\Sigma_n$'s. For every $n$, we denote by $g_n$ the Riemannian metric induced on the Cauchy surface $\Sigma_n$ by the Lorentzian metric $g$. We can write 
$$g_n=|\kappa_n|^\frac{1}{2}.\hat g_n$$
where $\hat g_n$ is a metric on $\Sigma_n$ with constant K-curvature $-1$. We denote by $[\hat g_n]$ the class of the metric $\hat g_n$ in the Teichm\"uller space $\mbox{Teich}(S)$. 

\begin{lemma}
\label{l.teichmuller} 
The sequence $([\hat g_n])_{n\in\NN}$ is relatively compact in the Teichm\"{u}ller space. 
\end{lemma}

\begin{proof} 
Since $(\Sigma_n)_{n\in\NN}$ is a decreasing sequence of Cauchy surfaces, all the $\Sigma_n$'s are in the past of $\Sigma_0$. Therefore all the $\Sigma_n$'s are in the past of a level set of the cosmological time of $M$, say the level set $C_a$.  By Proposition \ref{pro.ok1}, this implies that the sequence of the diameters of the $(\Sigma_n,g_n)$'s is bounded (by the diameter of the level set ${\mathcal C}_a$). Since $\kappa\neq 0$, deleting a finite number of elements of the sequence $(\Sigma_n)_{n\in\NN}$ if necessary, we can assume that the $\kappa_n$'s are bounded away from $0$. It follows that the sequence of the diameters of the $(\Sigma_n,\hat g_n)$'s is also bounded. By the collar neighbourhood lemma, this implies that the sequence of the systoles of the $(\Sigma_n,\hat g_n)$'s is bounded away from $0$. It follows that the sequence $(\Sigma_n,[\hat g_n])_{n\in\NN}$ is relatively compact in $\mbox{Teich}(S)$.  
\end{proof}

Lemma~\ref{l.teichmuller} can be reformulated as follows:

\begin{corollary}
\label{c.teichmuller}
For every $n\in\NN$, one can find a smooth embedding $f_n:\Sigma\hookrightarrow M$ such that $f_n(\Sigma)=\Sigma_n$ and such that the sequence $(f_n^*\hat g_n)_{n\in\NN}$ 
is relatively compact in the space of the Riemannian metrics on $\Sigma$ endowed with $C^\infty$ topology.\fin
\end{corollary}

Since the sequence $(\kappa_n)_{n\in\NN}$ converges, the sequence $(f_n^*g_n)_{n\in\NN}$ is also relatively compact in the space of the Riemannian metrics on $S$ endowed with $C^\infty$ topology. Extracting a subsequence if necessary, we get a sequence of embeddings (that we still denote by $(f_n)_{n\in\NN}$) of $\Sigma$ in $M$ such that the sequence of metrics $(f_n^*g_n)_{n\in\NN}$ converges in the $C^\infty$ topology towards a metric $g_\infty$ on $\Sigma$. 

Now, let us consider a point $x$ in $\Sigma$. For every $n$, let $y_n:=f_n(x)$.

\begin{lemma}
\label{l.relatively-compact-1}
The sequence of points $(y_n)_{n\in\NN}=(f_n(x))_{n\in\NN}$ has a convergent subsequence.
\end{lemma}

\begin{proof}
For every $n\in\NN$, the point $y_n$ belongs to the surface $\Sigma_n$. So, it is enough to prove that all the $\Sigma_n$'s are contained in a compact region of $M$.

On the one hand, since the sequence $(\Sigma_n)_{n\in\NN}$ is decreasing, all the $\Sigma_n$'s are contained in the past of the Cauchy surface $\Sigma_0$.  On the other hand, we have seen that the sequence of the systoles of the $(\Sigma_n,\hat g_n)$'s is bounded away from $0$ (proof of Lemma~\ref{l.teichmuller}). Since the sequence $(\kappa_n)_{n\in\NN}$ is bounded away from $0$ (deleting a finite number of surfaces $\Sigma_n$ if necessary), it follows that  the sequence of the systoles of the $(\Sigma_n,g_n)$'s is also bounded away from $0$ . By Theorem \ref{th.systole}, this implies that all the $\Sigma_n$'s are contained in the future of some level set $C_a$ of the cosmological time of $M$. As a consequence, all the $\Sigma_n$'s are contained in the compact subset $I^+(C_a)\cap I^-(\Sigma_0)$ of $M$. 
\end{proof}

\begin{lemma}
\label{l.relatively-compact-2}
The sequence of $1$-jets $(j^1f_n(x))_{n\in\NN}$ has a convergent subsequence. 
\end{lemma}

\begin{proof}
By Lemma~\ref{l.relatively-compact-1}, up to extracting a subsequence, we may assume that the sequence of points $(y_n)_{n\in\NN}$ converges towards a point $y\in M$. For every $n\in\NN$, the positive definite quadratic form $q_n:=(f_n^*g_n)_{|T_x \Sigma}$ is the pull back by $df_n(x)$ of the positive definite quadratic form $s_n:=g_{n|T_{y_n}\Sigma_n}$. On the one hand, the sequence $(q_n)_{n\in\NN}$ converges towards the positive definite quadratic form $q$ on $g_{\infty|T_{x}\Sigma}$. On the other hand, Corollary~\ref{c.uniformly-spacelike} implies that, up to extracting a subsequence, the sequence of subspaces $T_{y_n} \Sigma_n$ converges towards a spacelike plane $P$ in $T_y M$, and thus the sequence of quadratic forms $(s_n)_{n\in\NN}$ converges towards the positive definite quadratic form $s=g_{|P}$. Since the space of linear maps leaving invariant a positive definite quadratic form is compact, it follows that, up to extracting a subsequence, the sequence of linear maps $df_n(x)$ converges towards a linear map $d:T_x\Sigma\to P$ such that $q$ is the pull-back of $s$ by $d$.
 \end{proof}

Schlenker's result only concerns embeddings of discs in a simply connected Lorentzian space. So we need to lift everything to the universal covering. The universal covering of $\Sigma$ is the $2$-disc $\DD$. The universal covering $\wt M$ of $M$ is a convex open domain in $X=\mbox{Min}_3$, $\dS$ or $\wt\AdS$. Corollary~\ref{c.teichmuller} and Lemma~\ref{l.relatively-compact-2} implies that, for every $n\in\NN$, we can find a lift $\wt \Sigma_n$ of $\Sigma_n$ in $X$, a lift $\wt f_n:\DD\hookrightarrow X$ of $f_n$ such that:
\begin{itemize}
\item[--] the sequence of metrics $(\wt f_n^*\wt g_n)_{n\in\NN}$ converges towards a Riemannian metric $\wt g_\infty$ on $\DD$ in the $C^\infty$ topology (where $\wt g_n$ is the Riemannian metric induced on $\wt \Sigma_n$ by the Lorentzian metric of $X$),
\item[--] for every $\wt x\in\DD$ the sequence of $1$-jets $(j^1\wt f_n(\wt x))_{n\in\NN}$ has a convergent subsequence. 
\end{itemize}

Now, we argue by contradiction: \emph{we assume that the sequence of surfaces $\Sigma_n$ does not converge in the $C^\infty$ topology towards a smooth surface $S$ of constant curvature $\kappa$.} It follows in particular that the sequence of embeddings $(f_n)_{n\in\NN}$ does not converge in the $C^\infty$ topology. \emph{A fortiori}, the sequence of embeddings $(\wt f_n)_{n\in\NN}$ does not converge in the $C^\infty$ topology. So we are under the assumption of Schlenkers result:

\begin{theorem}[see \cite{Sch}, Th\'eor\`eme 5.6]
\label{th.Schlenker}
Let $(\wt f_n)_n\in\NN:\DD\to X$ be a sequence of uniformly elliptic immersions\footnote{An immersion $f$ of a $(n-1)$-dimensional manifold $N$ 
in a $n$-dimensional Lorentzian space $X$ of constant curvature $\Lambda$ such that $f(N)$ is spacelike and has constant K-curvature $\kappa<-\Lambda$ is a typical 
example of \emph{uniformly elliptic immersion}.} of a disc $\DD$ in a simply connected Lorentzian spacetime of constant curvature $(X,\wt g)$. On the one hand, assume 
that the metrics $\wt f_n^*\wt g$ converges in the $C^\infty$ topology towards a Riemannian metric $\wt g_\infty$ on $\DD$, and that there exists a point $x\in\DD$ 
such that the sequence of $1$-jets $(j^1\wt f_n(x))_{n\in\NN}$ converges. On the other hand, assume that the sequence $(\wt f_n)_{n\in\NN}$ does not converge in the 
$C^\infty$ topology in a neighbourhood of $x$. Then there exists a maximal geodesic $\gamma$ of $(\DD,\wt g_\infty)$ and a geodesic arc $\Gamma$ of $(X,\wt g)$ 
such that the sequence $(\wt f_{n|\gamma})_{n\in\NN}$ converges towards an isometry $f_\infty:\gamma\to\Gamma$.\fin
\end{theorem}

By this theorem, there exists a maximal geodesic $\gamma$ of $(\DD,\wt g_\infty)$ and a geodesic segment $\Gamma$ in $X$, such that $f_{n|\gamma}$ converges towards an 
isometry from $f_\infty:\gamma\to\Gamma$. Since $\wt g_\infty$ is obtained by lifting a Riemannian metric on a compact surface, it is geodesically complete. 
In particular, the geodesic $\gamma$ has infinite length. And, as a consequence, the geodesic arc $\Gamma$ also has infinite length. But since $f_n(\Sigma)\subset \wt M$ for every $n$, the geodesic arc $\Gamma$ must be contained in the closure of $\wt M$ in $X$. Morever, in the $\AdS$-case, $\Gamma$ must be contained in the closure of the complement of the lift of the convex core of $M$ (since all the $\Sigma_n$'s are contained in the complement of the convex core of $M$). This contradicts the following proposition:

\begin{proposition}
\label{nogeodesic}
Let $(M,g)$ be a non-elementary 3-dimensional MGHC spacetime with constant curvature $\Lambda$.
If $\Lambda\geq 0$ (flat case or locally de Sitter case), then there is no complete spacelike geodesic in $M$. If $\Lambda<0$ (locally anti-de Sitter case), then every complete spacelike geodesic of $M$ is contained in the convex core of $M$.
\end{proposition}

\begin{proof}
 In the flat case, we know that the spacetime is a quotient ${\bar{\Gamma}}\backslash{E}$ where $E$ is a \emph{proper} convex domain, i.e. contains no complete affine line (see \cite{Mes}):
the proposition follows. 

\medskip

In the locally anti-de Sitter case, we observe that any complete spacelike geodesic $c$ in $M$
lifts to a complete spacelike geodesic in $E(\Lambda_\rho^+) \subset \AdS$. Such a geodesic
admits two extremities in $\cQ = \partial\AA\DD\SS_3$. But these two extremities must
belong to the closure of $E(\Lambda_\rho^+)$ in $\AA\DD\SS_3$. The intersection between this
closure and $\cQ$ is reduced to $\Lambda_\rho^+$: hence, the spacelike geodesic under consideration
has extremities in $\Lambda_\rho^+$. It follows that $c$ is contained in the convex core.

\medskip

The remaining case is the locally de Sitter case. Let $c: \RR \to M$ be the spacelike geodesic, parametrized by arc-length, and let $\tilde{c}: \RR \to \wt{M}$ be its lifting in the universal covering. The composition
$\tilde{\zeta} = \cD \circ \tilde{c}$ with the developing map is a complete spacelike geodesic in $\dS$. 
In particular, the image of $\tilde{\zeta}$
is contained in a $2$-plane $P$ of $\dS \subset \RR^{1,3}$.

Reversing the time-orientation if necessary, we can assume that $(M,g)$ is future complete. Then $(\wt M,g)$ is isometric to the spacetime $\cB_0(S)$ associated to a simply conncted M\"obius surface $S$ (naturally homeomorphic to the universal cover of the compact $\Sigma$) (see \S\ref{ss.general-case}). It follows that the future of the image of $\tilde{c}$ in $\wt{M}$ is isometric to the universal covering $\wt\Omega(\Delta)$ - here
the geodesic $\Delta$ of $\HH^{3}$ is the intersection between $\HH^{3}$ and the orthogonal
$P^{\perp}$ in $\RR^{1,3}$.

Now we observe that $\wt\Omega(\Delta)$ can also be described as $\cB_0(\CC\PP^1 - \{0, +\infty \})$
if $0$, $\infty$ denotes the extremities of $\Delta$. In other words, we have an embedding
$F: \cB_0(\CC^\ast) \to \cB_0(S)$. According to Remark \ref{rk.functorial}, we have a map $f: \CC^\ast \to S$,
inducing a holomorphic map $\bar{f}: \CC^\ast \to \Sigma$. But since $\Sigma$ has genus $\geq 2$ such an holomorphic map must be constant: this is a contradiction.
\end{proof}

The contradiction we have obtained shows that our assumption was absurd. Therefore, the sequence of Cauchy surfaces $(\Sigma_n)_{n\in\NN}$ towards a surface $\Sigma_\infty$ in the $C^\infty$ topology. Clearly, this implies that $\Sigma_\infty$ is a convex Cauchy surface with constant K-curvature $\kappa=\lim \kappa_n$. This completes the proof of Theorem~\ref{th.sequence-constant} in the case where $\kappa_n\rightarrow\kappa>-\infty$.

\subsection{The $\kappa=-\infty$ case.} 

${}$

We are left to consider the case where $\kappa_n \to - \infty$. In this case, we want to prove that the sequence $(\Sigma_n)_{n\in\NN}$ is covering, i.e. that  $\bigcup_{n} I^+ (\Sigma_n)=M$. 

We argue by contradiction: we suppose that the convex domain $\Omega:= \bigcup_n I^+(\Sigma_n)$ is not the whole spacetime $M$. Then $\Sigma_\infty:=\partial\Omega$ is non-empty. As discussed in the beginning of \S\ref{section.convex}, $\Sigma_\infty$ is a topological surface. Consider an open set with compact closure $N$ in $\Sigma_\infty$, and choose a local time function on a neighbourhood of $N$. This local time function allows to decompose some neighbourhood (with compact closure) $W$ of $N$ as a product $W=N\times I$ where $I$ is an intervall in $\RR$ and $\{x\}\times I$ is timelike for every $x\in  N$. Using Lemma~\ref{le.autrelimit}, we see that this allows to write, for every $n$ large enough, the surface $\Sigma_n\cap W$ as a graph over $N$. By Corollary~\ref{c.uniformly-spacelike}, this sequence of graphs is uniformly spacelike. Obviously, this implies that the area of the local surface $\Sigma_n\cap W$ does not tend to $0$ when $n\to\infty$. In particular, the area of the Cauchy surface $\Sigma_n$ does not tend to $0$ when $n\to\infty$.
This contradicts the Gauss-Bonnet formula: up to a multiplicative constant (depending on the genus of the Cauchy surfaces of $M$) the area of $\Sigma_n$ equals $-\frac{1}{\Lambda + \kappa_n}$.

This completes the proof of Theorem \ref{th.sequence-constant}.


\section{Construction of barriers}
\label{s.barriers}

This section is essentially devoted to the proof of Theorem~\ref{th.asymptotic-barriers} and Proposition~\ref{p.perturbation}. The proof of Theorem~\ref{th.asymptotic-barriers} can be divided into two steps: 
\begin{itemize}
\item[--] first, we show that every 3-dimensional future complete non-elementary MGHC spacetime with constant curvature $\Lambda\geq 0$ contains a convex Cauchy surface $\Sigma$ with K-curvature strictly bounded from above by $-\Lambda$,
\item[--] then we push $\Sigma$ along the geodesics orthogonal to $\Sigma$ in order to get convex Cauchy surfaces whose K-curvature is arbitrarily close to $-\Lambda$.
\end{itemize}
Proposition~\ref{p.perturbation} will be an easy consequence of our estimates on the behaviour of the principal curvatures when one pushes a surface along the orthogonal geodesics. 

It is important at this stage to recall our conventions. Let $\Sigma$ be a spacelike surface in $(M,g)$. Recall that the second fundamental form of $\Sigma$ is defined by 
$II(X,Y)=-\langle \nabla_X n, Y \rangle$ where $n$ is the future oriented unit normal of $\Sigma$. The principal curvatures $\lambda_1,\lambda_2$ of $\Sigma$ are the eigenvalues of 
the second fundamental form, and the K-curvature of $\Sigma$ equals $-\lambda_1 \lambda_2$. For example, with these conventions, if $M$ is the Minkowski space $\RR^{1, 2}$ endowed 
with the lorentzian metric  $-dt^2 +dx^2 + dy^2$, the surface $\{-t^2+x^2+y^2 = - c^2, t >0\}$ has principal curvatures $\lambda = \mu = -c$.

\subsection{Existence of a Cauchy surface with controlled curvature.}

\subsubsection{The flat case.} We will prove the following result:

\begin{proposition}
\label{pro.init-flat}
Every non-elementary $3$-dimensional future complete flat MGHC spacetime contains a locally strictly convex Cauchy surface (in particular, a Cauchy surface with negative K-curvature). 
\end{proposition}

Equivalently, every non-elementary $3$-dimensional flat MGHC spacetime contains a Cauchy surface with negative principal curvatures. In order to prove this Proposition~\ref{pro.init-flat}, we will use the existence of constant mean curvature surfaces in flat MGHC spacetimes and the following proposition which is essentially due to A. Treibergs:

\begin{proposition}
\label{pro.treibergs}
Let $(M,g)$ be a non-elementary  MGHC flat spacetime. Then, every constant mean curvature Cauchy surface in $M$ has negative principal curvatures.
\end{proposition}

\begin{proof} 
Any Cauchy surface with constant mean curvature $\Sigma$ in $M$ lifts as a complete spacelike surface with constant mean curvature $S\simeq\wt\Sigma$ in the Minkowski space.  It is shown in \cite{treibergs} (see also \cite{Yau}) such a surface $S$ is either {\it strictly convex} (i.e. has negative principal curvatures), or splits as a direct product of a curve by an affine line. The splitting case is excluded in our context,  since we have assumed that our spacetime is non-elementary. The proposition follows.
\end{proof}

\begin{proof}[Proof of Proposition~\ref{pro.init-flat}]
In~\cite{And} and~\cite{ABBZ1}, it was shown that every future complete flat spacetime contains a Cauchy surface $S$ with constant mean curvature $-1$ (see \cite{And, ABBZ1}). Together with Proposition~\ref{pro.treibergs}, this proves Proposition~\ref{pro.init-flat}.
\end{proof}

\subsubsection{Transferring properties of CMC hypersurfaces from flat to $\dS$ and $\AdS$ spacetimes.} Now, we would like to get an analog of Proposition~\ref{pro.init-flat} in $\dS$ and $\AdS$-spacetimes. For this purpose, we will use a correspondance between CMC hypersurfaces in flat spacetimes, CMC hypersurface in dS spacetimes and CMC hypersurfaces in AdS spacetimes.

\begin{proposition}[See e.g. \cite{AMT}]
\label{pro.LH}
Let $\Sigma$ be a compact surface, $\eta$ be a Riemannian metric on $\Sigma$ with scalar curvature $R_{\eta}$, and  $h$ be  a quadratic differential on $\Sigma$.  Let $\Lambda$ and $H$ be two real numbers. There exists an isometric embedding of $(\Sigma,\eta)$ as a spacelike Cauchy surface with constant mean curvature $H$ having $h$ is the second fundamental form in {\em some} MGHC spacetime $(M,g)$ with constant curvature $\Lambda$ if and only if the two following conditions are satisfied:
\begin{enumerate}
\item the quadratic differential $h_{0} = h - H.\eta$ (i.e. the trace free part of $h$) is  holomorphic;
\item the Gauss equation $R_{\eta} + {\det}_{\eta}(h_{0}) = \Lambda - H^{2}$ is satisfied.
\end{enumerate}
\end{proposition}

\begin{remark}
If $(\Sigma,\eta,h)$ satisfies conditions (1) and (2), then:
\begin{itemize}
\item[--] one can take as a spacetime $(M,g)$ the Cauchy development of the solution of Einstein equation with a cosmological constant $\Lambda$, in a CMC gauge, with initial Cauchy data $(\Sigma, \eta, h)$,
\item[--] $(S,\eta,h)$ is called a \emph{CMC $(H, \Lambda)$-initial Cauchy data}. 
\end{itemize}
\end{remark}

The following immediate corollary of Proposition~\ref{pro.LH} will allow us to transfer properties of CMC hypersurfaces in flat spacetimes into properties of CMC hypersurfaces in $\dS$ and $\AdS$ spacetimes.

\begin{corollary}
\label{c.transfer}
A CMC $(H, \Lambda)$-initial data $(\Sigma, \eta, h=h_0+H\eta)$ gives rise to a CMC $(H', \Lambda')$-initial data of the form  $(\Sigma, \eta, h'=h_0+H'\eta)$ provided that
 $$\Lambda' - H'^{2} = \Lambda - H^{2}.$$
 \end{corollary}

\begin{remark}
\label{r.transfer-curvatures}
Assume that $(\Sigma, \eta, h=h_0+H\eta)$ is a CMC $(H, \Lambda)$-initial data and $(\Sigma, \eta, h'=h_0+H'\eta)$ is a  CMC $(H', \Lambda')$-initial data. Denote by $\lambda,\mu$ (resp. $\Lambda',\mu'$) the principal curvatures of $\Sigma$ seen as a CMC $(H,\Lambda)$-initial Cauchy data (resp. as a CMC $(H',\Lambda')$-initial Cauchy data). Then 
 $$\lambda^\prime = \lambda + (H^\prime - H)\quad\mbox{and}\quad\mu^\prime = \mu + (H^\prime - H).$$
 \end{remark}

\subsubsection{The $\dS$ case.} We will prove the following result:

\begin{proposition}
\label{pro.init-dS}
Every non-elementary $3$-dimensional future complete MGHC spacetime with constant curvature $\Lambda=+1$ contains a convex Cauchy surface with K-curvature strictly bounded from above by~$-\Lambda$. 
\end{proposition}

\begin{proof}
Consider a non-elementary $3$-dimensional future complete MGHC spacetime $(M,g)$ with constant curvature $+1$. It was proved in~\cite{ABBZ2} that $(M,g)$ admits a Cauchy hypersurface $\Sigma$ with constant mean curvature $-\sqrt{2}$. Denote by $\eta$ and $h$ the first and the second fundamental forms of $\Sigma$. Then $(\Sigma,\eta,h=h_0-\sqrt{2}\eta)$ is a CMC $(-\sqrt{2},1)$-initial Cauchy data. So, by Corollary~\ref{c.transfer}, $(\Sigma,\eta,h'=h_0-\eta)$ is a CMC $(-1,0)$-initial Cauchy data 
(i.e. $(\Sigma,\eta)$ admits an isometric embedding in a flat MGHC spacetime with constant mean curvature $-1$ and second fundamental form $h'$). Denote by $\lambda$ and $\mu$ (resp. $\lambda'$ and $\mu'$) the principal curvatures of $\Sigma$ seen as a CMC $(-\sqrt{2},+1)$ initial Cauchy data (resp. as a CMC $(-1,0)$ initial Cauchy data). According to Proposition~\ref{pro.treibergs}, the principal curvatures $\lambda'$ and $\mu'$ are negative. Using Remark~\ref{r.transfer-curvatures}, this implies that 
$$\lambda<-\sqrt{2}+1\quad\mbox{and}\quad \mu<-\sqrt{2}+1$$ 
(in particular, $\lambda$ and $\mu$ are negative, so that $\Sigma$ is convex in $M$.)
Moreover, since $\lambda+\mu=2H=-2\sqrt{2}$, the above inequality yields 
$$-\sqrt{2}-1<\lambda<-\sqrt{2}+1\quad\mbox{and}\quad -\sqrt{2}-1<\mu<-\sqrt{2}+1.$$ 
These bounds on $\lambda,\mu$ and the equality $\lambda+\mu=-2\sqrt{2}$ imply that
$$-\lambda.\mu<-1,$$
i.e. that the K-curvature of the Cauchy surface $\Sigma$ in $M$ is strictly bounded from above by $-\Lambda$, as wanted.
\end{proof}

\subsubsection{The $\AdS$ case.} The arguments developped above are not sufficient to get a directly a convex Cauchy surface in every $\AdS$ spacetime. Nevertheless, these argument provide us with a Cauchy surface with controlled principal curvatures: 

\begin{proposition}[See also Lemma 3.11 in \cite{KS}]
\label{pro.init-AdS}
Every non-elementary $3$-dimensional MGHC spacetime with constant curvature $\Lambda=-1$ contains a maximal Cauchy surface (i.e. a Cauchy surface with constant mean curvature $0$). Moreover, the principal curvatures of this maximal surface stay within the interval $(-1,1)$.
\end{proposition}

\begin{proof}
Consider a $3$-dimensional non-elementary  MGHC spacetime $(M,g)$ with constant curvature $-1$. It was proved in~\cite{BBZ} that $(M,g)$ admits a maximal Cauchy surface $\Sigma$. Denote by $\eta$ and $h$ the first and the second fundamental forms of $\Sigma$. Then $(\Sigma,\eta,h=h_0)$ is a CMC $(0,-1)$-initial Cauchy data. So, by Corollary~\ref{c.transfer}, $(\Sigma,\eta,h'=h_0-\eta)$ is a CMC $(-1,0)$-initial Cauchy data. Denote by $\lambda$ and $\mu$ (resp. $\lambda'$ and $\mu'$) the principal curvatures of $\Sigma$ seen as a CMC $(0,-1)$ initial Cauchy data (resp. as a CMC $(-1,0)$ initial Cauchy data). According to Proposition~\ref{pro.treibergs}, the principal curvatures $\lambda',\mu'$ are negative. Using Remark~\ref{r.transfer-curvatures}, this implies $\lambda<1$ and $\mu<1$. Since $H=-(\lambda+\mu)=0$, it follows that $-1<\lambda<1$ and $-1<\mu<1$.
\end{proof}

\begin{remark}
In subsection~\ref{ss.convex-concave-AdS}, we will explain how to show that every non-elementary MGHC spacetime with constant curvature $-1$ contains a convex (and a concave) Cauchy surface.
\end{remark}

\subsection{Pushing a Cauchy surface along the orthogonal geodesics}
\label{section.pushing}

${}$

Let $(M,g)$ be an (arbitrary) $3$-dimensional spacetime. Let $\Sigma$ be a spacelike surface in $M$. For every $x\in \Sigma$, denote by $n(x)$ the future oriented unit normal vector of $\Sigma$ at $x$. For $t$ small enough, the map 
 $$
 \begin{array}{cccl}
 \phi_{\Sigma}^t\;: & \Sigma & \to & M\\
 & x & \mapsto & \exp_{x}(t.n(x))
 \end{array}
 $$
 is well-defined, and is an embedding. For such a $t$, the surface 
 $$
 \Sigma^t:=\phi_{\Sigma}^t(\Sigma)
 $$ 
 is obviously a compact spacelike surface, hence a Cauchy surface. We say that the Cauchy surface $\Sigma^t:=\phi_\Sigma^t(\Sigma)$ is obtained by \emph{pushing $\Sigma$ along orthogonal geodesics for a time $t$}. 

In our situtation, since the ambient curvature $\Lambda$ is constant, one can compute explicitely the principal curvatures of the surface $\Sigma^t$:

\begin{proposition}
\label{pro.Gauss}
Pick a point $x\in \Sigma$ and a real number $t$. We assume that the map $\phi_\Sigma^t$ is well-defined (i.e. that the geodesics that are orthogonal to $\Sigma$ exists for a time at least $t$). We denote by $\lambda,\mu$ the principal curvature of $\Sigma$ at $x$. 

\smallskip

\noindent \textbf{1. The flat case.} Assume that $(M,g)$ is flat (i.e. locally isometric to $\Min$). If $\lambda.t\neq 1$ and $\mu.t\neq 1$  then $\phi_\Sigma^t$ is an embedding in the neighborhood of $x$, and the principal curvature of the surface $\Sigma^t$ at the point $\phi_t(x)$ are
$$
\lambda_{t } = \frac{\lambda}{1-\lambda t}\quad\mbox{and}\quad\mu_{t} = \frac{\mu}{1 -\mu t}.
$$

\noindent \textbf{2. The $\dS$-case.} Assume that $(M,g)$ has positive constant curvature (i.e. is locally isometric to $\dS$). If $\lambda \tanh(t) \neq 1$ and $\mu \tanh(t) \neq 1$ then $\phi_\Sigma^{t}$ is an embedding in the neighborhood of $x$, and the principal curvature of the surface $\Sigma^t$ at the point $\phi_t(x)$ are
$$
\lambda_{t} = \frac{\lambda - \tanh(t)}{1-\lambda\tanh(t)}\quad\mbox{and}\quad\mu_{t} = \frac{\mu - \tanh(t)}{1-\mu\tanh(t)}.
$$

\noindent \textbf{3. The $\AdS$-case.}  Assume that $(M,g)$ has negative constant curvature (i.e. is locally isometric to $\AdS$) If $\lambda \tan(t) \neq 1$ and $\mu \tan(t) \neq 1$ then $\phi_\Sigma^{t}$ is an embedding in the neighborhood of $x$, and the principal curvature of the surface $\Sigma^t$ at the point $\phi_\Sigma^t(x)$ are
$$
\lambda_{t} = \frac{\lambda + \tan(t)}{1-\lambda\tan(t)}\quad\mbox{and}\quad\mu_{t} = \frac{\mu + \tan(t)}{1-\mu\tan(t)}.
$$
\end{proposition}

\begin{proof}
Straightforward computation.
\end{proof}

As an immediate corollary, we get: 

\begin{corollary}
\label{cor.Gausscroit}
Pick a point $x\in \Sigma$. Denote by $H$ (resp. $\kappa$) the mean curvature (resp. the K-curvature) of the surface $\Sigma$ at $x$. For $t$ small enough, denote by $\kappa_t$ the  K-curvature of the surface $\Sigma^t$ at $\phi_\Sigma^t(x)$. Then,
$$\frac{\partial \kappa_t}{\partial t}_{| t=0} = 2(\kappa+\Lambda)H.$$
\end{corollary}

\begin{remark}
\label{rem.Gausscroit}
In particular, the K-curvature of $\Sigma^t$ at $\phi_\Sigma^{t}(x)$ increases with $t$ (for $t$ close to $0$) provided that the surface $\Sigma$ is convex and has a K-curvature $\kappa$ strictly bounded from above by $-\Lambda$.
\end{remark}

\subsection{Proof of Theorem~\ref{th.asymptotic-barriers}}

${}$

We consider a $3$-dimensional non-elementary MGHC spacetime $(M,g)$ with constant K-curvature $\Lambda\geq 0=0$. Reversing the time orientation if necessary, we assume that $(M,g)$ is future complete. 

\begin{proposition}
\label{p.pushing-convex}
If $\Sigma$ is a convex Cauchy surface in $M$ whose K-curvature is strictly bounded from above by $-\Lambda$, then:
\begin{enumerate}
\item for every $t\geq 0$, the map $\phi_\Sigma^t:\Sigma\to M$ is well-defined and is an embedding,
\item for every $t\geq 0$, the Cauchy surface $\Sigma^t:=\phi_\Sigma^t(\Sigma)$ is convex and has a K-curvature strictly bounded from above by $-\Lambda$,
\item the K-curvature of $\Sigma^t$ tends uniformly towards $-\Lambda$ when $t\to\infty$.
\end{enumerate} 
\end{proposition}

\begin{proof}
The map $\phi_\Sigma^t$ is defined for every $t\geq 0$ because the spacetime was assumed to be future complete. The other assertions follow from Proposition~\ref{pro.Gauss}.
\end{proof}

Proposition~\ref{pro.init-flat} (in the case $\Lambda=0$) or Proposition~\ref{pro.init-dS} (in the case $\Lambda=+1$) provide us with a convex Cauchy surface $\Sigma$ with K-curvature strictly bounded by $-\Lambda$.  Choose $\epsilon>0$ such that the K-curvature of $\Sigma$ is bounded from above by $-\Lambda-\epsilon$. Let $\kappa\in (-\Lambda-\epsilon,\Lambda)$. By item~1 and~2 of Proposition~\ref{p.pushing-convex}, for every $t\geq 0$, the surface $\Sigma^t=\phi_\Sigma^t(\Sigma)$ is well-defined, convex, and has K-curvature strictly bounded from above by $-\Lambda$. Moreover, by item~3 of Proposition~\ref{p.pushing-convex}, for $t>0$ large enough, the K-curvature of $\Sigma_t$ is strictly bounded from below by $\kappa$. Therefore, for $t>0$ large enough, $(\Sigma,\Sigma_t)$ is a pair of $\kappa$-barriers. This completes the proof of Theorem~\ref{th.asymptotic-barriers}.

\subsection{Convex and concave Cauchy surfaces in $\AdS$ spacetimes}
\label{ss.convex-concave-AdS}

\begin{proposition}
\label{pro.init-AdS-2}
Let $(M,g)$ be a non-elementary $3$-dimensional MGHC spacetime  of $\AdS$ type (i.e. with negative constant curvature). Then $M$ contains two Cauchy surfaces with constant K-curvature $-1$, one strictly convex and the other strictly concave. 
\end{proposition}

\begin{proof}
Let $\Gamma$ be the fundamental group of the Cauchy surfaces of $M$. By Theorem~\ref{t.ads-spacetimes}, there exists a representations $\rho:\Gamma\to \mbox{SL}(2,\RR)\times\mbox{SL}(2,\RR)$, a curve $\Lambda_\rho^+$ in $\partial\ADS$ associated to $\rho$, and a open set $E(\Lambda_\rho^+)$ in $\ADS\simeq\AdS$ such that $M\simeq \rho(\Gamma)\backslash E(\Lambda_\rho^+)$. We will need the following lemma:

\begin{lemma}
\label{p.reste-dans-domaine-invisible}
Let $\Sigma$ be a Cauchy surface in $M$, and $S$ be a lift of $\Sigma$ in $E(\Lambda_\rho^+)$. Let $t_0$ be a real number such that the map $\phi_S^t : S\to\AdS$ is an immersion for every $t$ between $0$ and $t_0$. Then $S^{t_0}=\phi_S^{t_0}(S)$ is contained in $E(\Lambda_\rho^+)$ and projects to a Cauchy surface $\Sigma^{t_0}$ in $M$. 
\end{lemma}

\begin{remark}
\emph{A posteriori}, the surface $\Sigma^{t_0}$ can of course be obtained by pushing $\Sigma$ along the orthogonal geodesics for a time $t_0$. Nevertheless, one cannot define $\Sigma^{t_0}$ directly in $M$ (without passing to the universal cover) since one does not know \emph{a priori} that the geodesics of $M$ that are orthogonal to $\Sigma$ exist for a long enough time (recall that $M$ is neither past nor future complete).
\end{remark}

Let us postpone the proof of Lemma~\ref{p.reste-dans-domaine-invisible}, and complete the proof of Proposition~\ref{pro.init-AdS-2}. According to Proposition~\ref{pro.init-AdS},  the spacetime $M$ admits a maximal Cauchy surface $\Sigma$ whose principal curvatures stay within the interval $(-1,1)$. Let $S$ be a lift of $\Sigma$ in $E(\Lambda_\rho^+)$. We denote by $\lambda(x),\mu(x)$ the principal curvatures of the surface $S$ at $x$. Since $|\lambda(x)|<1$ and $|\mu(x)|<1$ for every $x\in \Sigma$, Proposition~\ref{pro.Gauss} implies that the map $\phi^{t}_S$ is an immersion for every $t\in [-\pi/4,0]$. Therefore, Lemma~\ref{p.reste-dans-domaine-invisible} implies that $S^{-\pi/4}$ is embedded, contained in the domain $E(\Lambda_\rho^+)$, and projects to Cauchy surface $\Sigma^{-\pi/4}$ in $M$, which is obtained by pushing the surface $\Sigma$ along orthogonal geodesics for a time $(-\pi/4)$. Moreover, 
Proposition~\ref{pro.Gauss}  implies that, for every $x\in S$, the principal curvatures of the surface $S^{-\pi/4}$ at the point $\phi^{-\pi/4}_S(x)$ are 
$$\lambda_{-\pi/4}(x)=\frac{\lambda(x)-1}{1+\lambda(x)}\quad\mbox{and}\quad \mu_{-\pi/4}(x)=\frac{\mu(x)-1}{1+\mu(x)}.$$
In particular, the Cauchy surface $S^{-\pi/4}$  is strictly convex and has constant K-curvature $-1$. 
It follows immediately that the Cauchy surface $\Sigma^{-\pi/4}$ is also stricty convex, and has constant K-curvature $-1$. 

The same arguments show that the Cauchy surface $\Sigma^{\pi/4}$ is well-defined, strictly concave, and has constant K-curvature $-1$.
\end{proof}

We are left to prove Lemma~\ref{p.reste-dans-domaine-invisible}. The key point is the following~:

\begin{sublemma}
\label{l.immersion=plongement}
Let $\wt \Sigma$ be the (abstract) universal cover of the surface $\Sigma$ and $\phi:\widetilde \Sigma\to \AdS$ be a $\Gamma$-equivariant\footnote{By such, we mean that, for every $\wt x\in \widetilde \Sigma$ and every $\gamma\in\Gamma$, one has $\phi(\gamma.x)=\rho(\gamma).\phi(\wt x)$.} immersion, such that $\phi(\wt \Sigma)$ is spacelike. Then $\phi$ is a proper embedding, and $\phi(\widetilde \Sigma)$ is contained in $E(\Lambda_\rho^+)$ or in $E(\Lambda_\rho^-)$.
\end{sublemma}

\begin{proof}[Proof of Sub-lemma~\ref{l.immersion=plongement}]
Since $\phi(\widetilde \Sigma)$ is spacelike, the Lorentzian metric of $\AdS$ induces a Riemannian metric on $\phi(\widetilde \Sigma)$. Let us denote by $g$ the pull-back (by $\phi$) of this Riemannian metric  on $\widetilde \Sigma$. Since $\phi$ is $\Gamma$-equivariant, this Riemannian metric $g$ is $\Gamma$-invariant. And since $\Gamma$ is co-compact, $g$ is complete. Therefore, $\phi$ is a locally isometric immersion of the complete Riemannian surface $(\wt \Sigma,g)$ in $\AdS$. By a lemma of Mess (see~\cite[Lemma~6]{Mes}), this implies that $\phi$ is a proper embedding, and that $\phi(\widetilde \Sigma)$ is achronal\footnote{Remember that this means that every timelike curve in $\wt\AdS$ cannot intersect a lift of $\phi(\widetilde \Sigma)$ at two different points. We define this property in $\wt\AdS$ (rather than in $\AdS$) because $\AdS$ itself is not a causal space: any two points of $\AdS$ can be joined by a  timelike curve in $\AdS$.}.
In particular, $\phi(\widetilde \Sigma)$ is a closed achronal surface in $\AdS\simeq\ADS$. As explained in \S\ref{ss.AdS-spacetimes}, this implies that the boundary in $\ADS \cup \partial\ADS$ is either the curve $\Lambda_\rho^+$ or the curve~$\Lambda_\rho^-$ (see~\cite[Theorem~10.13]{barBTZ}). This implies that the Cauchy development $D(\phi(\widetilde \Sigma))$ is contained in $E(\Lambda_\rho^+)$ or $E(\Lambda_\rho^-)$ (see~\cite[Proposition~5.18]{BBZ}); in particular, the surface $\phi(\widetilde \Sigma)$ is contained in $E(\Lambda_\rho^+)$ or $E(\Lambda_\rho^-)$.
\end{proof}

\begin{proof}[Proof of Lemma~\ref{p.reste-dans-domaine-invisible}]
Let $\widetilde \Sigma$ be the (abstract) universal cover of $\Sigma$, and choose a $\Gamma$-equivariant homeomorphism $\phi_0:\widetilde \Sigma\to S$. For every $t$ between $0$ and 
$t_0$, the map $\phi_S^t:S\to S^t$ is $\rho(\Gamma)$-equivariant (since the group $\rho(\Gamma)$ acts by isometries of $\AdS$). Therefore, the map $\phi_t:=\phi_S^t\circ\phi_0:\widetilde \Sigma\to\AdS$ is a $\Gamma$-equivariant immersion, and $S^t=\phi_t(\widetilde \Sigma)$ is spacelike. By Lemma~\ref{l.immersion=plongement}, it follows that $S^t$ is properly embedded and contained in $E(\Lambda_\rho^+)$ or $E(\Lambda_\rho^-)$. Now, the open sets  $E(\Lambda_\rho^+)$ and $E(\Lambda_\rho^-)$ are disjoint. The surface $S^t$ depends continuously on $t$, and is contained in $E(\Lambda_\rho^+)$ for $t=0$. Hence, $S^t$ is contained in $E(\Lambda_\rho^+)$ for every $t$ between $0$ and~$t_0$. In particular, $S^{t_0}$ is contained in $E(\Lambda_\rho^+)$.
So, we have proved that the map $\phi_{t_0}:\widetilde \Sigma\to\AdS$ is a $\rho(\Gamma)$-equivariant embedding, and that the spacelike surface $S^{t_0}=\phi_{t_0}(\widetilde \Sigma)$ is  contained in $E(\Lambda_\rho^+)$. It follows that $S^{t_0}$ projects to a compact spacelike surface in $M\simeq \rho(\Gamma)\backslash E(\Lambda_\rho^+)$. To conclude the proof, we recall that every compact spacelike surface in a MGHC spacetime  is a Cauchy surface.
\end{proof}

\subsection{Proof of Proposition~\ref{p.perturbation}}

${}$

We consider a $3$-dimensional non-elementary MGHC spacetime $(M,g)$ with constant K-curvature $\Lambda$. We assume that $(M,g)$ admits a Cauchy surface $\Sigma$ which is strictly convex (i.e. has negative principal curvatures)  and has constant K-curvature $\kappa<-\Lambda$. Note that, since $\Sigma$ is strictly convex, the mean curvature of $\Sigma$ is negative. 

For $t$ small enough, we define the map $\phi_\Sigma^t$ and the surface $\Sigma^t$ as in Subsection~\ref{section.pushing}. For $t<0$, the Cauchy surface $\Sigma^t$ is of course in the past of $\Sigma$. For $t$ small enough, the principal curvatures of $\Sigma^t$ are close to those of $\Sigma$. In particular, for $t$ small enough, the Cauchy surface $\Sigma^t$ is strictly convex. Moreover, Corollary~\ref{cor.Gausscroit} implies that, for $t<0$ small enough, the K-curvature curvature of $\Sigma^t$ is strictly bounded from above by $\kappa$ (see also Remark~\ref{rem.Gausscroit}). This completes the proof of Proposition~\ref{p.perturbation}.
\fin

\begin{remark}
In the case here $\Lambda$ is negative ($\AdS$ spacetimes), there is an analog statement concerning concave surfaces: \\
\emph{Assume that $(M,g)$ admits a stricly concave Cauchy surface $\Sigma$ with constant K-curvature $\kappa$. Then, in the future of  $\Sigma$ (and arbitrarily close to $\Sigma$) one can find a strictly concave Cauchy surface whose K-curvature is strictly bounded from below by $\kappa$.}
\end{remark}


 \section{K-slicings of flat and $\dS$ spacetimes}
 \label{s.proof-main-1}

 We are now ready to prove the existence of K-slicings of non-elementary $3$-dimensional MGHC spacetimes of constant curvature curvature. In the present \S, we will only consider the cases of flat and locally de Sitter spacetimes (items~1 and~2 of Theorem~\ref{theo.main}). The case of locally anti-de Sitter spacetime (item~3 of Theorem~\ref{theo.main}) will be treated in \S\ref{end}. 

 All along this section, we consider a non-elementary $3$-dimensional MGHC spacetime $(M,g)$ of constant curvature curvature $\Lambda\geq 0$. Reversing time-orientation if necessary, we assume that $(M,g)$ is future complete. As explained in section~\ref{s.ingredients}, we will first get a local K-slicing (i.e. a K-slicing of some open subset of $M$), and then extend this local K-slicing to a global one.

\subsection{Local K-slicings.} 

\begin{definition}
A \emph{local K-slicing} in $M$ is a $1$-parameter family $(\Sigma_\kappa)_{\kappa\in I}$ of Cauchy surfaces in $M$ such that:
\begin{enumerate}
\item the parameter set $I$ is an interval included in $(-\infty,-\Lambda)$,
\item for every $\kappa\in I$,  the Cauchy surface $\Sigma_\kappa$ is convex and has constant K-curvature equal to $\kappa$.
\end{enumerate}
The set $U=\bigcup_{\kappa\in I} \Sigma_\kappa$ is called the \emph{support} of the local K-slicing. \end{definition}

Note that, in the above definition, one does not require the $\Sigma_\kappa$'s to be pairwise disjoint, or to depend continously on $\kappa$. Actually, these properties are automatically satisfied:

\begin{lemma}
\label{l.monotonicity}
Every local K-slicing $(\Sigma_\kappa)_{\kappa\in I}$ satisfies the following \emph{monotonicity property}: for every $\kappa_1,\kappa_2\in I$ such that $\kappa_1<\kappa_2$, the Cauchy surface $\Sigma_{\kappa_1}$ is strictly in the past of the Cauchy surface $\Sigma_{\kappa_2}$. 
\end{lemma}

In particular, if $(\Sigma_\kappa)_{\kappa\in I}$ is a local K-slicing, then the $\Sigma_\kappa$'s are pairwise disjoint.

\begin{proof}[Proof of Lemma~\ref{l.monotonicity}]
Consider a local K-slicing $(\Sigma_\kappa)_{\kappa\in I}$ in $M$, pick up two real numbers $\kappa_1,\kappa_2\in I$ such that $\kappa_1<\kappa_2$, and assume that $\Sigma_{\kappa_1}$ is not strictly in the past of $\Sigma_{\kappa_2}$.

Since $M$ is future complete, the map $\phi^t=\phi^t_{\Sigma_{\kappa_2}}$ does exists for every $t\geq 0$. Moreover, since $\Sigma_{\kappa_2}$ is convex and $\Lambda=0$ or $1$, Proposition~\ref{p.pushing-convex} implies that $\phi_{\Sigma_{\kappa_2}}^t$ is an embedding for every $t\geq 0$. So, for every $t\geq 0$,  $\Sigma_{\kappa_2}^t:=\phi_{\Sigma_{\kappa_2}}^t(\Sigma_{\kappa_2})$ is a smooth Cauchy surface. For $t>0$ large enough, the Cauchy surface $\Sigma_{\kappa_2}^t$ is strictly in the future of the Cauchy surface $\Sigma_{\kappa_1}$. Therefore, we may consider the infimum $t_{min}$ of all real numbers $t>0$ such that $\Sigma_{\kappa_2}^t$ is in the future of $\Sigma_{\kappa_1}$. Clearly, the surfaces $\phi_{t}(\Sigma_{\kappa_2})$ and $\Sigma_{\kappa_1}$ intersect at some point $x$ (otherwise, $t_{min}$ would not be minimal). From Proposition \ref{p.pushing-convex} and Remark~\ref{rem.Gausscroit}, the K-curvature of $\Sigma_{\kappa_2})$ is strictly bounded from below by $\kappa_2$. Since $\kappa_1<\kappa_2$, it follows in particular that the K-curvature of the surface $\Sigma_{\kappa_2}^t$ at $x$ is strictly bigger than those of the surface $\Sigma_{\kappa_1}$. This contradicts the maximum principle (see~\S\ref{ss.maximum}) since the surface $\Sigma_{\kappa_2}^t$ is in the future of the surface~$\Sigma_{\kappa_1}$. 
\end{proof}

\begin{remark}
\label{r.unique}
Actually, the argument of the above proof shows that: if $\Sigma,\Sigma'Ä$ are two convex Cauchy surfaces in $M$, whose K-curvatures are strictly bounded from above by $-\Lambda$, and such that the supremum of the K-curvature of $\Sigma$ is smaller than the infimum of the K-curvature of $\Sigma'$, then $\Sigma$ is in the past of $\Sigma'$.

In particular, for every $\kappa<-\Lambda$, there exists at most one convex Cauchy surface with constant K-curvature $\kappa$.
\end{remark}

\begin{lemma}
\label{l.continuous-foliation}
If $(\Sigma_\kappa)_{\kappa\in I}$ is a local K-slicing in $M$, then the surface $\Sigma_\kappa$ depends continuously on $\kappa$.
\end{lemma}

\begin{proof}
Consider $\kappa_0\in I$, and a neighbourhood $V$ of the Cauchy surface $\Sigma_{\kappa_0}$. We have to prove that, for $\kappa$ close enough to $\kappa_0$, the Cauchy surface $\Sigma_\kappa$ is contained in $V$.

By definition of a local K-slicing, the Cauchy surface $\Sigma_{\kappa_0}$ is convex, and has constant K-curvature $\kappa$.  
For $t$ small enough, the Cauchy surface $\Sigma_{\kappa_0}^t=\phi_{\Sigma_{\kappa_0}}^t(\Sigma_{\kappa_0})$ is well defined, 
contained in $V$, convex, has K-curvature strictly bounded from above by $-\Lambda$. Moreover, by Remark~\ref{rem.Gausscroit}, 
for $t>0$ (resp. $t<0$) small enough, the K-curvature Cauchy surface $\Sigma_{\kappa_0}^t$ is strictly bounded from below by $\kappa$ . 
Therefore, there exists $\eta>0$ and $\epsilon>0$, such that the surfaces $\Sigma_{\kappa_0}^{-\eta}$ and $\Sigma_{\kappa_0}^\eta$ are contained in $V$, and such that $(\Sigma_{\kappa_0}^{-\eta}, \Sigma_{\kappa_0}^\eta)$ is a pair of $\kappa$-barriers for every $\kappa\in (\kappa_0-\epsilon,\kappa_0+\epsilon)$. By Theorem~\ref{t.barriers}, it follows that, for every $\kappa\in (\kappa_0-\epsilon,\kappa_0+\epsilon)$, the neighbourhood $V$ contains a convex Cauchy surface with constant K-curvature $\kappa$. By the uniqueness explained in Remark~\ref{r.unique}, it follows that the Cauchy surface $\Sigma_\kappa$ is contained in $V$ for every $\kappa\in (\kappa_0-\epsilon,\kappa_0+\epsilon)$.
\end{proof}

Lemma~\ref{l.monotonicity} and~\ref{l.continuous-foliation} show that a local K-slicing is indeed a \emph{slicing} (i.e. a \emph{foliation}) of some open subset $U$ of $M$. Moreover, Lemma~\ref{l.continuous-foliation} implies that the support $U$ of a local K-slicing $(\Sigma_\kappa)_{\kappa\in I}$ has ``no hole": for every $\kappa_1,\kappa_2\in I$ such that $\kappa_1<\kappa_2$, the intersection $I^+(\Sigma_{\kappa_1})\cap I^-(\Sigma_{\kappa_2})$ is contained in $U$.

\subsection{Construction of  a local K-slicing}  
\label{monotone} 

${}$

Theorem~\ref{th.asymptotic-barriers} provides us with a non-empty open interval $I_0:=]-\Lambda-\epsilon,-\Lambda[$  such that, for every $\kappa\in I_0$, there exists a pair of $\kappa$-barriers. Using these barriers and Theorem~\ref{t.barriers}, we get obtain for every $\kappa\in I_0$ a convex Cauchy surface $\Sigma_\kappa$ with constant K-curvature equal to $\kappa$. The family $(\Sigma_\kappa)_{\kappa\in I_0}$ is by definition a local K-slicing. We denote by $U_0$ the support of this local K-slicing.

\begin{lemma}
The set $U_0$ is a neighbourhood of the future end of $M$, i.e. it contains the whole future of a Cauchy surface of $M$.
\end{lemma}

\begin{proof}
Thanks to Lemmas~\ref{l.monotonicity} and~\ref{l.continuous-foliation}, it is enough to prove that the Cauchy surface $\Sigma_\kappa$ ``escapes towards the future end of $M$" when $\kappa\to-\Lambda$: more precisely, if $K$ is a compact subset of $M$,  then $K$ is in the past of the surface $\Sigma_\kappa$ for every $\kappa$ close enough to $-\Lambda$.

In order to prove this, we consider any convex Cauchy surface $\widehat\Sigma$ in $M$ such that the curvature is strictly bounded from above by  $-\Lambda$ (for example, one might take $\widehat\Sigma$ to be a leaf of our local K-slicing). The map $\phi_{\widehat\Sigma}^t:\widehat\Sigma\to M$ is well-defined and is an embedding for every $t\geq 0$ (item~1 of  Proposition~\ref{p.pushing-convex}). For $t\geq 0$, denote by $\kappa_{max}(t)$ the supremum of the K-curvature of the surface $\widehat\Sigma^t=\phi_{\widehat\Sigma}^t(S)$. By item 2 of Proposition~\ref{p.pushing-convex}, 
one has $\kappa_{max}(t)<-\Lambda$ for every $t\geq 0$. By Remark~\ref{r.unique}, the surface $\Sigma_\kappa$ is in the future of $\widehat\Sigma^t$ for every $\kappa>\kappa_{max}(t)$. 

But of course, by definition of the surface $\widehat\Sigma^t$, it ``escapes towards the future end of $M$" when $t\to\infty$. Therefore, the Cauchy surface $\Sigma_\kappa$ escapes towards the future end of $M$ when $\kappa\to -\Lambda$. 
\end{proof}

 \subsection{Getting a global K-slicing} 
 \label{ss.global-flat-ds}

${}$

We say that a local K-slicing $({\Sigma'}_\kappa)_{\kappa\in I'}$ \emph{extends} a local K-slicing $({\Sigma}_\kappa)_{\kappa\in I}$ if $I\subset I'$ and ${\Sigma'}_\kappa={\Sigma}_\kappa$ for $\kappa\in I$. Our goal is to prove that our local K-slicing $(\Sigma_{\kappa})_{\kappa\in I_0}$ can be extended to get  K-slicing $(\Sigma_{\kappa})_{\kappa\in ]-\infty,-\Lambda[}$ whose support is the whole manifold $M$. 

For this purpose, we consider a local K-slicing $(\Sigma_\kappa)_{\kappa\in I}$ which extends our initial local K-slicing $(\Sigma_\kappa)_{\kappa\in I_0}$, and is \emph{maximal} among the local K-slicings extending $(\Sigma_\kappa)_{\kappa\in I_0}$. We denote by $U$ the support of $(\Sigma_\kappa)_{\kappa\in I}$. We have to prove that $I=(-\infty,-\Lambda)$ and $U=M$.

Assume that $I\neq (-\infty,-\Lambda)$. Then, there exists $\alpha\in ]-\infty,-\Lambda[$ such that $I=]\alpha,-\Lambda[$ or $I=[\alpha,-\Lambda[$. On the one hand, Theorem~\ref{th.sequence-constant} tells us that the first possibility cannot occur. Indeed, if $I=]\alpha,-\Lambda[$, then there exists a convex Cauchy surface $\Sigma_\alpha$ with constant K-curvature $\alpha$ such that $I^+(\Sigma_\alpha)=U$. Clearly, $(\Sigma_{\kappa})_{\kappa\in [\alpha,-\Lambda)}$ is a local K-slicing which extends the local K-slicing $(\Sigma_{\kappa})_{\kappa\in I}$, contradicting the maximality of $(\Sigma_{\kappa})_{\kappa\in I}$. On the other hand, Proposition~\ref{p.perturbation} implies that the second possibility cannot occur either. Indeed, if $I=[\alpha,-\Lambda[$, then Proposition~\ref{p.perturbation} provides us with a surface $\Sigma$ such that $(\Sigma,\Sigma_\alpha)$ is a pair of $\kappa$ barriers for every $\kappa$ smaller than $\alpha$ and close enough to $\alpha$. Using Theorem~\ref{t.barriers}, we get a convex Cauchy surface $\Sigma_{\kappa}$ with constant K-curvature $\kappa$ for every $\kappa\in ]\alpha',\alpha[$ for some $\alpha'<\alpha$. So, we get a local K-slicing $(\Sigma_\kappa)_{\kappa\in ]\alpha',-\Lambda[}$ which extends $(\Sigma_\kappa)_{\kappa\in I}$, contradicting the maximality of $(\Sigma_\kappa)_{\kappa\in I}$.

So, we have proved that $I=]-\infty,-\Lambda[$. Using once again Theorem~\ref{th.sequence-constant}, we see that this implies that $U=M$. Henceforth, $(\Sigma_\kappa)_{\kappa\in I}$ is a global K-slicing of our spacetime $M$. This completes the proof of of Theorem~\ref{theo.main} in the flat case and in the locally de Sitter case.


\section{K-slicings of $\AdS$-spacetime}
\label{end}
\label{s.proof-main-AdS}

The strategy used in \S\ref{s.proof-main-1} to get K-slicings on flat and $\dS$-spacetimes does not fully apply in the $\AdS$-case. The main problem is the lack of an analog of 
Proposition~\ref{p.pushing-convex} in the $\AdS$-setting\footnote{There can be no analog of Proposition~\ref{p.pushing-convex} in the $\AdS$-setting, simply because MGHC spacetimes of $\AdS$-type are neither future complete, nor past complete.}. To bypass this problem, we will develop an alternative argument based on the duality between convex and concave Cauchy surfaces in MGHC $\AdS$-spacetimes.

\subsection{Duality in $\AdS$-spacetimes}

\subsubsection{Duality between points and totally geodesic planes}
\label{sss.point-plane-ads}
We consider the quadratic form $Q_{2,2}=-x_1^2-x_2^2+x_3^2+x_4^2$ on $\RR^4$. Recall that $\AdS$ is the quadric $\{Q_{2,2}=-1\}$ endowed with the Lorentzian metric induced by $Q_{2,2}$. For every set $A\subset\RR^4$, we denote by $A^\perp$ the $Q_{2,2}$-orthogonal of $A$ in $\RR^4$.

Let $x$ be a point in $\AdS\subset \RR^4$. Then $x^\perp$ is an hyperplane in $\RR^4$, and the restriction of $Q_{2,2}$ to $x^\perp$ has signature $(-,+,+)$. It follows that $x^\perp\cap\AdS$ is the disjoint union of two spacelike totally geodesic planes in $\AdS$. We denote these totally geodesic planes by $x^\flat$ and $x^\sharp$, in such a way that~: if one starts at the point $x$, and follows a future-oriented (resp. past-oriented) timelike geodesic, then one crosses $x^\sharp$ before $x^\flat$ (resp. $x^\flat$ before $x^\sharp$)\footnote{The rough idea is that ``$x^\flat$ is in the past of $x$ and $x^\sharp$ is in the future of $x$". Unfortunately, things are not so simple since $\AdS$ is not a causal spacetime}. 

Conversely, let $P$ be a spacelike totally geodesic plane in $\AdS$. Then $P^\perp$ is a line in $\RR^4$, and the restriction of $Q_{2,2}$ to $P^\perp$ is negative definite. It follows that $P^\perp$ intersects $\AdS$ at two (antipodal) points.  Every geodesic of $\AdS$ orthogonal to $P$ passes through these two points. We denote these two points by $P^\flat$ and $P^\sharp$ in such a way that~: if one starts at a point of $P$, and follows a future-oriented (resp. past-oriented) timelike geodesic orthogonal to $P$, then one passes through $P^\sharp$ before passing through $P^\flat$ (resp. passes through $P^\flat$ before passing through $P^\sharp$)

\begin{remark}
\label{f.dual-and-Gauss}
It is easy to verify that every geodesic of $\AdS$ which is orthogonal $P$ passes through the two points $P^\flat$ and $P^\sharp$. Moreover, the length of a geodesic segment, 
orthogonal to $P$, and going from $P$ to $P^\flat$ or $P^\sharp$ is exactly $\pi/2$. In other words, $P^\sharp$ (resp. $P^\flat$) is obtained by pushing $P$ along the orthogonal geodesics for a time $\pi/2$ (resp. $-\pi/2$).
\end{remark}

It clearly follows from the definitions above that, for every point $x\in\AdS$ and every spacelike totally geodesic plane $P\subset\AdS$, one has 
\begin{equation}
\label{e.duality}
(x^{\flat})^\sharp=(x^\sharp)^\flat=x\quad\mbox{and}\quad(P^{\flat})^\sharp=(P^\sharp)^\flat=P
\end{equation}

\subsubsection{Duality between strictly convex and concave surfaces in $\AdS$.}
\label{sss.surfaces-ads}
Let $S$ be a (smooth) spacelike surface in $\AdS$. For every $x\in S$, we denote by $P_{S,x}$ the  totally geodesic plane in $\AdS$ which is tangent to $S$ at~$x$, and we recall that $(P_{S,x})^\flat$ and $(P_{S,x})^\sharp$ are two antipodal points in $\AdS$. Then we define the \emph{past dual set} $S^\flat$ and the \emph{future dual set} $S^\sharp$ of $S$ as follows:
$$S^\flat  = \{(P_{S,x})^\flat \mid x\in S\}\quad , \quad S^\sharp  = \{(P_{S,x})^\sharp \mid x\in S\}.$$
Clearly, $S^\flat$ and $S^\sharp$ are two subsets of $\AdS$, which are the image of each other under the antipodal map. Note that $S^\sharp$ and $S^\flat$ are subsets of $\AdS$ not surfaces in general; for example, if $S$ is a totally geodesic plane, then $S^\sharp$ (resp. $S^\flat$) is a single point. 

\begin{remark}
\label{r.support-plane}
A point $y$ is in $S^\flat$ (resp. $S^\sharp$) if and only if the totally geodesic plane $y^\sharp$ (resp. $y^\flat$) is tangent to $S$ at some point. This follows from the definitions of $S^\flat,S^\sharp$ and equalities~\eqref{e.duality}.
\end{remark}

\begin{proposition}
\label{p.dual-convex}
If $S$ is a strictly convex (resp. strictly concave) spacelike surface, then the sets $S^\sharp$ and $S^\flat$ are strictly concave (resp. strictly convex) spacelike surfaces. 
\end{proposition}

In order to prove this proposition, we introduce the set $\cP_{\AdS}$ of all spacelike totally geodesic planes in $\AdS$. Note that $\cP_{\AdS}$ has a natural structure of three-dimensional manifold.
For every spacelike surface $S$ in $\AdS$,  we consider the map
$$\begin{array}{lrll}
P_S : &S & \to & \cP_{\AdS}\\
& x & \mapsto & P_{S,x}
\end{array}$$
The following lemma is a simple remark, but plays a crucial role in the proof of Proposition~\ref{p.dual-convex}~:

\begin{lemma}
\label{l.immersion}
Given a spacelike surface $S$ in $\AdS$ and a point $x\in S$, the second fundamental form of $S$ at $x$ is non-degenerate if and only if the derivative at $x$ of the map $P_S$ has maximal rank (i.e. rank $2$).
\end{lemma}

\begin{proof}
Let $\mbox{T}^{-1}\AdS$ be the set of all pairs $(x,n)$ where $x\in\AdS$ and $n$ is a future-oriented unit timelike vector in $\mbox{T}_x\AdS$. Let $\mbox{T}^\cP\AdS$ be the set of all pairs $(x,P)$ where $x\in\AdS$ and $P$ is a spacelike totally geodesic plane containing $x$. For every $x\in\AdS$ and every  spacelike totally geodesic plane $P$ containing $x$, let $n_{P,x}$ be the future-oriented unit timelike vector orthogonal to $P$ at $x$. The map $(x,P)\mapsto (x,n_{P,x})$ is clearly a diffeomorphism between $\mbox{T}^\cP\AdS$ and $\mbox{T}^{-1}\AdS$. For every spacelike surface $S$, this diffeomorphism induces an identification between the derivative of the map $P_S$ and the Weingarten operator of $S$. Therefore, for every $x\in S$, the derivative at $x$ of $P_S$ has maximal rank if and only if the Weingarten operator of $S$ at $x$ has maximal rank.
\end{proof}

\begin{proof}[Proof of Proposition~\ref{p.dual-convex}]
We assume $S$ is a strictly convex spacelike surface in $\AdS$. We will prove that $S^\sharp$ is a strictly concave spacelike surface. The same arguments apply for $S^\flat$ and/or in the case where $S$ is strictly concave.

\medskip

\noindent\textit{First step: $S^\sharp$ is an immersed surface.}
Since $S$ is strictly convex, for every $x\in S$, the second fundamental form of $S$ at $x$ is non-degenerate. Together with Lemma~\ref{l.immersion}, this shows that, for every $x\in S$, the derivative at $x$ of the map $P_S$ has maximal rank. Therefore, the map $P_S:x\mapsto P_{S,x}$ is an immersion of $S$ in $\cP_{\AdS}$. Moreover, the map $P\mapsto P^\sharp$ is obviously a local diffeomorphism between $\cP_{\AdS}$ and $\AdS$. Therefore the map 
$$\begin{array}{rcl}
S & \to & \AdS \\
 x & \mapsto & (P_{S,x})^\sharp
 \end{array}$$
 is an immersion of $S$ in $\AdS$. The range of this map is the set $S^\sharp$ (by the very definition of $S^\sharp$). Therefore $S^\sharp$ is an immersed surface.

\medskip

\noindent\textit{Second step: $S^\sharp$ is an injectively immersed spacelike concave surface.} 
Let $y\in S^\sharp$. By definition of $S^\sharp$, there exists a point $x\in S$ such that  
$y=(P_{S,x})^\sharp$. We will prove that the totally geodesic plane $x^\sharp$ meets $S^\sharp$ at $y$, and that $S^\sharp$ is contained in the past of $x^\sharp$ in a neighbourhood of $y$. 

Since the point $x$ belongs to the totally geodesic plane $P_{S,x}$, the point $y=(P_{S,x})^\sharp$ belongs to the totally geodesic plane $x^\sharp$. In other words, the totally geodesic plane $x^\sharp$ meets the surface $S^\sharp$ at $y$. Now let us consider a sequence $(y_k)_{k\in\NN}$ of points of $S^\sharp$ converging to $y$ when $k\to\infty$. By Remark~\ref{r.support-plane}, for every $k\geq 0$, the totally geodesic plane $y_k^\flat$ is tangent to the surface $S$ at some point $x_k$. Since $S$ is strictly convex, this implies that $S$ is in the future of the plane $y_k^\flat$ in some neighbourhood $U_k$ of the point $x_k$ (and the size of the neighbourhood $U_k$ does not depend of $k$ provided that the $x_k$'s stay in a compact subset of $S$).  Moreover, if $k$ is large, the totally geodesic plane $y_k^\flat$ is close to the totally geodesic plane $y^\flat=((P_{S,x})^\sharp)^\flat=P_{S,x}$. Since $S$ is \emph{strictly} convex, this implies that $x_k$ is close to $x$ when $k$ is large. It follows that, for $k$ large enough, the point $x$ is in the neighbourhood $U_k$. In particular, for $k$ large enough, the point $x$ is in the future of the totally geodesic plane $y_k^\flat$. It follows that, for $k$ large enough, the point $y_k=(y_k^\flat)^\sharp$ is in the past of the totally geodesic plane $x^\sharp$.  

So, we have proved that the totally geodesic plane $x^\sharp$ meets the immersed surface $S^\sharp$ at $y$, and that there exists a neighbourhood $U$ of the point $y$ in $S^\sharp$ such that $S^\sharp$ is contained in the past of $x^\sharp$ in $U$. It follows that $x^\sharp$ is the unique totally geodesic plane which is tangent to $S^\sharp$ at $y$, and that $S^\sharp$ is ``locally contained" in the past of this totally geodesic plane. Since $y$ is an arbitrary point in $S^\sharp$, this shows that $S^\sharp$ is injectively immersed, spacelike and concave (see \S\ref{sub.concan}). 

\medskip

\noindent\textit{Third step: $S^\sharp$ is strictly concave.} Since we already know that $S^\sharp$ is concave, we only need to prove that the second fundamental form of $S^\sharp$ is non-degenerate at each point of $S^\sharp$. According to Lemma~\ref{l.immersion}, it is equivalent to prove that the map $P_{S^\sharp}:y\mapsto P_{S^\sharp,y}$ is an immersion of $S^\sharp$ in $\cP_{\AdS}$.

We have proved during the second step that, for every point $y\in S^\sharp$, if $x$ is the unique point of the surface $S$ such that $y=(P_{S,x})^\sharp$, then $x^\sharp$ is  the unique totally geodesic plane tangent to $S^\sharp$ at $y$, that is $P_{S^\sharp,y}=x^\sharp$. Moreover, we have proved in the first and second step above that the map $x \mapsto (P_{S,x})^\sharp$
 is an injective immersion of the surface $S$ in $\AdS$, and that the range of this immersion is the surface $S^\sharp$. It follows that the map $(P_{S,x})^\sharp \mapsto x$ is an immersion of $S^\sharp$ in $\AdS$.
Since the map $x\mapsto x^\sharp$ is obviously a local diffeomorphism from $\AdS$ to $\cP_{\AdS}$, this shows that the map 
  $$P_{S^\sharp} : 
  \begin{array}[t]{cll}
  S^\sharp & \to & \cP_{\AdS} \\
  y=(P_{S,x})^\sharp & \mapsto  & x^\sharp=P_{S^\sharp,y}
  \end{array}$$
   is an immersion. Therefore, the surface $S^\sharp$ is stricly convex.
\end{proof}

During the above proof of Proposition~\ref{p.dual-convex}, we have seen that, for every point $y\in S^\sharp$, if $x$ is the unique point of the surface $S$ such that $y=(P_{S,x})^\sharp$, then $P_{S^\sharp,y}=x^\sharp$, and therefore $(P_{S^\sharp,y})^\flat=(x^\sharp)^\flat=x$. This shows that~:

\begin{corollary}
\label{c.inverse}
If $S$ is a strictly convex (resp. strictly concave) spacelike surface, then 
$$(S^\sharp)^\flat = S$$ 
and the map $\left |\begin{array}{lll} S^\sharp & \to & (S^\sharp)^\flat \\ y & \mapsto & (P_{S,y})^\flat\end{array}\right.$ is the inverse of the map $\left |\begin{array}{lll} S & \to & S^\sharp \\ x & \mapsto & (P_{S,x})^\sharp\end{array}\right.$.
\end{corollary}

Of course, similar arguments show that:

\begin{corollary}
If $S$ is a strictly convex (resp. strictly concave) spacelike surface, then 
$$(S^\flat)^\sharp = S$$ 
and the map $\left |\begin{array}{lll} S^\flat & \to & (S^\flat)^\sharp \\ y & \mapsto & (P_{S,y})^\sharp\end{array}\right.$ is the inverse of the map $\left |\begin{array}{lll} S & \to & S^\flat \\ x & \mapsto & (P_{S,x})^\flat\end{array}\right.$.
\end{corollary}

\begin{proposition}
\label{p.curvature-dual}
Let $S$ be a strictly convex or strictly concave spacelike surface. Let $x_0$ be a point in $S$, and $y_0=(P_{S,x_0})^\sharp$ (resp. $y_0=(P_{S,x_0})^\flat$) be the corresponding point in $S^\sharp$ (resp. $S^\flat$). If  $\lambda,\mu$ are the principal curvatures of~$S$ at $x_0$, then the principal curvatures of $S^\sharp$ (resp. $S^\flat$) at $y_0$ are $-\lambda^{-1},-\mu^{-1}$. 
\end{proposition}

\begin{proof}
For every $x\in S$, let $n_{S,x}\in T_x\AdS$ be the future-directed unit normal vector of $S$ at $x$. Since $\AdS$ is a quadric in $\RR^4$, we can see $n_{S,x}$ as an element of $T_x\RR^4$. The key point of the proof is the following observation~:

\medskip

\noindent \textit{Claim. For every $x\in S$, the canonical isomorphism between $T_x\RR^4$ and $\RR^4$ maps the vector $n_{S,x}\in T_x\RR^4$ to the point $(P_{S,x})^\sharp=-(P_{S,x})^\flat\in\RR^4$.}  

\medskip

\noindent Let $y\in\RR^4$ be the image of the vector $n_{S,x}$ under the canonical isomorphism between $T_x\RR^4$ and $\RR^4$. For every subspace $V_x$ of $T_x\RR^4$, denote by $\overline{V_x}\subset \RR^4$ the image of $V_x$ under this isomorphism. For every $A\subset\RR^4$, denote by $A^\perp$ the $Q_{2,2}$ orthogonal of $A$. Since $P_{S,x}$ is a totally geodesic plane of $\AdS$, there exists a hyperplane $H$ of $\RR^4$ such that $P_{S,x}$ is a connected component of $H\cap\AdS$. Clearly, $H$ is the vector subspace of $\RR^4$ spanned by $P_{S,x}$, and $H=\RR.x \oplus \overline{T_x P_{S,x}}$.  The vector $n_{s,x}$ belongs to $T_x\AdS$. Hence the point $y$ belongs to $\overline{T_x\AdS}=x^\perp$. Moreover the vector $n_{s,x}$ is orthogonal to $T_x S= T_x P_{S,x}$. Hence the point $y$ belongs to $(\overline{T_x P_{S,x}})^\perp$. So we have proved that $y$ belongs to $(\RR.x\oplus \overline{T_x P_{S,x}})^\perp = H^\perp = (P_{S,x})^\perp$. Moreover the vector $n_{S,x}$ has norm $-1$, and the Lorentzian metric on $\AdS$ is induced by the quadratic form $Q_{2,2}$. Hence the point $y$ belongs to $\AdS=\{z\in\RR^4\mid Q_{2,2}(z)=-1\}$. So we have proved $y$ belongs to $(P_{S,x})^\perp\cap\AdS=\{(P_{S,x})^\flat , (P_{S,x})^\sharp\}=\{(P_{S,x})^\flat , (P_{S,x})^\sharp\}$. Finally, the fact that $n_{S,x}$ is future-oriented means that $y$ is equal to $(P_{S,x})^\sharp$ rather than $(P_{S,x})^\flat$. This completes the proof of the claim.

\medskip

Now, for every $x\in\AdS$, we identify $T_x\RR^4$ with $\RR^4$ (using the canonical isomorphism between these two spaces). The above claim shows that the 
Weingarten map of $S$ at $x_0$ is identified with the derivative at $x_0$ of the map $x\mapsto (P_{S,x})^\sharp$ 
(which also the opposite of the derivative of the map $x\mapsto (P_{S,x})^\flat$, since $(P_{S,x})^\sharp$ and $(P_{S,x})^\flat$ are two antipodal points). 
Of course the same is true if we replace $S$ by $S^\sharp$~: the Weingarten map of $S^\sharp$ at $y_0$ is identified with the derivative at $y_0$ of the map 
$y\mapsto (P_{S^\sharp,y})^\sharp$  (which also the opposite of the derivative of the map $y\mapsto (P_{S^\sharp,y})^\flat$). Together with Corollary~\ref{c.inverse}, this shows that the Weingarten map of $S$ at $x_0$ is identified with the opposite of the inverse of the Weingarten map of $S^\sharp$ at $y_0$. The proposition follows.
\end{proof}

\begin{corollary}
\label{c.dual-constant-curvature}
If $S$ is a spacelike surface with constant K-curvature $\kappa\neq 0$, then $S^\sharp$ and $S^\flat$ are spacelike surfaces with constant K-curvature $\kappa^{-1}$.
\end{corollary}

\begin{proof}
Every surface with non-zero constant K-curvature is either strictly convex or strictly concave. Therefore the corollary is an immediate consequence of Proposition~\ref{p.curvature-dual}.
\end{proof}

It is interesting to observe that the future dual $S^\sharp$~and the past dual $S^\flat$ of a spacelike surface can also 
be defined by pushing along orthogonal geodesics~:

\begin{proposition}
\label{p.dual-Gauss-flow}
For every spacelike surface $S$ in $\AdS$, the future dual $S^\sharp$ (resp. the past dual $S^\flat$) is 
obtained by pushing $S$ along orthogonal geodesic for a time $\pi/2$ (resp. a time $-\pi/2$).
\end{proposition}

\begin{proof}
Let $S$ be a spacelike surface in $\AdS$. For every $x\in S$, we denote by $n_{S,x}$ the future-directed unit normal vector of $S$ at $x$. For every~$t\in\RR$, let
$$S^{t}:=\left\{\exp_x\left(t.n_{S,x}\right) \mid x\in S\right\}$$ the set of points at distance $t$ from $S$ along an orthogonal geodesic
($S^t$ is not necessarily a surface when $|t|$ is large, but this is not relevant here). 
We shall prove that~$S^{\pi/2}=S^\sharp$ (the same arguments apply to show that~$S^{-\pi/2}=S^\flat$).

Consider a point $y$ in $S^\sharp$. By definition of $S^\sharp$, there exists a point $x\in S$ such that $y=P_{S,x}^\sharp$. Since the surface $S$ and the totally geodesic plane $P_{S,x}$ are tangent at $x$, the vector $n_{S,x}$ is orthogonal to $P_{S,x}$. Using Remark~\ref{f.dual-and-Gauss}, it follows that the point $y=P_{S,x}^\sharp$ coincides with the point $\exp_x\left(\frac{\pi}{2}.n_{S,x}\right)$. In particular, $y\in S^{\pi/2}$.  Conversely, consider a point $y\in S^{\pi/2}$. There exists a point $x\in S$ such that $y=\exp_x\left(\frac{\pi}{2}.n_{S,x}\right)$. The same argument as above shows that $(P_{S,x})^\sharp=\exp_x(\frac{\pi}{2}.n_{S,x})=y$. In particular, $y\in S^\sharp$.
\end{proof}

\subsubsection{Duality for convex and concave surfaces in MGHC $\AdS$-spacetimes.}
Now we consider a MGHC spacetime $M$ of constant curvature $-1$. According to Theorem~\ref{t.ads-spacetimes}, there exists a representation 
$\rho:\Gamma\to\op{SL}(2, \RR) \times \op{SL}(2, \RR)$ (where $\Gamma$ is the fundamental group of the Cauchy surfaces of $M$) and an open set $E(\Lambda^+_\rho)\subset\AdS$ such that $M$ is isometric to the quotient $M(\rho):=\rho(\Gamma_g)\backslash E(\Lambda^+_\rho)$.

Let $\Sigma$ be a strictly convex Cauchy surface in $M$. Consider the lift $S$ of $\Sigma$ in $E(\Lambda^+_\rho)\subset\AdS$. Then $S$ 
is a strictly convex spacelike surface in $\AdS$. So we can consider the future dual $S^\sharp$ of $S$ as defined in the previous paragraph. 
By Proposition~\ref{p.dual-convex},  $S^\sharp$  is a strictly concave spacelike surface in $\AdS$. Since the surface $S$ is $\rho(\Gamma)$-invariant,  so is the surface $S^\sharp$ (the duality is clearly a $\mbox{Isom}(Q_{2,2})$-equivariant operation). Moreover, since $S$ is strictly convex, Proposition~\ref{pro.Gauss} shows that $\phi_S^t:S\to\AdS$ is an immersion for every $t\in [0,\pi/2)$, and step~1 of the proof of Proposition~\ref{p.dual-convex} shows  that $\phi_S^{\pi/2}:S\to\AdS$ is an immersion. Hence, Lemma~\ref{p.reste-dans-domaine-invisible} shows that the surface $S^\sharp=S^{\pi/2}$ is contained in $E(\Lambda^+_\rho)$, and projects to a Cauchy surface $\Sigma^\sharp=\Sigma^{\pi/2}$ in
 $M\simeq \Gamma\backslash E(\Lambda^+_{\rho})$. This is a compact strictly concave spacelike surface in $M$, i.e. equivalently, a strictly concave Cauchy surface in $M$. We say that $\Sigma^\sharp$ is the \emph{future dual} of the Cauchy surface $\Sigma$. 

Similarly, one can define  the \emph{past dual} $\Sigma^\flat$ of a strictly concave Cauchy surface $\Sigma$ in $M$. 

\begin{remark}
If $\Sigma$ is a strictly convex Cauchy surface in $M$, and $S$ is a lift of $\Sigma$ in $E(\Lambda^+_\rho)$, then the past dual $S^\flat$ of $S$ is well-defined. Yet, $S^\flat$ is not contained in $E(\Lambda^+_\rho)$ (actually it is contained in $E(\Lambda^-_\rho)$). In particular, $S^\flat$ cannot be projected in $M$. This is the reason why the past dual of $\Sigma$ does not exist. Similarly, the future dual of a strictly concave Cauchy surface in $M$ does not exist.
\end{remark} 

By Corollary~\ref{c.inverse}, for every  strictly convex Cauchy surface $\Sigma$ in $M$, one has $(\Sigma^\sharp)^\flat=\Sigma$. Similarly, for  every  strictly concave Cauchy surface $\Sigma$ in $M$, one has $(\Sigma^\flat)^\sharp=\Sigma$. By Corollary~\ref{c.dual-constant-curvature}, if $\Sigma$ is a strictly convex (resp. concave) Cauchy surface with constant K-curvature $\kappa$, then $\Sigma^\sharp$ (resp. $\Sigma^\flat$) has constant K-curvature $\kappa^{-1}$.

\subsection{Proof of Theorem~\ref{theo.main} in the $\AdS$-case}

${}$

Let $(M,g)$ be a non-elementary MGHC spacetime with constant curvature $\Lambda=-1$. We denote by $M^-$ and $M^+$ the past tight region and the future tight region of $M$. In order to prove Theorem~\ref{theo.main}, we have to construct a K-slicing on $M^-$, and  a K-slicing on $M^+$. The starting point of our construction will be the Cauchy surfaces $\Sigma^-$ and $\Sigma^+$ provided by Proposition~\ref{pro.init-AdS-2}. Recall that $\Sigma^-$ is a convex Cauchy surface and  $\Sigma^+$ is a convex Cauchy surface, both with constant K-curvature $-1$.

\subsubsection{Construction of the K-slicings on $J^-(\Sigma_-)$ and $J^+(\Sigma_+)$}

The Cauchy surface $\Sigma^-$ is convex and has constant K-curvature $-1$. Using Proposition~\ref{p.dual-convex}, we get a convex Cauchy surface $\Sigma^{--}$ in the past of $\Sigma^-$ and a real number $\epsilon>0$, such that $(\Sigma^{--},\Sigma^-)$ is a pair of $\kappa$-barriers for every $\kappa\in (-1-\epsilon,-1)$. Together with Theorem~\ref{t.barriers}, this shows the existence, for every $\kappa\in (-1-\epsilon,-1)$, of a convex Cauchy surface $\Sigma_\kappa^-$ with constant K-curvature $\kappa$. We set $\Sigma_1^-:=\Sigma^-$. Then $(\Sigma_\kappa^-)_{\kappa \in (-1-\epsilon,-1]}$ is a local K-slicing. Now, we consider a local K-slicing $(\Sigma_\kappa)_{\kappa\in I}$ which  extends  $(\Sigma_\kappa)_{\kappa\in  (-1-\epsilon,-1]}$, and is maximal for this property. The same arguments as in \S\ref{ss.global-flat-ds} show that $(-\infty,-1]\subset I$ and that the support of $(\Sigma_\kappa)_{\kappa\in I}$ contains the past $J^-(\Sigma^-)$ of the Cauchy surface $\Sigma^-$. So, we have proved the following: 

\begin{proposition}
\label{p.AdS-local-past}
There exists a local K-slicing $(\Sigma_\kappa^-)_{\kappa\in (-\infty,-1]}$ whose support is the past $J^-(\Sigma^-)$ of the Cauchy surface $\Sigma^-$.\fin
\end{proposition}

Of course, similar arguments show that:

\begin{proposition}
\label{p.AdS-local-future}
There exists a local K-slicing $(\Sigma_\kappa^+)_{\kappa\in (-\infty,-1]}$ whose support is the future $J^+(\Sigma^+)$ of the Cauchy surface $\Sigma^+$.\fin
\end{proposition}

\subsubsection{Construction of the K-slicings on $M^-\setminus J^-(\Sigma^-)$ and $M^+\setminus J^+(\Sigma^+)$}
Proposition~\ref{p.AdS-local-past} provides us with a local K-slicing  $(\Sigma_\kappa^-)_{\kappa\in (-\infty,-1]}$ whose support is the past of the Cauchy surface $\Sigma^-$. To complete the proof of Theorem~\ref{theo.main} in the $\AdS$-case, we need to extend this local K-slicing, and get a K-slicing $(\Sigma_\kappa^-)_{\kappa\in (-\infty,0)}$  whose support is  the whole past tight region $M^-$. For this purpose, we use the duality between convex and concave Cauchy surfaces~: for every $\kappa\in (-1,0)$, we set 
$$\Sigma^-_\kappa:=\left(\Sigma^+_{1/\kappa}\right)^\flat.$$
(observe that this definition is not chasing its own tail since the surface $\Sigma^+_{1/\kappa}$ has already been defined for $\kappa\in (0,1)$; see Proposition~\ref{p.AdS-local-future}).

\begin{claim}
\label{c.AdS-K-slicing}
The families of Cauchy surfaces $(\Sigma^-_\kappa)_{\kappa\in (-\infty,0)}$ is a local K-slicing whose support is included in the past tight region $M^-$. 
\end{claim}

\begin{proof}
For $\kappa\leq -1$, we already know that  the Cauchy surface $\Sigma^-_\kappa$ is a convex and has constant K-curvature $\kappa$. Moreover, for $\kappa\in (-1,0)$, 
we know that the Cauchy surface $\Sigma^+_{1/\kappa}$ is strictly concave and has constant K-curvature $\frac{1}{\kappa}$. Together with Proposition~\ref{p.curvature-dual}, this implies that, for $\kappa\in (-1,0)$, the Cauchy surface $\Sigma^-_\kappa:=(\Sigma^+_{1/\kappa})^\flat$ is a convex 
and has constant K-curvature $\kappa$. This shows that $(\Sigma^-_\kappa)_{\kappa\in (-\infty,0)}$ is a local K-slicing. Since every convex Cauchy surface in $M$ 
is contained in $M^-$, the support of this K-slicing must be contained in $M^-$.
\end{proof}

\begin{claim}
\label{c.AdS-support}
For every point $x\in M^-$, there exists $\kappa\in (-\infty,0)$ such that  $x$ is in the past of the Cauchy surface $\Sigma_\kappa^-$.
\end{claim}

\begin{proof}
We denote by $\check\tau:M\to (0,\pi)$ the reverse cosmological time of $M$. Recall that  the past tight region $M^-$ of $M$ can be characterized as follows:
\begin{equation}
\label{e.reverse-time-tight-region}
M^-=\{x\in M \mid \check\tau(x) > \pi/2\}.
\end{equation}
The support of the local K-slicing $(\Sigma_{1/\kappa}^+)_{\kappa\in [-1,0)}=(\Sigma_\kappa^+)_{\kappa\in (-\infty,-1]}$ is a neighbourhood of the future end of $M$ (recall that this means that it contains the future of a Cauchy surface). Since $\check\tau$ is regular, this implies that 
\begin{equation}
\label{e.limite}
\lim_{\kappa\to 0}\;\sup_{x\in\Sigma_{1/\kappa}^+}\;\check\tau(x) = 0.
\end{equation}
Now, recall that, for every $\kappa\in [-1,0)$, the surface $\Sigma^-_\kappa=(\Sigma^+_{1/\kappa})^\flat$ is obtained by pushing the surface 
$\Sigma^+_{1/\kappa}$ under the time $(-\pi/2)$ along orthogonal geodesics. In particular, every point of $\Sigma^-_\kappa$ is the past end of a geodesic segment orthogonal to $\Sigma^+_{1/\kappa}$ of length $-\pi/2$. This implies that, for every $\kappa\in [-1,0)$, the length of a timelike curve joining the surface $\Sigma^-_\kappa$ to the surface $\Sigma^+_{1/\kappa}$ is at most $\pi/2$ (recall that, in a globally hyperbolic spacetime, the supremum of the lengths of the timelike curves joining a surface $\Sigma$ to a point $p$ is realized by a timelike geodesic orthogonal to $\Sigma$). Together with~\eqref{e.limite}, this implies that 
\begin{equation}
\label{e.limite-2}
\lim_{\kappa\to 0}\;\sup_{x\in\Sigma_\kappa^-}\;\check\tau(x) \leq \frac{\pi}{2}
\end{equation}
(actually, this inequality turns out to be an equality since, for evey $\kappa$, the surface $\Sigma_\kappa^-$ is convex, thus contained in $M^-=\{x\in M \mid \check\tau(x) > \pi/2\}$). Inequality~\eqref{e.limite-2} and equality~\eqref{e.reverse-time-tight-region} show that for every point $x\in M^-$, there exists $\kappa\in  (-1,0)$ such that $x$ is in the past of the Cauchy surface $\Sigma_\kappa^-$. 
\end{proof}

Claim~\ref{c.AdS-support} together with Lemma~\ref{l.continuous-foliation} show that the support of the local K-slicing $(\Sigma_\kappa^-)_{\kappa\in (-\infty,0)}$ is the whole past tight region $M^-$. Of course, similar arguments show that the support of the K-slicing $(\Sigma_\kappa^+)_{\kappa\in (-\infty,0)}$ is the whole future tight region $M^+$. This completes  the proof of Theorem~\ref{theo.main} in the locally anti-de Sitter case.


\section{K-slicings of hyperbolic ends}
\label{s.hyperbolic-ends}

In this section, we explain how Theorem~\ref{theo.ends} can be seen as a corollary of Theorem~\ref{theo.main} (more precisely, of the existence of K-slicings in non-elementary MGHC $\dS$-spacetimes). 
We will not go into too many details for several reasons:
\begin{itemize}
\item[--] Theorem~\ref{theo.ends} is not a new result; it was proved by F. Labourie some fifteen years ago (using techniques rather different from ours);
\item[--] the main tool which allows to translate a result concerning  non-elementary MGHC $\dS$-spacetimes into a result concerning hyperbolic ends is the so-called $\dS\leftrightarrow\HH^3$ duality. This duality is completely similar to the $\AdS\leftrightarrow\AdS$ duality described in the previous section.
\item[--] this duality was already observed and used by Mess in \cite[\S~6]{Mes} to prove that locally de Sitter MGHC spacetimes admit K-slicings using Labourie's result. Here we adopt the reverse point of view: we explain how to deduce Labourie's result from the existence of K-slicings on locally de Sitter MGHC spacetimes.
\end{itemize}

\subsection{$\dS\leftrightarrow\HH^3$ duality}

\subsubsection{Duality between points and totally geodesic planes}
\label{sss.point-plane-dS-H}
Consider the quadratic form $Q_{1,3}=-x_1^2+x_2^2+x_3^2+x_4^2$ on $\RR^4$. Recall that $\dS$ is the hyperboloid $\{Q_{1,3}=1\}$ endowed with the Lorentzian metric induced by $Q_{1,3}$. Also recall the each of the two sheets of the hyperboloid  $\{Q_{1,3}=-1\}$ is a copy of the $3$-dimensional hyperbolic space; we denote them by $H^3_-$ and $H^3_+$. For every set $A\subset\RR^4$, we denote by $A^\perp$ the $Q_{2,2}$-orthogonal of $A$ in $\RR^4$. We choose an orientation on $\RR^4$. 

Let $x$ be a point in $\dS\subset \RR^4$. Then $x^\perp$ is an hyperplane in $\RR^4$, and the restriction of $Q_{1,3}$ to $x^\perp$ has signature $(+,+,+)$.  It follows that $x^\perp\cap H_3^+$ is a totally geodesic plane in $H_3^+$. We denote this totally geodesic plane by $x^*$. The orientation of $\RR^4$ and the choice of $x$ (rather than $-x$) defines an orientation on $x^\perp$, and subsequently on $x^*$.  Conversely, let $P$ be an oriented totally geodesic plane in $H^3_+$. Then $P^\perp$ is a oriented line in $\RR^4$, and the restriction of $Q_{1,3}$ to $P^\perp$ is negative definite. It follows that $P^\perp$ intersects $\dS$ at two (antipodal) points, and the orientation on $P^\perp$ allows to distinguish choose one of these two points that we denote by $P^*$. Clearly, for every point $x\in\dS$ and every totally geodesic plane $P\subset H^3_+$, one has $(x^*)^*=x$ and $(P^*)^*=P$. This defines a duality between the points of $\dS$ and the totally geodesic planes in $H^3_+\simeq\HH^3$.

Now, let $x$ be a point in $H^3_+$. One can check that $x^\perp\cap\dS$ is a spacelike totally geodesic space in $\dS$; we denote this totally geodesic plane by $x^*$. Conversely, let $P$ be a spacelike totally geodesic plane in $\dS$. The intersection $P^\perp\cap H^3_+$ is reduced to a point, that we denote by $P^*$. Clearly,  for every point $x\in H^3_+$ and every spacelike totally geodesic plane $P\subset\dS$, one has $(x^*)^*=x$ and $(P^*)^*=P$. This defines a duality between the points of $H^3_+\simeq\HH^3$ and the spacelike totally geodesic planes in $\dS$.

\subsubsection{Duality between convex surfaces in $\dS$ and convex surfaces in $H^3_+$.}

Let $S$ be a  strictly convex spacelike surface in $\dS$. Using exactly the same construction as in \S\ref{sss.surfaces-ads} (but now using the duality between points and planes 
described in \S\ref{sss.point-plane-dS-H} instead of those described in \S\ref{sss.point-plane-ads}), one can to define a strictly convex $S^*\subset H_3^+$ such that the support planes of $S$ (resp. $S^*$) are the duals of the points of $S^*$ (resp. $S$).  Conversely, given  a strictly convex surface in $H^3_+$, one can define a strictly convex spacelike surface in $\dS$ such that the support planes of $S$ (resp. $S^*$ are the duals of the points of $S^*$ (resp. $S$). Then, for every strictly convex (resp. strictly convex spacelike) surface $S$ in $H^3_+$ (resp. $\dS$), one has $(S^*)*=S$. Moreover, the same arguments as in the proof of Proposition~\ref{p.curvature-dual} allow to prove:

\begin{proposition}
Let $S$ be a strictly convex (resp. strictly convex spacelike) surface in $H^3_+$ (resp. $\dS$). If $S$ has constant K-curvature $\kappa$, then $S^*$ has constant curvature $-\kappa^{-1}$.
\end{proposition}

\subsubsection{Duality between MGHC $\dS$-spacetimes and hyperbolic ends}

\begin{definition}
A (geometrically finite) \emph{hyperbolic end} is a  hyperbolic $3$-manifold $(M,g)$ such that:
\begin{itemize}
\item[--] $M$ is homeomorphic to $\Sigma \times (0, +\infty)$ where $\Sigma$ is a closed surface, 
\item[--] if $(\overline{M},\bar g)$ is the metric completion of $M$, and $\bar{\Sigma} = \overline{M} \setminus M$, then $(\bar{\Sigma},\bar g)$ is a hyperbolic surface homeomorphic to $\Sigma$,
\item[--] $(\bar{\Sigma},\bar g)$ is concave, i.e. there is no geodesic in $M$ connecting two elements
of $\bar{\Sigma}$.
\item[--] $(\bar{\Sigma},\bar g)$ is a pleated surface.
\end{itemize}
\end{definition}

Let $(M,g)$ be a hyperbolic end. It admits a conformal boundary homeomorphic to $\Sigma$. This conformal
boundary is naturally endowed 
with a M\"obius structure. As explained in \S\ref{ss.general-case}, every compact surface with negative Euler caracteristic endowed with a M\"obius structure defines a non-elementary MGHC $\dS$-spacetime. We shall denote by $(M^*,g^*)$ the 
non-elementary MGHC $\dS$-spacetime associated to $(\bar{\Sigma},\bar{g})$. We say that $(M^*,g^*)$ is the dual of the hyperbolic end $(M,g)$.

Conversely, for every non-elementary MGHC $\dS$-spacetime $(M,g)$, one can define a geometrically finite hyperbolic end $(M^*,g^*)$. We say that $(M^*,g^*)$ is the dual of $(M,g)$.

\subsubsection{Duality between convex surfaces in MGHC $\dS$-spacetimes and convex surfaces in hyperbolic ends}

Consider a non-elementary MGHC $\dS$-spacetime $(M,g)$ and the hyperbolic $(M^*,g^*)$ which is dual to $(M,g)$.

Let $\Sigma$ be a strictly convex Cauchy surface in $M$. Then $\Sigma$ lifts to a strictly convex immersed spacelike surface $S\subset\dS$. The dual surface $S^*$ is a strictly convex immersed surface in $H^3_+$. It can be checked that $S^*$ projects to a compact (strictly convex) surface $\Sigma^*$ in $M^*$. 

Conversely, if $\Sigma$ is a strictly convex surface in $M^*$, one can lift $\Sigma$ to a strictly convex surface $S$ in $H^3_+$, consider the dual $S^*$ of $S$ , and project $S^*$ to a strictly convex Cauchy surface $\Sigma^*$ in $((M^*)^*,(g^*)^*)=(M,g)$.

For every strictly convex Cauchy (resp. compact) surface $\Sigma$ in $(M,g)$ (resp. in $(M^*,g^*)$), one has $(\Sigma^*)^*=\Sigma$. Moreover, if $\Sigma$ has constant K-curvature $a$, then $\Sigma^*$ has constant K-curvature $-1/a$.

\subsection{Sketch of proof of Theorem~\ref{theo.ends}}

${}$

Consider a hyperbolic end $(M,g)$. Consider the future complete non-elementary MGHC $\dS$-spacetimes  $(M^*,g^*)$ dual to the hyperbolic end $(M,g)$. According to Theorem~\ref{theo.main}, $(M^*,g^*)$ admits a K-time $\kappa^*:M^*\to (-\infty,-1)$. For every $a\in (-\infty,0)$, the level set $\Sigma^*_a:=(\kappa^*)^{-1}(a)$ is a strictly convex Cauchy surface in $M^*$ with constant K-curvature equal to $a$. Then $\Sigma_a:=(\Sigma^*_a)^*$ is a strictly convex compact surface in $M$ with constant K-curvature $-1/a$. Then $\{\Sigma_a\}_{a\in ]-\infty,-1[}$ is a family of compact surfaces with constant K-curvature in $M$. Using the fact that $\{\Sigma^*_a\}_{a\in ]-\infty,-1[}$ is a trivial foliation
of $M^*$, it is quite easy to prove that $\{\Sigma_a\}_{a\in ]-\infty,-1[}$ is a trivial foliation of $M$. Theorem~\ref{theo.ends} follows.


\section{Surfaces with prescribed K-curvature}
\label{s.prescribed}

In this section, we prove Theorem~\ref{cor.prescri} and Corollary~\ref{coro.prescri}. This theorem will follow from : 
\begin{itemize}
\item[--]  the existence K-times on MGHC spacetimes with constant curvature (Theorem~\ref{theo.main}) which will provide us with a pair of barriers,
\item[--] a generalisation of Theorem~\ref{t.barriers} (Theorem~\ref{t.barriers-prescribed} below),  which asserts that surfaces with prescribed K-curvature exist as soon as barriers exist. 
\end{itemize}

\begin{definition}
Let $(M,g)$ be a 3-dimensional globally hyperbolic spacetime with compact Cauchy surfaces. Let $f:M\to (-\infty,0)$ be a smooth function. A \emph{pair of $f$-barriers} is a pair of disjoint strictly convex Cauchy surfaces $\Sigma^-,\Sigma^+$ in $M$, such that: 
\begin{enumerate}
\item[a.] $\Sigma^-$ is in the past of $\Sigma^+$, 
\item[b.] $\kappa^{\Sigma^-}(x)\leq f(x)$ for every $x\in\Sigma^-$, 
\item[c.] $\kappa^{\Sigma^+}(x)\geq f(x)$ for every $x\in\Sigma^+$. 
\end{enumerate}
\end{definition}

\begin{theorem}[Gerhardt, \cite{Ger0}]
\label{t.barriers-prescribed}
Let $(M,g)$ be a 3-dimensional spatially compact globally hyperbolic spacetime, and $f:M\to ]-\infty,0[$ be   a smooth function.  Assume that $M$ admits a pair of $f$-barriers $(\Sigma^-,\Sigma^+)$. Then $M$ admits a strictly convex Cauchy surface $\Sigma$ such that  $\kappa^{\Sigma}(x)= f(x)$ for all~$x\in\Sigma.$
\end{theorem}

\begin{remark}
\label{r.sign-convention-Gerhardt-2}
We recall that the sign conventions of Gerhardt are different from ours (see Remark~\ref{r.sign-convention-Gerhardt}). This is the reason why the function $f$ is required to be strictly positive in~\cite{Ger0}, whereas it is required to be strictly negative in the statement above.
\end{remark}

\begin{proof}[Proof of Theorem~\ref{cor.prescri}]
Consider a 3-dimensional non-elementary MGHC spacetime $(M,g)$ with constant curvature $\Lambda$. Recall that in the $\Lambda\geq 0$ case we assume that $(M,g)$ is future complete (see Remark~\ref{rk.futurepast}). Let $f:M_0\to \RR$ be a smooth function such that the range of $f$ is contained in a compact interval $[a,b] \subset ]-\infty\,,\,\min(0,-\Lambda)[$. Define $M_0$ as follow~: $M_0=M$ is $\Lambda\geq 0$ (flat and locally de Sitter case), and $M_0$ is the past of the convex core of $M$ if $\Lambda<0$ (anti de Sitter case).  By Theorem~\ref{theo.main}, there exists a K-time $\kappa:M_0\to ]-\infty, \min(-\Lambda,0)[$. Consider the level sets of $\kappa$:
$$\Sigma^-:=\kappa^{-1}(a)\quad\mbox{and}\quad\Sigma^+:=\kappa^{-1}(b).$$
By definition of a K-time, $\Sigma^-,\Sigma^+$ are two strictly convex Cauchy surfaces with constant K-curvarture respectively equal to $a$ and $b$, and $\Sigma^-$ is in the past of $\Sigma^+$. In particular, $(\Sigma^-,\Sigma^+)$ is a pair of $f$-barriers. Therefore Theorem~\ref{t.barriers-prescribed} applies in our situation and provides us with a strictly convex Cauchy surface $\Sigma$ such that $\kappa^{\Sigma}(x)=f(x)$ for every $x$ in $\Sigma$.
\end{proof}

\begin{proof}[Proof of Corollary~\ref{coro.prescri}]
Write $M$ as a topological product $\Sigma_0\times\RR$, consider the function $f:M\to\RR$ defined by $f(x,t)=f_0(x)$ where $x\in\Sigma_0$ and $t\in\RR$, and apply Theorem~\ref{cor.prescri} to the function~$f$. 
\end{proof}


\section{The Minkowski problem}
\label{s.Minkowski}

The purpose of this section is to prove Theorem~\ref{Minkowski}. As for the proof of Theorem~\ref{cor.prescri}, we will use Theorem~\ref{theo.main} to get a pair of barriers, and then, a theorem of Gerhardt which asserts that surfaces with prescibed K-curvature exist as soon as barriers exist. 
As a preliminary step, we will need to translate Theorem~\ref{Minkowski} into a statement concerning a function defined on the unit tangent bundle of a non-elementary MGHC flat spacetime. 

\bigskip

We begin by stating Gerhardt's result:

\begin{theorem}[Gerhardt, see~\cite{Ger1}; see also Remarks~\ref{r.sign-convention-Gerhardt} and~\ref{r.sign-convention-Gerhardt-2}]
\label{t.barrieres-Minkowski}
Let $(M,g)$ be a spatially compact globally hyperbolic spacetime. Let $\Omega$ be an open subset of $M$ bounded by two disjoint strictly convex Cauchy surfaces $\Sigma^-$ and $\Sigma^+$, where $\Sigma^-$ is assumed to be in the past of $\Sigma^+$. 
Denote by $T^{-1} \Omega$ the bundle of future-oriented unit timelike tangent vectors over $\Omega$:
$$
T^{-1} \Omega = \{(x,\nu)\in TM \mid x\in \Omega\,,\,\nu\mbox{ is future-oriented  and }g(\nu,\nu)=-1\}.$$ 
Consider a smooth function $\Phi : T^{-1} \Omega\to\RR$ with the two following properties:
\begin{enumerate}
\item[(i)]  there is a negative constant $c_1$ such that, for all $(x,\nu)\in T^{-1}\Omega$,
 $$\Phi(x,\nu)\leq c_1<0.$$ 
\item[(ii)] there are some constants $c_2,c_3\in\RR$ such that, for all $(x,\nu)\in T^{-1} \Omega$,
$$|||d_x\Phi(x,\nu)|||\leq c_2 \left(1+\|\nu\|^2\right) \quad\mbox{and}\quad |||d_\nu\Phi(x,\nu)|||\leq c_3 \left(1+\|\nu\|\right)$$
where $d_x\Phi$ and $d_\nu\Phi$ are the derivatives of $\Phi$ with respect to $x$ and $\nu$, where $|||.|||$ is the operator norm associated to an arbitrary auxiliary Riemannian metric on $M$, and $\|\nu\|$ is the norm of $\nu$ for this Riemannian metric\footnote{More precisely, we choose a Riemannian metric $\eta$ on $M$, we observe that for every $(x,\nu)\in TM$ there is a canonical identification between $T_{(x,\nu)} (T_xM)\simeq T_x M$, and we set: 
\begin{eqnarray*}
\|v\| & := & \sqrt{\eta(v,v)}\\
|||d_x\Phi(x,\nu)||| & := & \sup_{v\in T_x M} \frac{|d_x\Phi(x,\nu).v|}{\|v\|}\\
|||d_\nu\Phi(x,\nu)||| & := & \sup_{v\in T_{(x,\nu)}(T^{-1}_x\Omega)\subset T_{(x,\nu)}(T_xM)\simeq T_x M} \frac{|d_\nu\Phi(x,\nu).v|}{\|v\|}
\end{eqnarray*}}.
\end{enumerate}
Assume that $(\Sigma^-,\Sigma^+)$ is a pair of $\Phi$-barriers, that is:
\begin{enumerate}
\item[--] for every $x\in \Sigma^-$, one has $\kappa^{\Sigma^-}(x)\leq \Phi(x,\nu^{\Sigma^-}(x))$, 
\item[--] for every $x\in \Sigma^+$, one has $\kappa^{\Sigma^+}(x)\geq \Phi(x,\nu^{\Sigma^+}(x))$. 
\end{enumerate}
Then there exists a strictly convex Cauchy surface $\Sigma$ in $M$ such that,  for every $x\in \Sigma$, one has
$$\kappa^{\Sigma}(x)=\Phi\left(x,\nu^\Sigma(x)\right).$$ 
\end{theorem}

In order to show that Theorem~\ref{t.barrieres-Minkowski} applies in our situation, we shall need the following elementary lemma:

\begin{lemma}  
\label{growth}
Let $\Gamma$ is a co-compact Fuchsian group in $\mbox{SO}(1,2)$, and $f: \HH^2 \to \RR$ be a $\Gamma$-invariant function. See $\HH^2$ as the upper sheet of the hyperboloid $x_1^2+x_2^2-x_3^2=-1$ in $\RR^3$, and consider an arbitrary euclidian  norm $\| . \|$ on $\RR^3$. 
Then there exists a constant $c$ such that, for every $\mathrm{n}\in\HH^2$, 
$$|||df(\mathrm{\nu})|||\leq c$$
where $|||.|||$ is the operator norm\footnote{More precisely, $\displaystyle |||df(\nu)|||=\sup_{v\in T_\nu\HH^2\subset T_\nu\RR^3\simeq\RR^3} \frac{|df(\nu).v|}{\|v\|}$.} associated to the euclidian norm $\|.\|$.
\end{lemma}

\begin{proof}
Let $\|.\|_{Lor}:= \sqrt{x_1^2 + x_2^2-x_3^2}$ be the Lorentzian pseudo-norm on $\RR^3$ which is preserved by $\mbox{SO}(1,2)$, and $\langle.,.\rangle_{Lor}$ be the associated Lorentzian pseudo-scalar product. 
We will prove the lemma for the Euclidean norm $\|.\|_{Euc}:=\sqrt{x_1^2 + x_2^2+x_3^2}$; the general case of an arbitrary Euclidean norm will follow, up to a change of the constant $c$.

We denote by $\nabla f$ the Lorentzian gradient of $f$ (i.e. the gradient with respect to the hyperbolic norm on $\HH^2$ induced by $\|.\|_{Lor}$). First observe that, for every $\nu\in\HH^2$, and every vector $v\in T_\nu\HH^2\subset \RR^3$, we have:
$$\Vert v \Vert_{Euc} \geq \Vert v \Vert_{Lor}$$
Now, observe that, for every $\nu\in\HH^2$, the restriction of $\langle.,.\rangle_{Lor}$ to the plane $T_\nu \HH^2$ is positive definite. Hence, for every $\nu\in\HH^2$ and every $v\in T_\nu\HH^2$,  the Cauchy Schwarz inequality for the restriction of  $\langle.,.\rangle_{Lor}$ to $T_\nu \HH^2$ yields:
$$| \langle (\nabla f)(\nu), v \rangle_{Lor} | \leq \Vert (\nabla f)(\nu) \Vert_{Lor} \; \Vert v \Vert_{Lor}.$$
As a consequence,  for every $\nu\in\HH^2$ and every $v\in T_\nu\HH^2$, we have:
\begin{eqnarray*}
| df(\nu).v | & =  & | \langle (\nabla f)(\nu), v \rangle_{Lor} | \\
            & \leq & \Vert (\nabla f)(\nu) \Vert_{Lor} \; \Vert v \Vert_{Lor} \\
            & \leq & \Vert (\nabla f)(\nu) \Vert_{Lor} \; \Vert v \Vert_{Euc}
            \end{eqnarray*}
But since $\nabla f(\nu)$ is $\Gamma$-equivariant, where $\Gamma$ is co-compact, its hyperbolic norm $\Vert \nabla f(\nu) \Vert_{Lor}$ is uniformly bounded from above. The lemma follows.
\end{proof}

Now we are in a position to prove Theorem~\ref{Minkowski}. We consider a co-compact Fuchsian subgroup $\Gamma$ in $\mbox{SO}(1,2)$, a subgroup $\mathbf{\Gamma}$ of $\mbox{SO}(1,2)\ltimes\RR^3$ which projects bijectively on $\Gamma$, and a $\Gamma$-invariant function $f:\HH^2\to\RR$. 

Theorem~\ref{th.Mess-flat} states that there exists a future-complete regular domain $E=E(\mathbf{\Gamma})$ in $\Min$ such that the action of $\mathbf{\Gamma}$ on $E$ is free and properly discontinuous, and such that $M:=\mathbf{\Gamma}\backslash E$ is a non-elementary future-complete MGHC flat spacetime. 
We consider the bundle 
$$
T^{-1} \Min =  \{(x,\nu) \in T\Min \mid \nu\mbox{ future-oriented, }\langle\nu,\nu\rangle=-1\},
$$ 
and we canonically identify $T^{-1}\Min$ with $\Min\times\HH^2$. Then we consider the function $F:T^{-1}\Min\simeq \Min\times\HH^2\to (-\infty,0)$ defined by 
$$F(x,\nu):=f(\nu)$$ 
The group $\mathbf{\Gamma}$ acts on $T^{-1}\Min$ by 
$$\mathrm{g}.(x,\nu)=(\mathbf{g}.(x),d\mathbf{g}.x(\nu))=(\mathbf{g}.(x), \mathrm{g}(\nu))$$
(where $\mathbf{g}\in\mathbf{\Gamma}\subset\mbox{SO}(1,2)\ltimes\RR^3$ and $g\in\Gamma\subset\mbox{SO}(1,2)$ is its linear part) and 
$$\mathbf{\Gamma}\backslash (T^{-1}\Min) = T^{-1} M =  \{(x,\nu) \in TM \mid \nu\mbox{ future-oriented, }\langle\nu,\nu\rangle=-1\}.$$ 
Since the function $f$ is $\Gamma$-invariant, the function $F$ is $\mathbf{\Gamma}$-invariant. Hence $F$ induces a function $\Phi:T^{-1} M\to (-\infty,0)$. 

Consider a Cauchy surface $\Sigma$ in $M=\mathbf{\Gamma}\backslash E$. Then $\Sigma$ lifts to a $\mathbf{\Gamma}$-invariant spacelike surface $S$ in $E\subset\Min$.  
Let $\nu^S$, $\nu^\Sigma$, $\kappa^S$, $\kappa^\Sigma$ be respectively the Gauss map of $S$, the Gauss map of $\Sigma$, the K-curvature of $S$ and the K-curvature of $\Sigma$. We have the following equivalences 
$$\begin{array}{lrcll}
& f(\nu) & = & \kappa^S\circ (\nu^S)^{-1} (\nu) & \mbox{for every }\nu\in \mbox{range}(\nu^S)\subset\HH^2 \\ 
\Longleftrightarrow & F(x,\nu^S(x)) & = & \kappa^S(x) &  \mbox{for every }x\in S\\
\Longleftrightarrow & \Phi(x,\nu^\Sigma(x)) & = & \kappa^\Sigma(x) &  \mbox{for every }x\in\Sigma
\end{array}$$
Therefore, in order to prove Theorem~\ref{Minkowski}, it is enough to find a strictly convex Cauchy surface $\Sigma$ in $M$ such that 
\begin{equation}
\label{e.Minkowski}
\Phi\left(x,\nu^\Sigma(x)\right)=\kappa^\Sigma(x)\quad\mbox{for every }x\in \Sigma
\end{equation}

We want to use Theorem~\ref{t.barrieres-Minkowski} to get such a  Cauchy surface $\Sigma$. So, we have to prove that the function $\Phi$ satisfies the conditions~(i) and (ii) in the statement of Theorem~\ref{t.barrieres-Minkowski}, and we have to find a pair of $\Phi$-barriers. 

The function $f:\HH^2\to (-\infty,0)$ is continuous and $\Gamma$-invariant. The group $\Gamma$ is co-compact. Hence, the range of $f$ is a compact interval $[a,b]\subset (0,+\infty)$. The range of the function $\Phi:T^{-1} M\to (-\infty,0)$ is the same compact interval $[a,b]$. In particular, condition~(i) of Theorem~\ref{t.barrieres-Minkowski} is satisfied. 

By Theorem~\ref{theo.main}, the spacetime $M$ admits a K-time $\kappa:M\to (-\infty,0)$. We consider the Cauchy surfaces $\Sigma^-:=\kappa^{-1}(a)$ and $\Sigma^+:=\kappa^{-1}(b)$. By definition of a K-time, the Cauchy surface $\Sigma^-$ is strictly convex and has constant K-curvature $a$, the surface $\Sigma^+$ is strictly convex Cauchy and has constant K-curvature $b$, and $\Sigma^-$ is in the past of $\Sigma^+$. In particular, $(\Sigma^-,\Sigma^+)$ is a pair of $\Phi$-barriers in the sense defined in the statement of Theorem~\ref{t.barrieres-Minkowski}. 

We denote by  $\Omega$ the open subset of $M$ bounded by the Cauchy surfaces $\Sigma^-$ and $\Sigma^+$. Observe that $\Omega$ is relatively compact in $M$. We denote by $O$ the lift of $\Omega$ in $M$.

Since the function $F$ is independant of $x$, we have $d_x F(x,\nu)=0$ for every $(x,\nu)\in T^{-1}\Min$. Hence we have $d_x \Phi(x,\nu)=0$ for every $(x,\nu)\in T^{-1}\Min$. In  particular, the first inequality in  condition~(ii) of Theorem~\ref{t.barrieres-Minkowski} is satisfied. 

Let $\eta$ be a Riemannian metric on $M$. One can lift $\eta$ to a $\mathbf{\Gamma}$-invariant Riemannian metric $h$ on $E\subset\Min$. For every $(x,\nu)\in T^{-1}M$, if $(\tilde x,\tilde\nu)\in T^{-1}\Min\simeq\Min\times \HH^2$ is a lift of $(x,\nu)$, then 
\begin{equation}
\label{e.derivees}
d_\nu \Phi(x,\nu) = d_{\tilde\nu} F(\tilde x,\tilde \nu) = df(\tilde\nu)
\end{equation}
According to Lemma~\ref{growth}, for every $\wt x\in O$, there exists a constant $c_{\tilde x}$, such that, for every $\tilde\nu\in\HH^2$,
$$|||df(\tilde \nu)|||_{\tilde x}\leq c_{\bar x}$$ 
where $|||.|||_{\tilde x}$ is the operator norm associated to the euclidian norm $h_{\tilde x}$ on $\RR^3\simeq T_{\tilde x}\Min$. Now observe that $\sup_{\tilde\nu\in \HH^2}|||df(\tilde \nu)|||_{\bar x}$ depends in a bounded way of $\tilde x$ as far as $\tilde x$ stays in $O$; indeed the Riemannian metric $h$ and the derivative $df$ are $\mathbf{\Gamma}$-equivariant and $\mathbf{\Gamma}\backslash O=\Omega$ is relatively compact. Hence, there exists a constant $c$ such that, for every $\tilde x\in O$ and every $\tilde\nu\in\HH^2$,
$$|||df(\tilde \nu)|||_{\tilde x}\leq c.$$
Using equality~\eqref{e.derivees}, this implies that, for every $(x,\nu)\in T^{-1}\Omega$, 
$$|||d_\nu \Phi(x,\nu)|||\leq c$$
where $|||d_\nu \Phi(x,\nu)|||$ is the operator norm of $d_\nu \Phi(x,\nu)$ associated to the Riemannian metric $\eta$. In particular, the second inequality in condition (ii) of Theorem~\ref{t.barrieres-Minkowski} is satisfied.

So all the hypotheses of Theorem~\ref{t.barrieres-Minkowski} are satisfied. Hence this theorem provides us with a strictly convex Cauchy surface $\Sigma$ in $M$ satisfying~\eqref{e.Minkowski}. In order to achieve he proof of Theorem~\ref{Minkowski} the only remaining point is to prove the uniqueness of $\mathbf{\Gamma}$-solution for a given uniform lattice $\mathbf{\Gamma}$ of 
$\mbox{SO}(1,2)\ltimes\RR^3$.

Let $S_1$, $S_2$ be two such $\mathbf{\Gamma}$-invariant solutions. For any $t$, let $S^t_i$ be the surface obtained by pushing $S_i$ along the normal geodesics during the time $t$. Let $t_1$ be the minimal time such that $S^{t_1}_1$ lies in the future of $S_2$. Since $S_1$ and $S_2$ projects in the quotient as compact Cauchy surfaces, $t_1$ is well-defined and finite. Exchanging $S_1$ with $S_2$ if necessary, one can assume $t_1 \geq 0$. Then $S_2$ and $S_1^{t_1}$ are tangent at a common point $x$. This point is at lorentzian distance $t_1$ from a point $y$ in $S_1$ along a timelike geodesic orthogonal
to $S_1$ at $y$. This geodesic is also orthogonal to $S^{t_1}$, and thus to $S_2$, at $x$. Hence 
$\nu^{S_1}(y) = \nu^{S_2}(x)$. Since they are both solutions of the same Minkowski problem, 
$\kappa^{S_1}(y) = \kappa^{S_2}(x)$. But the K-curvature strictly increases by pushing along normal geodesics (Remark~\ref{rem.Gausscroit}): at one hand, the K-curvature of $S^{t_1}_1$ is bounded from above by $\kappa^{S_2}$ (since it lies in the future of $S_2$); on the other hand, if $t_1 > 0$, 
it should be strictly bigger than $\kappa^{S_1}(y)$. It follows that $t_1$ is $0$: 
$S_1$ lies in the future of $S_2$. In particular, $t_2 \geq 0$. Permuting the role of $S_2$ and $S_1$, we prove similarly that $S_2$ lies in the future of $S_1$. Hence $S_1 = S_2$. Theorem~\ref{Minkowski} follows.

 \end{document}